\documentclass[12pt]{article}

\usepackage[utf8]{inputenc}
\usepackage[T1]{fontenc}
\usepackage[american]{babel}
\usepackage{mathpazo}
\usepackage[margin=1in]{geometry}
\usepackage{scrextend}
\usepackage{setspace}
\usepackage{amsmath,amssymb,amsthm,mathtools}
\usepackage{enumitem}
\usepackage[round]{natbib}
\setcitestyle{aysep={}}
\usepackage{bibunits}
\usepackage{booktabs}
\usepackage{tabularx}
\usepackage{longtable}
\usepackage{pgfplots}
\pgfplotsset{compat=1.16}
\usetikzlibrary{arrows.meta}
\usepackage{microtype}
\usepackage{xcolor}
\usepackage{hyperref}
\hypersetup{colorlinks=true, linkcolor=blue!50!black, citecolor=blue!50!black, urlcolor=blue!50!black, pdftitle={Are Revealed Preferences All You Need for AI Alignment?}, pdfauthor={Joshua S. Gans and Richard Holden}}
\theoremstyle{plain}
\newtheorem{theorem}{Theorem}
\newtheorem{proposition}{Proposition}
\newtheorem{assumption}{Assumption}
\newtheorem{applemma}{Lemma}

\newtheorem{supplemma}{Lemma}[section]
\newtheorem{suppclaim}[supplemma]{Claim}

\newcommand{\R}{\mathbb{R}}
\newcommand{\BR}{\mathrm{BR}}
\newcommand{\one}{\mathbf{1}}

\newcommand{\E}{\mathbb{E}}
\newcommand{\cl}{\operatorname{cl}}
\DeclareMathOperator*{\argmax}{arg\,max}
\DeclareMathOperator{\spn}{span}
\DeclareMathOperator{\aff}{aff}
\DeclareMathOperator{\conv}{conv}

\setlist{itemsep=0.2em, topsep=0.4em}

\title{Are Revealed Preferences All You Need for AI Alignment?}
\author{Joshua S. Gans\thanks{Rotman School of Management, University of Toronto, and NBER, \href{mailto:joshua.gans@utoronto.ca}{joshua.gans@utoronto.ca}.} \and Richard Holden\thanks{UNSW Business School, \href{mailto:richard.holden@unsw.edu.au}{richard.holden@unsw.edu.au}. Thanks to Jacob Dorn and seminar participants at Cornell for helpful comments. Thanks to Refine.ink, Claude Opus 5.5 and Gemini Argon 4 for valuable research and the SSHRC for funding. Responsibility for all errors is our own. A video explainer of this paper is available here: \url{https://youtu.be/vM-Nsd4mG_g}.}}
\date{November 13, 2026}

\begin{document}

\maketitle

\begin{abstract}
\setstretch{1}
\noindent AI agents learn what principals want from what principals choose. We offer a criterion for when a principal's decisions, at every belief an AI might hold, are all the AI needs. For acting, they are. Cautious actions, such as deferring to a person, typically divide the principal's actions into blocks, and for choosing costless evidence, decisions are enough exactly when the same evidence is best in every block. Otherwise, the AI acts as its principal would yet can gather the wrong evidence, and greater capability can turn that error into incorrect actions. Choices between two mistakes reveal what decisions miss.

\bigskip
{\raggedright\noindent\textit{Keywords:} artificial intelligence, alignment, revealed preference, identification, value of information, delegation, incomplete contracts.

\par}
\end{abstract}

\newpage

\section{Introduction}\label{sec:intro}

AI agents are often charged with gathering their own evidence. An agent that maintains a firm's servers decides which files to inspect before deciding what to do with them. A coding agent decides which tests to run before committing a change. A research assistant decides which sources to read before answering a question. Their principals don't stipulate what evidence these agents gather. Instead, the principal's preferences are intended to be learned before deployment. These may come from direct contingent preferences specified by principals or, for anything beyond the very simple, extracted from records of human choices. That is demonstrations of what a person would do, and comparisons that record which of two responses a person prefers \citep{christiano2017,ouyang2022}, what economists call \textit{revealed preferences}. A prominent view in AI safety holds that this is as it should be. In the formulation of \citet{russell2019}, an AI's only objective should be to realize human preferences, and it should learn those preferences from human behavior.

Revealed preference looks equal to the task. A principal that shows what it would do at every belief an AI might hold, and not only when the facts are known, reveals more than how it ranks outcomes, because choices under uncertainty reveal how strongly one outcome is preferred to another \citep{vnm1947}. And evidence is valuable only insofar as it changes decisions \citep{blackwell1953}, so an AI agent that takes the action its principal would take at every belief might be expected to seek the evidence its principal would seek. Our question is as follows: are the principal's decisions, at every belief the AI might hold, all that the AI needs?

For acting, as Blackwell reasoning anticipates, they are. For choosing evidence, it depends. Specifically, the answer depends on the decisions and on the evidence that the AI can gather, and we offer a criterion that takes both as given. The criterion turns on actions that \emph{separate} others. An action separates two others when the principal's choice cannot pass from one to the other, as its belief changes, unless the separating action is optimal on the way, just as ``I~don't know'' lies between ``no'' and ``yes.'' Someone who can say ``I~don't know'' never has to guess between yes and no, and so never reveals how an incorrect yes compares with an incorrect no. The principal's decisions then reveal the stakes on each side of the separating action only up to a scale for that side, and nothing about how the scales compare.\footnote{When no action separates others, decisions reveal the principal's payoffs up to the units in which they are measured. The units matter only when evidence has a cost that the AI must weigh against its value, a case taken up in Section~\ref{sec:test}. Choices between two mistakes, which supply the comparison of scales that decisions miss (Section~\ref{sec:records}), leave the units open too.} In general, the principal's actions divide into \emph{blocks}, the largest sets of actions that no action separates, and each block has a scale of its own. The actions that separate are typically cautious ones: declining to answer \citep{chow1970,kalai2025}, deferring to a person \citep{hadfieldmenell2017,mozannar2020}, or, as we will explore below, quarantining a file. These options make an AI agent safe to deploy, and a principal takes them wherever the choice between the other actions is close, so its decisions never show how one mistake compares with another.

Whether the missing comparison matters depends on the evidence. The criterion is as follows. The principal's decisions are all that the AI needs for a choice among sources of evidence if and only if the source that is best when only the actions of one block are available is also best when only the actions of each other block are available. An AI whose objective reproduces its principal's decisions values evidence within each block as the principal does, up to the scale of the block. So a designer can check the criterion with the AI's objective alone, without knowing the principal's payoffs. When the criterion fails, the gap does not show in what the AI does with its evidence. The AI takes the action its principal would take at every belief, so no test of its actions, however thorough, detects an incorrect scale. But it can run the wrong tests, read the wrong sources, and ask too many questions or too few. And when the AI becomes capable enough to reach beliefs at which blocks merge, the wrong scale produces wrong actions as well, and the principal can be worse off than it was with the less capable AI. What decisions leave out is nevertheless small. It is one number fewer than the number of blocks, and a principal that says where it would be indifferent between two mistakes, neither of which it would make, supplies such a number.

\subsection{An example}\label{sec:example}

Consider an AI that maintains a firm's servers and finds a suspicious file, which is either malicious or legitimate. The AI can delete the file, leave it, or quarantine it, which moves it somewhere safe and flags it for a person to review. Deleting a legitimate file costs the firm $10$, as does leaving a malicious one, and quarantining costs $1$ either way, after which the file is handled correctly. If $p$ is the AI's belief that the file is malicious, deleting costs $10(1-p)$ in expectation, leaving costs $10p$ and quarantining costs $1$. A firm that held the AI's belief would therefore leave the file when $p\leq t_1=1/10$, delete it when $p\geq t_2=9/10$ and quarantine it in between. We call a pair of actions between which the firm switches a \emph{margin}. Leaving and quarantining form the lower margin, and quarantining and deleting form the upper one.

Many objectives make exactly these decisions. On the upper margin, the firm gains $1$ by deleting a malicious file rather than quarantining it and loses $9$ by deleting a legitimate one, and it switches at the belief at which the two balance. An objective that counted the gain as $3$ and the loss as $27$ would switch at the same belief. We call the objective that does so, and otherwise agrees with the firm, the \emph{tripled objective}.\footnote{It keeps the firm's payoffs for leaving and quarantining and values deleting at $2$ when the file is malicious and at $-28$ when it is legitimate. Under it, leaving and deleting are equally good only at $p=7/10$, where quarantining is better than both, so it makes the firm's decisions at every belief.} No record of the firm's decisions, however complete, tells it apart from the firm's own payoffs.

Now suppose that, before it acts, the AI can run one of two tests. A test costs nothing and changes nothing but the AI's belief, so the firm's ranking of tests is implied by its payoffs from leaving, quarantining and deleting, and it has no separate preferences over tests to convey.\footnote{Gathering evidence can also have consequences of its own, as when probing a system damages it. It is then an action whose stakes must be learned like those of any other action (Section~\ref{sec:capability}).} A \emph{clearing test} sometimes proves that a legitimate file is legitimate: it reports ``clear'' with probability $2/5$ when the file is legitimate and never when it is malicious. A \emph{confirming test} sometimes proves that a malicious file is malicious: it reports ``confirm'' with probability $1/5$ when the file is malicious and never when it is legitimate. Let $p=1/2$. The clearing test reports ``clear'' with probability $1/5$, and the firm then leaves the file instead of quarantining it, which saves $1$. The confirming test reports ``confirm'' with probability $1/10$, and the firm then deletes the file instead of quarantining it, which also saves $1$. Without a report, the belief moves only to $5/8$ or to $4/9$ and the firm still quarantines. The firm therefore values the clearing test at $1/5$ and the confirming test at $1/10$, and it would run the clearing test. The tripled objective agrees about the clearing test, but it counts deleting a malicious file as worth $3$ more than quarantining it, so it values the confirming test at $3/10$ and runs that one. Whichever test the AI runs, the action it then takes is exactly the one that the firm would take.

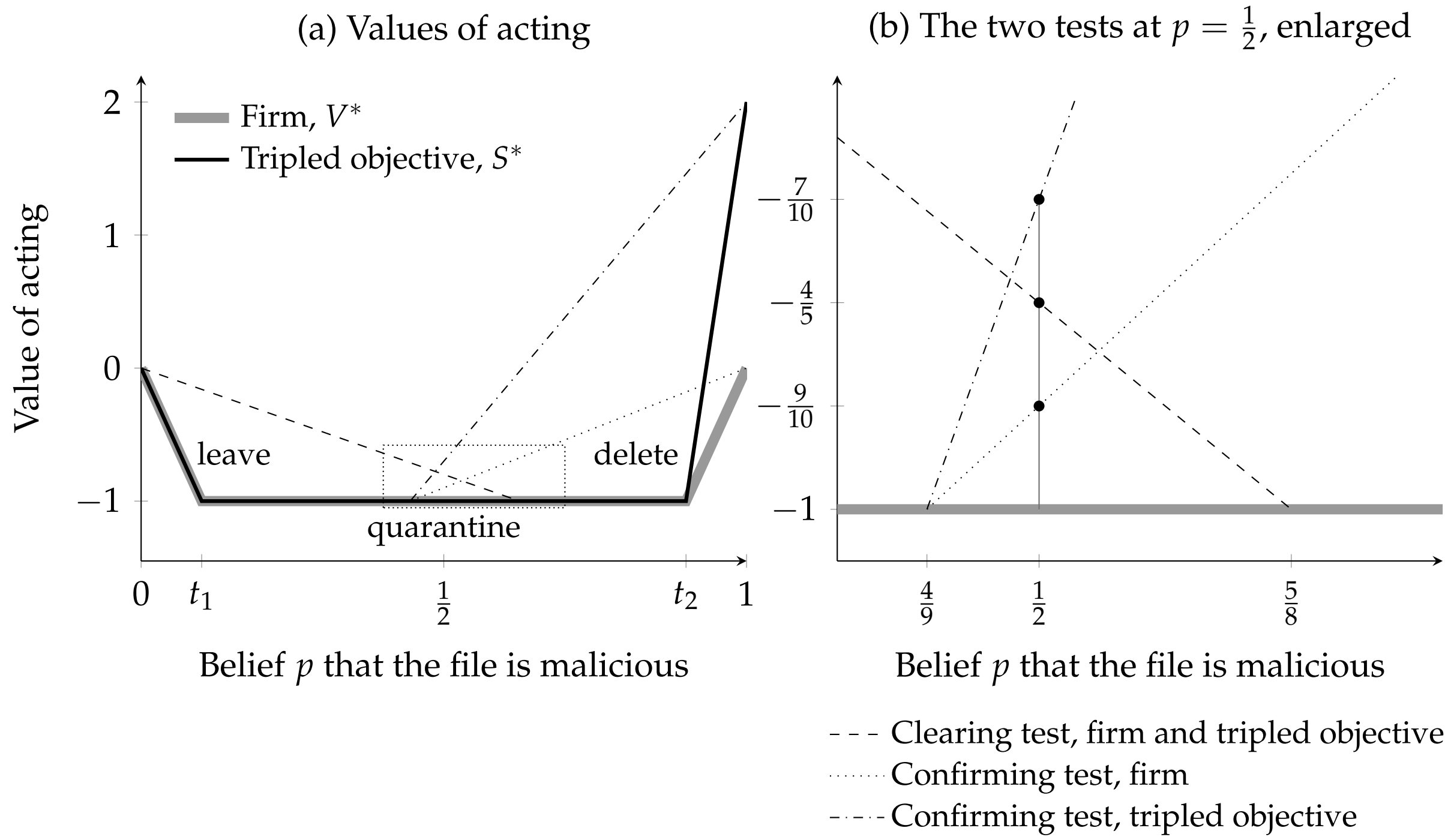
\begin{figure}[t]
\centering
\begin{tikzpicture}
\begin{axis}[
  name=va, width=0.5\textwidth, height=0.42\textwidth,
  xmin=0, xmax=1, ymin=-1.45, ymax=2.2,
  xtick={0,0.1,0.5,0.9,1},
  xticklabels={$0$,$t_1$,$\frac12$,$t_2$,$1$},
  ytick={-1,0,1,2},
  tick label style={font=\small}, label style={font=\small},
  xlabel={Belief $p$ that the file is malicious},
  ylabel={Value of acting},
  axis x line=bottom, axis y line=left,
  title={\small (a) Values of acting}, title style={yshift=-2pt},
  legend style={draw=none, fill=none, font=\footnotesize, at={(0.03,0.97)}, anchor=north west},
  legend cell align=left]
\addplot[black!40, line width=3.2pt] coordinates {(0,0) (0.1,-1) (0.9,-1) (1,0)};
\addlegendentry{Firm, $V^*$}
\addplot[black, line width=1.2pt] coordinates {(0,0) (0.1,-1) (0.9,-1) (1,2)};
\addlegendentry{Tripled objective, $S^*$}
\addplot[black, thin, dashed, forget plot] coordinates {(0,0) (0.625,-1)};
\addplot[black, thin, dotted, forget plot] coordinates {(0.444444,-1) (1,0)};
\addplot[black, thin, dashdotted, forget plot] coordinates {(0.444444,-1) (1,2)};
\draw[densely dotted] (axis cs:0.4,-1.05) rectangle (axis cs:0.7,-0.58);
\node[anchor=west, font=\footnotesize, inner sep=1pt] at (axis cs:0.086,-0.64) {leave};
\node[font=\footnotesize, inner sep=1pt] at (axis cs:0.5,-1.22) {quarantine};
\node[anchor=east, font=\footnotesize, inner sep=1pt] at (axis cs:0.894,-0.64) {delete};
\end{axis}
\begin{axis}[
  at={(va.east)}, anchor=west, xshift=1.0cm,
  width=0.5\textwidth, height=0.42\textwidth,
  xmin=0.4, xmax=0.7, ymin=-1.05, ymax=-0.58,
  xtick={0.444444,0.5,0.625},
  xticklabels={$\frac49$,$\frac12$,$\frac58$},
  ytick={-1,-0.9,-0.8,-0.7},
  yticklabels={$-1$,$-\frac{9}{10}$,$-\frac{4}{5}$,$-\frac{7}{10}$},
  tick label style={font=\small}, label style={font=\small},
  xlabel={Belief $p$ that the file is malicious},
  axis x line=bottom, axis y line=left,
  title={\small (b) The two tests at $p=\frac12$, enlarged}, title style={yshift=-2pt},
  legend style={draw=none, font=\footnotesize, at={(0.5,-0.30)}, anchor=north, legend columns=1},
  legend cell align=left]
\addplot[black!40, line width=3.2pt, forget plot] coordinates {(0.4,-1) (0.7,-1)};
\addplot[black, thin, dashed, domain=0.4:0.625, samples=2] {-1.6*x};
\addlegendentry{Clearing test, firm and tripled objective}
\addplot[black, thin, dotted, domain=0.444444:0.7, samples=2] {-1+(x-0.444444)*1.8};
\addlegendentry{Confirming test, firm}
\addplot[black, thin, dashdotted, domain=0.444444:0.5185, samples=2] {-1+(x-0.444444)*5.4};
\addlegendentry{Confirming test, tripled objective}
\draw[black!55, thin] (axis cs:0.5,-1) -- (axis cs:0.5,-0.7);
\addplot[only marks, mark=*, mark size=1.5pt, forget plot] coordinates {(0.5,-0.8) (0.5,-0.9) (0.5,-0.7)};
\end{axis}
\end{tikzpicture}
\caption{What decisions reveal in the file example. \footnotesize (a) The firm's value of acting $V^*$ and the tripled objective's $S^*$ have kinks at the same beliefs, $t_1=1/10$ and $t_2=9/10$, so they make the same decisions, which label the three segments, but the upper kink is three times as sharp under the tripled objective. A test at $p=1/2$ is worth the height at $1/2$ of the chord that joins the values of acting at its posteriors, measured above the value of acting there, which is $-1$. The thin lines are the chords of the two tests. (b) The dotted rectangle in panel (a), enlarged. Both objectives value the clearing test at $1/5$, but the firm values the confirming test at $1/10$ and the tripled objective at $3/10$.}
\label{fig:example}
\end{figure}

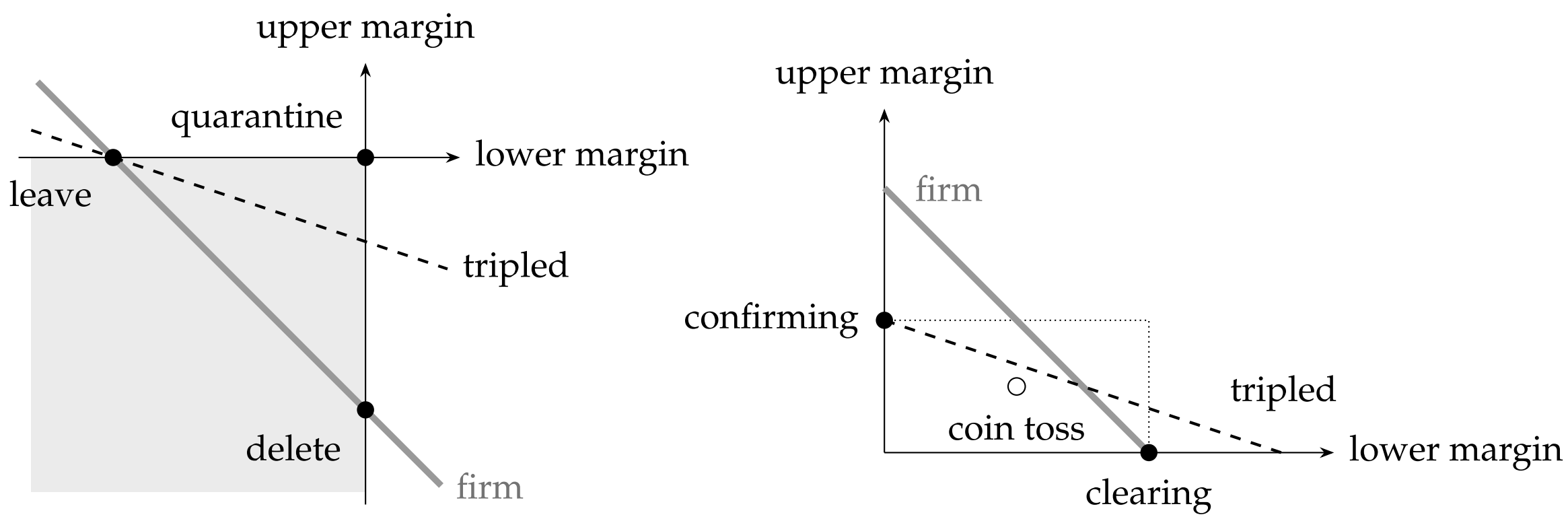
\begin{figure}[t]
\centering
\begin{tikzpicture}[every node/.style={font=\footnotesize}]
\begin{scope}[x=0.62cm, y=0.62cm]
\fill[black!8] (-5.3,-5.3) rectangle (0,0);
\draw[-{Stealth[length=4pt]}] (-5.5,0) -- (1.5,0);
\draw[-{Stealth[length=4pt]}] (0,-5.5) -- (0,1.5);
\node[anchor=south] at (0,1.5) {upper margin};
\node[anchor=west] at (1.5,0) {lower margin};
\draw[black!40, line width=1.8pt] (-5.2,1.2) -- (1.2,-5.2);
\draw[thick, dashed] (-5.3,0.4333) -- (1.3,-1.7667);
\fill (0,0) circle (2.4pt);
\fill (-4,0) circle (2.4pt);
\fill (0,-4) circle (2.4pt);
\node[anchor=south east] at (-0.12,0.08) {quarantine};
\node[anchor=north east] at (-4.1,-0.12) {leave};
\node[anchor=north east] at (-0.15,-4.15) {delete};
\node[anchor=west, text=black!55] at (1.2,-5.2) {firm};
\node[anchor=west] at (1.3,-1.7667) {tripled};
\node at (-2.0,-6.6) {\small (a) Actions, relative to quarantining};
\end{scope}
\begin{scope}[shift={(5.1cm,-2.9cm)}, x=13cm, y=13cm]
\draw[-{Stealth[length=4pt]}] (0,0) -- (0.34,0);
\draw[-{Stealth[length=4pt]}] (0,0) -- (0,0.26);
\node[anchor=south] at (0,0.26) {upper margin};
\node[anchor=west] at (0.34,0) {lower margin};
\draw[densely dotted] (0,0.1) -- (0.2,0.1) -- (0.2,0);
\draw[black!40, line width=1.8pt] (0,0.2) -- (0.2,0);
\draw[thick, dashed] (0,0.1) -- (0.3,0);
\fill (0.2,0) circle (2.4pt);
\fill (0,0.1) circle (2.4pt);
\draw[fill=white] (0.1,0.05) circle (2.4pt);
\node[anchor=north] at (0.2,-0.008) {clearing};
\node[anchor=east] at (-0.008,0.1) {confirming};
\node[anchor=north] at (0.1,0.04) {coin toss};
\node[anchor=west, text=black!55] at (0.012,0.2) {firm};
\node[anchor=south west] at (0.25,0.02) {tripled};
\node at (0.17,-0.092) {\small (b) Tests: value on each margin};
\end{scope}
\end{tikzpicture}
\caption{Why decisions cannot reveal the relative scale and choices of evidence can. \footnotesize The belief is $p=1/2$. (a) Each action is placed at its payoff relative to quarantining on each margin. Leaving loses $4$ on the lower margin and deleting loses $4$ on the upper one. Both lie in the shaded region, where an action is no better than quarantining on either margin, so quarantining is chosen however the margins are weighted. The thick gray line is an indifference line of the firm, which weights the margins equally, and the dashed line is one of the tripled objective, which weights the upper margin three times as heavily. The two rank leaving and deleting differently, but neither would choose either. (b) Each test is placed at its value on each margin. Neither test is better on both margins, so the choice between them depends on the weights: the firm prefers the clearing test and the tripled objective the confirming test. The dotted lines mark the best value on each margin, and a coin toss between the tests secures half of it on both.}
\label{fig:dominance}
\end{figure}

Figure~\ref{fig:example} shows what the firm's decisions do and do not reveal. The firm's value of acting on its belief, $V^*(p)=\max_aV(a,p)$, is flat where the firm quarantines and has a kink at each switching belief, and a test is worth the amount by which it raises the expected value of acting. The decisions fix where the kinks are. They do not fix how sharp each kink is, and what a test is worth depends on the sharpness of the kinks that its posteriors straddle. The tripled objective's value of acting has the same kinks, the upper one three times as sharp. In other words, the value of acting is the payoff of a portfolio of call options on the belief, with strikes at the switching beliefs, and decisions reveal the strikes but not the sizes of the positions.\footnote{For a single yes-or-no question, this point is known from proper scoring rules, which can elicit the same forecasts while rewarding information differently \citep{osband1989,li2022}, and from classifiers trained on utility-weighted losses \citep{autor2025}. For an earlier exploration of these issues, see \citet{gans1996impossibility}.}

Why can no decision reveal how sharp one kink is relative to the other? The firm's problem is in effect two questions that share an answer. Should the file be left or quarantined, and should it be quarantined or deleted? The firm leaves the file when leaving is the correct answer to the first, deletes it when deleting is the correct answer to the second, and quarantines it when quarantining is the correct answer to both, and at no belief is leaving the correct answer to the first while deleting is the correct answer to the second.

The firm therefore never has to weigh one question against the other. Counting the second question three times as heavily, as the tripled objective does, changes by how much its winner wins but not which option wins (Figure~\ref{fig:dominance}a). The tests are different. The clearing test has value only for the first question and the confirming test only for the second, so ranking them requires weighing one question against the other (Figure~\ref{fig:dominance}b). The firm's decisions thus never trade a gain on one margin against a loss on the other, whereas its choice of evidence does.

With deleting unavailable, the firm and the tripled objective both run the clearing test, and with leaving unavailable both run the confirming test, so the two disagree only about how the questions compare. When one test is best for both questions, as the confirming test is at beliefs below $1/13$ and the clearing test at beliefs above $27/31$, every objective that makes the firm's decisions runs it, and the decisions are then enough (Section~\ref{sec:onequestion}).

A choice that does weigh one margin against the other reveals what is missing. Asked at $p=1/2$ whether it would rather delete the file or leave it, the firm, which would do neither, is indifferent, because each mistake costs it $4$ more than quarantining. The tripled objective counts the shortfall from deleting as $12$ and strictly prefers to leave. Such a \emph{choice between mistakes} compares the two margins directly, and the belief at which the firm is indifferent between the two mistakes pins down the relative scale.

\begin{figure}[t]
\centering
\begin{tikzpicture}[scale=5.0, every node/.style={font=\footnotesize}]
\begin{scope}
\fill[black!12] (0,0) -- (1,0) -- (0.9,0.173205) -- (0.1,0.173205) -- cycle;
\fill[black!32] (0.5,0.866025) -- (0.3,0.519615) -- (0.533333,0.808290) -- cycle;
\draw[thin, densely dotted] (0.166667,0.288675) -- (0.625,0.649519);
\draw (0.1,0.173205) -- (0.9,0.173205);
\draw (0.3,0.519615) -- (0.533333,0.808290);
\draw (0,0) -- (1,0) -- (0.5,0.866025) -- cycle;
\node[below] at (0,0) {routine};
\node[below] at (1,0) {critical};
\node[above] at (0.5,0.866025) {malicious};
\node at (0.5,0.085) {leave};
\node at (0.53,0.34) {quarantine};
\node[left] at (0.25,0.76) {delete};
\draw[-{Stealth[length=3pt]}, thin] (0.25,0.76) -- (0.445,0.745);
\begin{scope}[shift={(0.5,-0.27)}, every node/.style={font=\scriptsize}]
\node[draw, rounded corners=2pt, inner sep=2pt] (L) at (-0.36,0) {leave};
\node[draw, rounded corners=2pt, inner sep=2pt] (Qa) at (0,0) {quarantine};
\node[draw, rounded corners=2pt, inner sep=2pt] (D) at (0.36,0) {delete};
\draw[thick] (L) -- (Qa);
\draw[thick] (Qa) -- (D);
\end{scope}
\node at (0.5,-0.47) {\small (a) $\kappa=2$: two blocks, two scales};
\end{scope}
\begin{scope}[shift={(1.55,0)}]
\fill[black!12] (0,0) -- (1,0) -- (0.75,0.433013) -- (0.35,0.433013)
  -- (0.166667,0.288675) -- cycle;
\fill[black!32] (0.5,0.866025) -- (0.166667,0.288675) -- (0.35,0.433013)
  -- (0.583333,0.721688) -- cycle;
\draw (0.166667,0.288675) -- (0.35,0.433013) -- (0.75,0.433013);
\draw (0.35,0.433013) -- (0.583333,0.721688);
\fill (0.35,0.433013) circle[radius=0.012];
\draw (0,0) -- (1,0) -- (0.5,0.866025) -- cycle;
\node[below] at (0,0) {routine};
\node[below] at (1,0) {critical};
\node[above] at (0.5,0.866025) {malicious};
\node at (0.56,0.20) {leave};
\node[right] at (0.80,0.62) {quarantine};
\draw[-{Stealth[length=3pt]}, thin] (0.80,0.62) -- (0.61,0.535);
\node[left] at (0.20,0.64) {delete};
\draw[-{Stealth[length=3pt]}, thin] (0.20,0.64) -- (0.41,0.625);
\begin{scope}[shift={(0.5,-0.33)}, every node/.style={font=\scriptsize}]
\fill[black!15] (-0.27,0) -- (0.27,0) -- (0,0.15) -- cycle;
\node[draw, rounded corners=2pt, fill=white, inner sep=2pt] (L2) at (-0.27,0) {leave};
\node[draw, rounded corners=2pt, fill=white, inner sep=2pt] (D2) at (0.27,0) {delete};
\node[draw, rounded corners=2pt, fill=white, inner sep=2pt] (Q2) at (0,0.15) {quarantine};
\draw[thick] (L2) -- (D2);
\draw[thick] (L2) -- (Q2);
\draw[thick] (Q2) -- (D2);
\end{scope}
\node at (0.5,-0.47) {\small (b) $\kappa=5$: one block, one scale};
\end{scope}
\end{tikzpicture}
\caption{The cost of caution determines what decisions reveal. \footnotesize Each point of a triangle is a belief over the three states, and the regions show the firm's choice. Below each triangle is the margin graph, which joins two actions when the firm switches between them at some belief. (a) Deleting and leaving are equally good on the dotted line, which lies inside the region where the firm quarantines. Quarantining separates the beliefs, and the two margins have scales of their own. (b) All three actions are optimal at the belief $(1/2,1/10,2/5)$, marked by the dot. The three margins meet there, and decisions fix the firm's payoffs up to a single scale.}
\label{fig:regions}
\end{figure}
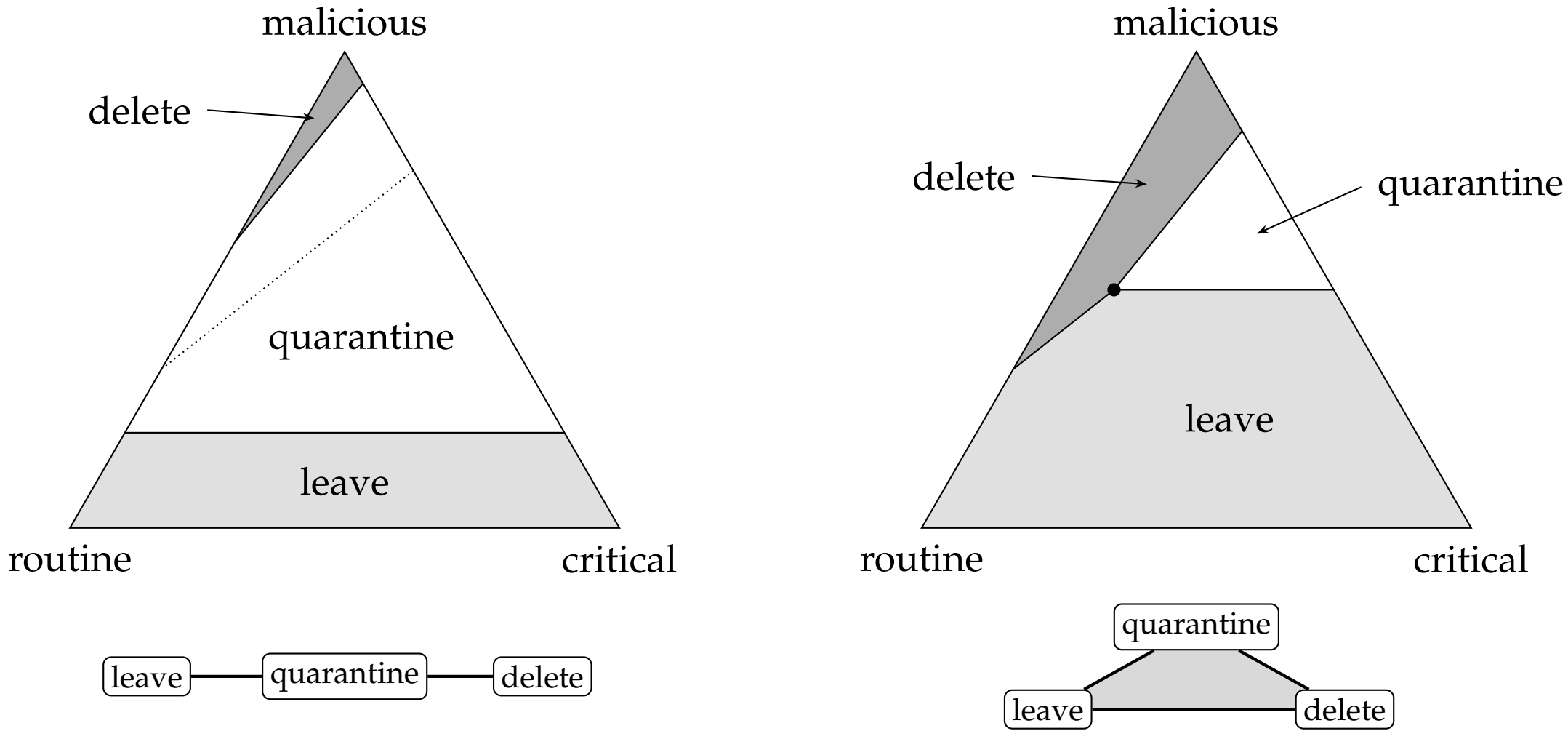

Is the gap inherent in learning from decisions? It is not, and what closes it shows from where it comes. Suppose that a legitimate file is either critical or routine, that deleting a critical file costs $30$ and deleting a routine one costs $5$, and that quarantining costs $\kappa$. With three states of the world, beliefs vary in two dimensions (Figure~\ref{fig:regions}). When $\kappa=5$, there is a belief at which the firm is indifferent among all three actions, and near it are beliefs at which the firm switches between each pair of them. The stakes on the three pairs add up, since the gain from deleting rather than leaving is the gain from deleting rather than quarantining plus the gain from quarantining rather than leaving. An objective can therefore reproduce all three switches only by multiplying the stakes on all three by the same number. Decisions then reveal the firm's payoffs up to a single scale, and an AI that learns them ranks every test as the firm would. When $\kappa=2$, quarantining is cheap enough that the firm quarantines wherever deleting and leaving are equally good. The three-way tie disappears. Quarantining separates the beliefs at which the firm deletes from those at which it leaves, and the two margins are again free to have scales of their own. The cost of caution thus determines whether the firm's decisions reveal its payoffs.

\subsection{Results}\label{sec:general}

Theorem~\ref{thm:main} shows that the example is general. For any finite set of actions, any convex set of beliefs that the AI can hold and payoffs in general position (Assumption~\ref{ass:acting}), the principal's actions divide into blocks, the largest sets of actions no two of which are separated by another action. The decisions reveal the stakes within each block up to a positive scale for the block, and they reveal nothing else. One direction of the theorem is immediate. Each switch between two actions fixes the stakes between them up to a scale, and a belief at which three actions tie forces three such scales to agree, as the tie does when $\kappa=5$. The difficulty is the converse, that scales chosen freely for each block never disturb a decision at any belief, a global property that these local conditions must deliver. It holds because the blocks and the separating actions that join them form a tree, and because on a convex set of beliefs every cycle of switches is built from three-way ties, a local-to-global step of the kind that makes local incentive constraints sufficient on convex domains \citep{carroll2012,archerkleinberg2014}.

Theorem~\ref{thm:test} provides the criterion. At a given belief, the principal's decisions \emph{settle} the AI's choice among experiments in the following sense. Some experiment is optimal for every objective that reproduces the decisions if and only if one experiment is best in every block, where an experiment's value in a block is its value when only the actions of that block are available. \citet{blackwell1953} asks when one experiment is at least as valuable as another in every decision problem, and knowing the principal's decisions reduces the problems that must be compared to the $K$ blocks. Because an experiment's value in a block is the same, up to the scale of the block, under every objective that reproduces the decisions, the criterion can be checked with the objective that the AI already has. When it fails, an experiment's \emph{guaranteed share}, the smallest fraction of the best attainable value that it secures in any block, bounds what the choice can lose. Section~\ref{sec:contexts} finds the blocks in three contexts: evidence that bears on one question, a cautious middle course, and an option to decline to answer.

The remaining results concern what to do when the criterion fails. A comparison that includes the principal's own choice reveals nothing that decisions do not, and comparisons of an AI's own responses approach this case as the AI improves. A choice between two mistakes on opposite sides of a separating action, such as the principal would make if that action were unavailable, tests the relative scale of two blocks against any number that a designer likes, and when two experiments differ in value only in those two blocks, one such choice shows which the principal prefers (Proposition~\ref{prop:records}). With $K$ blocks, $K-1$ such choices at beliefs of indifference identify every relative scale, and fewer never do. Capability matters in two opposite ways (Proposition~\ref{prop:capability}). As the AI's evidence becomes richer, the beliefs it can hold expand and blocks can merge. An objective fitted to decisions at the poorer beliefs then takes wrong actions unless it gives the merging blocks the same scale, and the principal can be worse off than it was before the AI improved. But decisions at the richer beliefs reveal the relative scale of the blocks that merge.

Two ends of this picture are known. For a single yes-or-no question, proper scoring rules that elicit the same forecasts can reward information differently \citep{osband1989,li2022}, and payoff functions with the same best responses have been characterized by positive weights on pairs of actions \citep{morrisui2004}. Our contribution is the structure in between and a criterion that uses it. We show which of these weights decisions tie together on any set of beliefs, that decisions leave the ratio of two weights free exactly when an action separates others, and that the leading case of a separating action is the cautious option on which safe deployment relies. For designers of AI systems, the results provide a procedure that uses only the objective that the AI has (Section~\ref{sec:practice}).

Throughout, we take records of the principal's choices to be complete and made at known beliefs, the AI's beliefs to be correct, and training to install the learned objective exactly. These assumptions favor learning preferences from records, so the limits we find are limits of what such records contain. Appendix~\ref{app:proofs} proves the two theorems and the two propositions, and the Supplemental Appendix proves the remaining claims.

\subsection{Related literature}\label{sec:lit}

The result for acting is close to the single-agent case of best-response equivalence. \citet[Proposition~4]{morrisui2004} characterize payoff functions with the same best responses, for beliefs anywhere in the simplex and under a genericity condition, by a positive weight on each pair of actions that are both best responses at some belief, and \citet{liu1996}, as reported by \citet{tewoldeconitzer2024}, shows that only some of these pairs need be related. For ordered actions, \citet[Propositions~7 and~8]{morrisui2004} show that the weights on adjacent pairs can be chosen independently exactly when payoffs are unimodal, and that unimodality suffices with a continuum of actions. Theorem~\ref{thm:main} determines which weights must coincide and which are free on any range of beliefs, and shows what the free ones govern. Its decomposition of the value of evidence by block extends the representation of proper scoring rules for binary events as mixtures of cost-weighted classification losses \citep{schervish1989}.

Scoring rules that induce the same reports can differ in the weights they place on thresholds, and so in the incentives they give to acquire information \citep{osband1989,li2022}, which is our point for a binary event, where every margin is a block. \citet{azrielilehrer2008} and \citet{frankelkamenica2019} characterize the valuations of information that arise from some decision problem, whereas we ask which features of a given problem its decisions reveal. Theorem~\ref{thm:test} compares experiments over a restricted class of decision problems, as does the order of \citet{atheylevin2018} for monotone problems, with the class given by a principal's decisions rather than by a property of payoffs. \citet{lu2019} and \citet{caplindean2015} recover beliefs, utilities and costs of information from choices, whereas we take beliefs as known and ask what each kind of choice identifies.

Theories of delegation take the agent's preferences as given and ask which actions a principal should permit \citep{holmstrom1984,alonsomatouschek2008}, whereas here the principal shapes the agent's preferences. In the theory of delegated expertise, conflict over what an expert learns arises from a difference in preferred actions or from the expert's private cost of learning \citep{demskisappington1987,szalay2005,deimenszalay2019}. Here the AI's preferred action coincides with its principal's at every belief and the AI bears no private cost, so the conflict arises only from the scales, as with incongruent performance measures in multitask agency \citep{holmstrommilgrom1991,baker2002}. In Bayesian persuasion, a sender needs to know only what the receiver would do at each posterior \citep{kamenicagentzkow2011}, so an AI that wanted only to steer its principal's actions would need nothing beyond the principal's decisions. An AI that gathers evidence for its principal also needs the stakes.

In computer science, results on which transformations of a reward preserve optimal policies \citep{ng1999} and on what behavior identifies about a reward \citep{ngrussell2000,skalse2023} are counterparts of the result for acting, and \citet{brown2021} define exact value alignment, as we do, by the inclusion of an AI's optimal choices in a person's. Objectives that agree with the intended one in training can diverge in deployment \citep{langosco2022}, more so as agents become more capable \citep{pan2022}, and Section~\ref{sec:capability} locates where this happens when the objective is learned from decisions. Work on eliciting preferences chooses queries for their effect on the decision to be made \citep{boutilier2006,lindner2021}, and our results say which queries can carry information about the relative scale. In economics, a growing literature studies how an AI whose objective is given or uncertain should be overseen \citep{bkm2026,gansholden2026} and how training choices shape what a prediction system learns \citep{gans2026preferences}. In the terms of \citet{agrawal2019judgment}, the AI supplies prediction and the objective supplies judgment, and we ask how much judgment each task requires. \citet{hadfieldmenell2019} frame alignment as incomplete contracting by analogy, and Section~\ref{sec:conclusion} relates our results to the foundations of incomplete contracts.

\section{Model}\label{sec:model}

A principal delegates a choice to an AI. The facts that determine which action is best form a finite set $\Omega$ of at least two \emph{states} of the world, which the AI does not observe. For instance, in answering questions the state is the correct answer, and for an AI that manages a computer system it is the condition of the system. The AI chooses from a finite set $A$ of actions, no two of which have the same payoffs in every state, and the principal's payoff from action $a$ in state $\omega$ is $V(a,\omega)$. The AI acts on a belief $q\in\Delta(\Omega)$, its posterior after whatever evidence it has gathered, such as a language model's credence over candidate answers. We take this belief to be correct, so that the only question is what the AI does with it.\footnote{The predictive distribution of a pretrained language model is close to calibrated, but later training can distort it \citep{openai2023}. Miscalibrated beliefs are a separate source of misalignment, which we set aside.} For any payoff function $W:A\times\Omega\to\R$, we write $W(a,q)=\sum_{\omega}q(\omega)W(a,\omega)$ for the expected payoff, $\BR_W(q)=\argmax_{a\in A}W(a,q)$ for the set of best actions and $W^*(q)=\max_{a\in A}W(a,q)$ for the \emph{value of acting} on belief $q$. With two states we identify a belief with $p=q(\omega_1)$.

The AI maximizes an \emph{objective} $S:A\times\Omega\to\R$, the formal counterpart of a reward function, which a designer installs before deployment, and it chooses from $\BR_S(q)$.\footnote{Reinforcement learning from human feedback maximizes the installed reward less a penalty on the divergence from a reference policy, and maximizing the expected installed payoff is the limit as the weight on the penalty goes to zero \citep{ouyang2022}. With a positive weight, the AI takes each action with a probability that rises with its payoff under the objective \citep{rafailov2023}, so the scales that we find to be free matter for acting as well (Supplemental Appendix~\ref{supp:penalty}). We also take the objective to be an expected payoff at the AI's belief, as for an agent that pairs a calibrated predictor with a reward for each action in each state. A reward model that scores actions directly from the context need not be linear in the belief, and decisions then do not pin down its valuation of evidence even within a block, so this assumption also favors learning from records.} The AI is \emph{aligned at belief $q$} if $\BR_S(q)\subseteq\BR_V(q)$, so that every action it might choose is one that the principal would choose with the AI's belief. The benchmark is the principal's choice at the AI's belief because the principal delegates in order to use the AI's evidence.

The vector $g_{ab}=V(a,\cdot)-V(b,\cdot)\in\R^\Omega$ records the principal's gain from $a$ over $b$ in each state. We call these vectors the principal's \emph{stakes}, and the vectors $e_{ab}=S(a,\cdot)-S(b,\cdot)$ the objective's stakes, and we write $\xi\cdot\xi'$ for the inner product on $\R^\Omega$, so that $V(a,q)-V(b,q)=g_{ab}\cdot q$. Two features of the principal's payoffs never affect a choice among actions. If $S=\lambda V+\beta$ for some $\lambda>0$ and $\beta:\Omega\to\R$, then $S(a,q)=\lambda V(a,q)+\beta\cdot q$, where $\beta\cdot q$ is the same for every action, so $\BR_S(q)=\BR_V(q)$ at every belief. The most that choices among actions could reveal about $V$ is therefore $V$ up to a positive scale and an amount added to every action's payoff in each state.

\paragraph{Evidence.} The designer installs the objective, and the AI then chooses what evidence to gather, observes it and takes an action. The AI's evidence is an \emph{experiment}: a finite set of signals together with, for each state $\omega$, a probability distribution $f(\cdot\mid\omega)$ over them. At a belief $q$, a signal $s$ with positive probability $P(s)=\sum_\omega q(\omega)f(s\mid\omega)$ leads to the posterior $q_s(\omega)=q(\omega)f(s\mid\omega)/P(s)$. The \emph{value} of experiment $E$ at belief $q$ under a payoff function $W$ is
\begin{equation}\label{eq:value}
I_W(E,q)=\sum_sP(s)\,W^*(q_s)-W^*(q),
\end{equation}
the expected gain from choosing an action after seeing the signal rather than before. It is nonnegative because $W^*$ is convex and the posteriors average to $q$. The AI values experiments by $I_S$ and the principal by $I_V$. If $S=\lambda V+\beta$, then $I_S=\lambda I_V$, so such an objective also ranks every experiment as the principal does.

\paragraph{The range of beliefs.} A principal rarely knows the exact belief its AI will hold, but it may know which beliefs the AI's evidence can produce. A \emph{range} is a nonempty convex set $Q\subseteq\Delta(\Omega)$ that is open relative to its affine hull $\aff Q$, and the AI's belief can be any belief in the closure of $Q$. We say that a property holds \emph{on the range} if it holds at every belief in the closure of $Q$, and that an experiment \emph{keeps beliefs on the range} at $q$ if its posteriors at $q$ lie in that closure. A principal that knows nothing about the AI's evidence takes the range to be the set of beliefs with full support, and then every experiment keeps beliefs on the range. An AI whose evidence bears only on whether a file is malicious has a range that is a segment, however many states there are, and a more capable AI has a larger range.

\paragraph{Records.} The designer does not know $V$ and fits the objective to records of the principal's choices. The leading record is the principal's \emph{decisions}, the actions $\BR_V(q)$ that it would choose at each belief on the range, as demonstrations at known beliefs would show.\footnote{Real records are finite and rarely include beliefs. Judges and physicians systematically mispredict \citep{kleinberg2018,mullainathan2022}, so a designer that reads their decisions at correct beliefs attributes their mistakes to the stakes. \citet{rambachan2024} identifies mistaken beliefs by restricting which characteristics enter the decision maker's payoffs, and records that pair choices with elicited beliefs \citep{agarwal2023} meet the assumption that beliefs are known.} An objective that reproduces this record is aligned on the range, and the question is what else the record determines.

\section{What decisions reveal}\label{sec:acting}

\subsection{Separating actions and blocks}\label{sec:decisions}

Let $A_Q$ be the set of actions that are optimal for the principal at some belief in $Q$. A pair of actions is a \emph{margin} if, at some belief in $Q$, they are the principal's only optimal actions, and three actions form a \emph{three-way tie} if, at some belief in $Q$, they are the principal's only optimal actions. We maintain the following assumption, which rules out ties that arise only because payoffs satisfy particular equalities.

\begin{assumption}\label{ass:acting}
(i) The principal's payoff function $V$ is regular on $Q$: every action in $A_Q$ is optimal for the principal at every belief in some nonempty subset of $Q$ that is open relative to $\aff Q$, and no two actions in $A_Q$ have equal expected payoffs at every belief in $\aff Q$. (ii) Whenever two or three actions are all optimal for the principal at some belief in $Q$, there is a belief in $Q$ at which exactly those actions are optimal. (iii) No action outside $A_Q$ is optimal for the principal at any belief on the range.\footnote{The assumption excludes, for instance, four actions that tie at a single belief when beliefs vary in two dimensions. It holds for almost every payoff function when the range is the relative interior of a polytope, as the set of beliefs with full support is (Supplemental Appendix~\ref{supp:generic}).}
\end{assumption}

What decisions reveal about a single margin $\{a,b\}$ can be seen directly. An aligned objective must be indifferent between $a$ and $b$ where the principal switches between them, and it must rank them as the principal does on either side. Because expected payoffs are linear in beliefs, this fixes the objective's stakes on the margin as a positive multiple of the principal's, $e_{ab}=\lambda_{ab}\,g_{ab}$, in the directions in which beliefs on the range can move.\footnote{Precisely, $e_{ab}\cdot\xi=\lambda_{ab}\,g_{ab}\cdot\xi$ for every $\xi$ in the linear span of $Q$ (Lemma~\ref{lem:local}). Decisions on a range say nothing about payoffs in directions in which the AI's beliefs never move, a point to which Section~\ref{sec:capability} returns.} We call $\lambda_{ab}$ the \emph{scale} of the margin. Decisions are not informative about this since multiplying the stakes on a margin by a positive number leaves the switch unchanged.

Whether decisions relate the scales of two margins depends on how the principal's choice moves among its actions as the belief changes. An action $c$ \emph{separates} two other actions $a$ and $b$ in $A_Q$ if the principal's choice cannot pass from $a$ to $b$, as the belief moves within the range, unless $c$ is optimal on the way. Formally, $c$ separates $a$ and $b$ if no connected set of beliefs in $Q$ at which $c$ is not optimal contains both a belief at which $a$ is optimal and a belief at which $b$ is. Two actions that $c$ separates lie on opposite sides of $c$. An action \emph{separates the range} if it separates some pair of actions in $A_Q$, which is the case exactly when the beliefs in $Q$ at which it is not optimal form a disconnected set. A \emph{block} is a set of at least two actions in $A_Q$, no two of which are separated by another action, that is contained in no larger set with this property. The margins of a block are the margins between its actions, and we write $K$ for the number of blocks. In the file example, quarantining separates leaving and deleting, so there are two blocks, one made up of leaving and quarantining and the other of quarantining and deleting. An action that separates the range belongs to more than one block, and when $A_Q$ has at least two actions, every other action in $A_Q$ belongs to exactly one (Lemma~\ref{lem:blocks}).

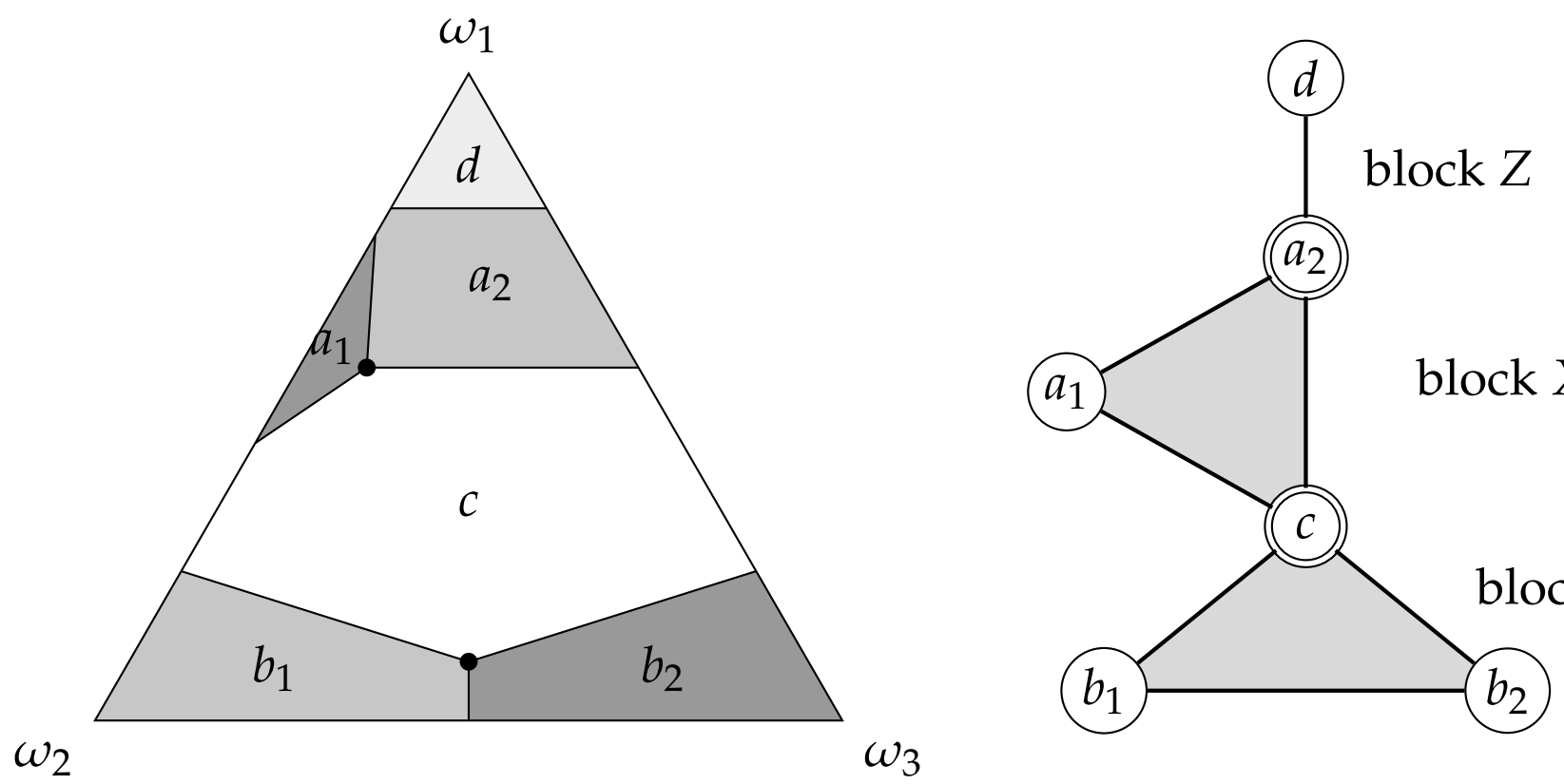
\begin{figure}[t]
\centering
\begin{tikzpicture}[scale=5.2, every node/.style={font=\footnotesize}]
\begin{scope}
\fill[black!7] (0.5,0.8660) -- (0.3958,0.6856) -- (0.6042,0.6856) -- cycle;
\fill[black!22] (0.3958,0.6856) -- (0.3750,0.6495) -- (0.3636,0.4724) -- (0.7273,0.4724) -- (0.6042,0.6856) -- cycle;
\fill[black!40] (0.3750,0.6495) -- (0.2143,0.3712) -- (0.3636,0.4724) -- cycle;
\fill[black!22] (0.1154,0.1999) -- (0,0) -- (0.5,0) -- (0.5,0.0787) -- cycle;
\fill[black!40] (0.5,0) -- (1,0) -- (0.8846,0.1999) -- (0.5,0.0787) -- cycle;
\draw (0.3958,0.6856) -- (0.6042,0.6856);
\draw (0.3750,0.6495) -- (0.3636,0.4724);
\draw (0.2143,0.3712) -- (0.3636,0.4724) -- (0.7273,0.4724);
\draw (0.1154,0.1999) -- (0.5,0.0787) -- (0.8846,0.1999);
\draw (0.5,0) -- (0.5,0.0787);
\fill (0.3636,0.4724) circle[radius=0.012];
\fill (0.5,0.0787) circle[radius=0.012];
\draw (0,0) -- (1,0) -- (0.5,0.8660) -- cycle;
\node[above] at (0.5,0.8660) {$\omega_1$};
\node[below left] at (0,0) {$\omega_2$};
\node[below right] at (1,0) {$\omega_3$};
\node at (0.5,0.745) {$d$};
\node at (0.53,0.585) {$a_2$};
\node at (0.318,0.50) {$a_1$};
\node at (0.5,0.29) {$c$};
\node at (0.24,0.07) {$b_1$};
\node at (0.76,0.07) {$b_2$};
\node at (0.5,-0.15) {\small (a) The principal's choice at each belief};
\end{scope}
\begin{scope}[shift={(1.62,0.06)}]
\fill[black!15] (0,0.56) -- (-0.32,0.38) -- (0,0.20) -- cycle;
\fill[black!15] (0,0.20) -- (-0.27,-0.02) -- (0.27,-0.02) -- cycle;
\node[circle, draw, fill=white, inner sep=1.4pt, minimum size=0.52cm] (d) at (0,0.80) {$d$};
\node[circle, draw, double, double distance=1pt, fill=white, inner sep=1.4pt, minimum size=0.52cm] (a2) at (0,0.56) {$a_2$};
\node[circle, draw, fill=white, inner sep=1.4pt, minimum size=0.52cm] (a1) at (-0.32,0.38) {$a_1$};
\node[circle, draw, double, double distance=1pt, fill=white, inner sep=1.4pt, minimum size=0.52cm] (c) at (0,0.20) {$c$};
\node[circle, draw, fill=white, inner sep=1.4pt, minimum size=0.52cm] (b1) at (-0.27,-0.02) {$b_1$};
\node[circle, draw, fill=white, inner sep=1.4pt, minimum size=0.52cm] (b2) at (0.27,-0.02) {$b_2$};
\draw[thick] (d) -- (a2);
\draw[thick] (a2) -- (a1);
\draw[thick] (a2) -- (c);
\draw[thick] (a1) -- (c);
\draw[thick] (c) -- (b1);
\draw[thick] (c) -- (b2);
\draw[thick] (b1) -- (b2);
\node[anchor=west] at (0.05,0.68) {block $Z$};
\node[anchor=west] at (0.12,0.40) {block $X$};
\node[anchor=west] at (0.20,0.12) {block $Y$};
\node at (0,-0.21) {\small (b) The margin graph};
\end{scope}
\end{tikzpicture}
\caption{Separating actions and blocks in an example with six actions. \footnotesize (a) The principal's choice at each belief with full support over three states, when the payoffs in those states are $(\frac45,-\frac35,-\frac{12}5)$ for $a_1$, $(1,-\frac65,-\frac65)$ for $a_2$, $(-2,\frac35,-\frac15)$ for $b_1$, $(-2,-\frac15,\frac35)$ for $b_2$, $(0,0,0)$ for $c$ and $(2,-5,-5)$ for $d$. The dots mark the two three-way ties. (b) The margin graph, which joins two actions when they form a margin. The action $c$ separates $a_1$, $a_2$ and $d$ from $b_1$ and $b_2$, and $a_2$ separates $d$ from the other actions, so there are three blocks. The separating actions are drawn with double borders, and the two blocks whose margins a three-way tie joins are shaded.}
\label{fig:blocks}
\end{figure}

Figure~\ref{fig:blocks} shows an example with six actions and three blocks. The blocks and their actions form a tree, with each action joined to the blocks that contain it (Lemma~\ref{lem:structure}).

Within a block, the margins share a scale, and the reason is the three-way tie. If $a$, $b$ and $c$ tie at some belief, then near it all three pairs are margins. Stakes add up, $g_{ac}=g_{ab}+g_{bc}$, for the objective as for the principal, so
\[
\lambda_{ac}\,(g_{ab}+g_{bc})=\lambda_{ab}\,g_{ab}+\lambda_{bc}\,g_{bc}.
\]
The stakes $g_{ab}$ and $g_{bc}$ are not proportional there, so the three scales must be equal. In other words, the principal's switch between $a$ and $c$ is among its decisions, and where that switch lies depends on how the stakes between $a$ and $b$ compare with the stakes between $b$ and $c$. Any two margins of a block are joined by a chain of margins in which consecutive margins belong to a common three-way tie (Lemma~\ref{lem:blocks}), so a block has a single scale. Between blocks there is no such chain. The principal never switches between two actions that a third separates, so the belief at which it would do so, which is what would relate the scales on the two sides, is not among its decisions.

Finally, we call an objective \emph{exact on the range} if there are $\lambda>0$ and $\beta:\Omega\to\R$ such that $S(a,q)=\lambda V(a,q)+\beta\cdot q$ for every action $a\in A_Q$ at every belief $q$ on the range. An exact objective is the principal's payoff function on the range, up to the two normalizations that no choice among actions can reveal. For a payoff function $W$ and a block $X$, let $W^*_X(q)=\max_{a\in X}W(a,q)$ be the value of acting under $W$ when only the actions of $X$ are available. For an experiment $E$ that keeps beliefs on the range at $q$, let
\[
\phi^W_X(E,q)=\sum_sP(s)\,W^*_X(q_s)-W^*_X(q)
\]
be the value of $E$ in that smaller problem. We write $\phi_X$ for $\phi^V_X$, and we call $\phi(E,q)\in\R^K$ the \emph{block values} of $E$ at $q$.

\subsection{The representation}\label{sec:theorem}

The following theorem states what decisions reveal.

\begin{theorem}\label{thm:main}
Let Assumption~\ref{ass:acting} hold.
\begin{enumerate}[label=(\alph*)]
\item An objective $S$ is aligned on the range if and only if no action outside $A_Q$ is optimal for $S$ at any belief on the range and there are scales $\lambda_X>0$, one for each block, such that, for all actions $a$ and $b$ of each block $X$ and every belief $q$ on the range,
\begin{equation}\label{eq:rep}
S(a,q)-S(b,q)=\lambda_X\bigl[V(a,q)-V(b,q)\bigr].
\end{equation}
The scales are then unique, and $\BR_S(q)=\BR_V(q)$ at every belief on the range. Every vector of positive scales arises from some aligned objective.
\item If $S$ is aligned on the range with scales $\lambda_X$, then every experiment $E$ that keeps beliefs on the range at a belief $q\in Q$ has
\[
I_S(E,q)=\sum_X\lambda_X\,\phi_X(E,q)\qquad\text{and}\qquad I_V(E,q)=\sum_X\phi_X(E,q),
\]
with $\phi_X(E,q)\geq0$ for every block $X$.
\item There are at least two blocks if and only if some action separates the range. Consequently, every objective that is aligned on the range is exact on the range if and only if no action separates the range.
\end{enumerate}
\end{theorem}

\noindent Theorem~\ref{thm:main} says that the principal's decisions reveal its stakes block by block, each block up to a scale of its own, and nothing else. That an aligned objective must satisfy \eqref{eq:rep} follows from the argument above. The converse holds because the blocks behave like separate decisions that share options. At every belief each block has a best action, and following these best actions from block to block leads to the principal's choice (Lemma~\ref{lem:structure}). Multiplying the stakes within a block by a positive number changes by how much its best action wins and never which action wins, so the choice is unchanged.\footnote{The same freedom arises, more transparently, when an AI makes several decisions whose payoffs add up, such as what to do with a file and whether to install an update. Scaling each decision's payoffs by its own positive number changes no choice but changes the value of evidence. That freedom does not survive interaction: if beliefs can move each of two yes-or-no decisions across its margin separately, a slight complementarity or substitutability where both are marginal puts all four actions in one block, although the beliefs at which decisions then reveal the relative scale form a set that shrinks with the interaction (Supplemental Appendix~\ref{supp:separable}). What a separating action adds is that the scales can be free within a single decision.} No decision weighs a gain in one block against a loss in another, so decisions cannot reveal how the principal would weigh the two. We call the ratios of the scales of different blocks the \emph{relative scale}. Two actions belong to a common block if and only if every aligned objective ranks them as the principal does at every belief in $Q$ (Supplemental Appendix~\ref{supp:records}).

Part (b) says that an aligned AI values evidence as if each block were a decision of its own. On the range, the principal's value of acting is the sum of the values of acting in the blocks, less a linear function of the belief that drops out of the value of every experiment, and the objective's value of acting is the same sum with each block weighted by its scale.\footnote{Precisely, $V^*(q)=\sum_XV^*_X(q)-\sum_{c\in A_Q}(n_c-1)V(c,q)$ at every belief on the range, where $n_c$ is the number of blocks that contain $c$ (Lemma~\ref{lem:decomp}). In convex geometry, the principal's value of acting on the range is the support function of a Minkowski sum of one polytope for each block, and three-way ties play the role of the triangular faces that make a polytope indecomposable \citep{shephard1963,kallay1982}.} With two states, every margin is a block, the value of acting in the block of the $i$th margin is a linear function plus $d_i(p-t_i)^+$, where $t_i$ is the switching belief and $d_i$ the increase in the slope of $V^*$ there, and part (b) reads
\begin{equation}\label{eq:kinks}
\begin{aligned}
I_V(E,p)&=\sum_id_i\Bigl[\sum_sP(s)\,(p_s-t_i)^+-(p-t_i)^+\Bigr],\\
I_S(E,p)&=\sum_i\lambda_i\,d_i\Bigl[\sum_sP(s)\,(p_s-t_i)^+-(p-t_i)^+\Bigr].
\end{aligned}
\end{equation}
Each term is the sharpness of a kink times the expected distance by which the experiment carries the belief across it. Decisions reveal the switching beliefs $t_i$ and nothing about how the sharpnesses $d_i$ compare. In the file example, $d_1=d_2=10$, the clearing test has block values $(1/5,0)$ at $p=1/2$ and the confirming test $(0,1/10)$, and the tripled objective has scales $(1,3)$, which gives the values in Figure~\ref{fig:dominance}.

Part (c) says when decisions reveal everything. If no action separates the range, every aligned objective is exact on the range and values every experiment that keeps beliefs on the range at a fixed multiple of the principal's value, so the principal's decisions are all that the AI needs to rank evidence. If some action separates the range, every relative scale is consistent with the decisions, and whether the AI needs more depends on the evidence that it can gather (Section~\ref{sec:test}).

The reason is that a margin identifies stakes only up to scale. In models that recover levels from paired comparisons, such as two-way fixed effects for wages or the ranking of items from contests between pairs, a path of comparisons fixes each difference in turn, and levels are identified within connected sets \citep{akm1999,bradleyterry1952}. Here a path of margins fixes nothing about relative scales, identification needs cycles of margins through three-way ties, and no such cycle runs from one side of a separating action to the other. Decisions leave the most scales free when no three actions tie. Every margin is then a block, and the principal's expected payoffs are single-peaked on the tree that the margins form \citep{black1948,demange1982}, and concave in a suitably spaced action when that tree is a line.\footnote{If the \emph{margin graph} has the actions in $A_Q$ as its vertices and the margins as its edges, an action separates the range exactly when removing it disconnects the graph, and when $A_Q$ has at least two actions, the blocks are the graph's biconnected components. Supplemental Appendix~\ref{supp:structure} proves these statements and those on single-peaked payoffs.}

\section{When decisions are enough}\label{sec:test}

An AI whose objective is aligned on the range takes the actions that its principal would take, so what remains is its choice of evidence. Fix a belief $q\in Q$. A \emph{menu} at $q$ is a finite set $M$ of costless experiments that keep beliefs on the range at $q$, from which the AI runs one before it acts. By part (b) of Theorem~\ref{thm:main}, the AI values an experiment $E$ at $\sum_X\lambda_X\phi_X(E,q)$, whereas the principal values it at $\sum_X\phi_X(E,q)$. The AI therefore overvalues experiments that bear on blocks with large scales relative to experiments that bear on blocks with small ones, and the principal's decisions do not say which scales are too large.\footnote{An aligned objective ranks two experiments and their dilutions, which report the signal only part of the time, as the principal does exactly when the average of its scales, weighted by the shares of an experiment's value that come from each block, is the same for both (Supplemental Appendix~\ref{supp:corollary}).}

We say that the principal's decisions \emph{settle} the choice from $M$ at $q$ if some experiment on $M$ is optimal on $M$ at $q$ for every objective that is aligned on the range. The choice is then the same whatever payoffs lie behind the decisions. When the decisions do not settle it, every experiment on the menu is suboptimal for some principal with those decisions, so no rule for choosing from the menu, whether or not it maximizes an expected payoff, is right for all of them. Let $m_X=\max_{E\in M}\phi_X(E,q)$ be the largest value in block $X$ that an experiment on the menu attains. An experiment $E\in M$ is \emph{best in block $X$} if $\phi_X(E,q)=m_X$, and its \emph{guaranteed share} is
\[
G(E)=\min\bigl\{\phi_X(E,q)/m_X:\ m_X>0\bigr\},
\]
with $G(E)=1$ if no experiment on the menu has positive value. The guaranteed share is the fraction of the best attainable value that $E$ secures in the block in which it falls furthest short, and it equals one exactly when $E$ is best in every block.

\begin{theorem}\label{thm:test}
Let Assumption~\ref{ass:acting} hold, let $q\in Q$, and let $M$ be a menu at $q$.
\begin{enumerate}[label=(\alph*)]
\item The principal's decisions settle the choice from $M$ at $q$ if and only if some experiment on $M$ is best in every block. In that case, an experiment on $M$ is optimal at $q$ for an objective that is aligned on the range if and only if it is best in every block, and every such experiment is optimal for the principal.
\item For every experiment $E$ on $M$ and every objective $W$ that is aligned on the range,
\[
I_W(E,q)\geq G(E)\max_{E'\in M}I_W(E',q),
\]
and if some experiment on $M$ has positive value, no number larger than $G(E)$ has this property.
\item If an objective $S$ is aligned on the range with scales $\lambda_X$, then $\phi^S_X(E,q)=\lambda_X\,\phi_X(E,q)$ for every block $X$ and every experiment $E$ on $M$. The experiments that are best in each block and the guaranteed share of every experiment are therefore the same whether they are computed from $V$ or from $S$.
\end{enumerate}
\end{theorem}

\noindent Theorem~\ref{thm:test} says that decisions settle a choice of evidence exactly when the blocks agree about it. Intuitively, an aligned objective values an experiment at a weighted sum of its values in the blocks, with positive weights that the decisions leave free. An experiment that is best in every block is best however the blocks are weighted. If no experiment is best in every block, then for each experiment there is a block in which another does better, and an objective that weights that block heavily enough prefers the other. 

Part (c) makes the criterion one that a designer can check. Restricting the actions to those of one block removes the scales of the other blocks from the AI's valuation of evidence. The scale that remains multiplies the value of every experiment alike, so the experiment that the AI would run with only the actions of a block available is the one that the principal would run. In the file example at $p=1/2$, the clearing test is best in the block of the lower margin and the confirming test in the block of the upper one, so the choice is unsettled (Figure~\ref{fig:dominance}b).

The criterion is a comparison of experiments in the sense of \citet{blackwell1953}, restricted to the decision problems that the principal might have. At a belief with full support, one experiment is at least as valuable as another in every decision problem only if the second is a garbling of the first, and under every payoff function that reproduces the principal's decisions exactly when it is at least as valuable in each of the $K$ blocks. Decisions therefore settle the choice when the menu contains an experiment of which every other is a garbling, when the principal's payoffs satisfy single crossing and one experiment is more accurate than the others \citep{lehmann1988,quahstrulovici2009}, when the AI can run its experiments together, and when the experiments have positive value in at most one block (Supplemental Appendix~\ref{supp:choose}). An experiment has positive value in a block only if one of its signals changes which action of that block is best, so an unsettled choice requires evidence that can change the choice in two blocks.

When decisions do not settle the choice, part (b) says how much an experiment can lose. An experiment $E$ secures at least the share $G(E)$ of the best value on the menu however the blocks are weighted, and a principal whose stakes lie almost entirely in the block in which $E$ falls furthest short gets little more. In the file example, each test is worth nothing in one block, so each has a guaranteed share of zero: the firm's decisions are consistent with stakes under which either test secures an arbitrarily small share of the value of the other. Tossing a coin between them guarantees half (Figure~\ref{fig:dominance}b), and in general a lottery over the experiments that are best in each block guarantees the share $1/K'$, where $K'$ is the number of blocks in which some experiment on the menu has positive value.\footnote{This is diversification under ambiguity in the sense of \citet{manski2009}. No lottery guarantees more than $1/K'$ when each experiment has value in a single block, and in general the best lottery solves a zero-sum game against an opponent who chooses the block. Less is lost when the relative scale is wrong by a limited amount. If $\lambda_{\min}$ and $\lambda_{\max}$ are the smallest and largest of the AI's scales among the blocks in which some experiment on the menu has positive value, the experiment that the AI prefers is worth at least $\lambda_{\min}/\lambda_{\max}$ of the principal's choice, and the bound is attained whenever there are two blocks. With costly experiments, the expected loss is of second order in the error in the scales when the menu is the same at every belief and the AI's belief has a bounded density (Supplemental Appendix~\ref{supp:misdirection}).}

\paragraph{Paying for evidence.} Evidence usually has a cost, in computation, in time or in a person's attention, and then the levels of the scales matter as well as their ratios. If running $E$ has a cost that the objective counts as the principal does, the AI runs $E$ when $\sum_X\lambda_X\phi_X(E,q)$ exceeds the cost, and the principal would run it when $\sum_X\phi_X(E,q)$ does. No ranking of actions or of experiments reveals the level of the scales, and the level matters even with a single block, so decisions never determine how much evidence the AI should pay for. Because the blocks from which an experiment draws its value change with the belief, the same AI can gather too much evidence at some beliefs and too little at others. Suppose that the AI can run the file in a sandbox, which reveals whether the file is malicious, at a cost of $1/2$. The firm would sandbox a file when $1/20\leq p\leq19/20$. An AI whose objective has the scales $(1/2,3/2)$, which is half the tripled objective, sandboxes only from $p=1/12$, leaving untested files that the firm would sandbox, and continues up to $p=27/28$, sandboxing files that the firm would simply delete, although every action that it then takes is the firm's.\footnote{\citet{mullainathan2022} find that emergency physicians test some patients at low risk of a heart attack and leave untested some at high risk, and attribute the pattern to errors in predicting risk, because a private benefit from testing would produce errors in one direction only. A wrong relative scale can also produce errors in both directions, but it leaves every final action optimal at the correct posterior, which mistaken beliefs do not. The decomposition of Theorem~\ref{thm:main} extends to plans that gather evidence in sequence and stop at beliefs on the range, because beliefs form a martingale (Supplemental Appendix~\ref{supp:sequential}).}

\section{Three decision contexts}\label{sec:contexts}

Applying the criterion requires the blocks, and in three common contexts a designer can find them from features of the principal's decisions that are easy to see.

\begin{figure}[t]
\centering
\begin{tikzpicture}
\begin{axis}[
  name=ca, width=0.5\textwidth, height=0.38\textwidth,
  xmin=0, xmax=1, ymin=0, ymax=0.15,
  xtick={0,0.1,0.5,0.9,1},
  xticklabels={$0$,$t_1$,$\frac12$,$t_2$,$1$},
  ytick={0,0.05,0.1},
  yticklabels={$0$,$\frac1{20}$,$\frac1{10}$},
  tick label style={font=\small}, label style={font=\small},
  xlabel={Belief $t$},
  ylabel={Expected distance across $t$},
  axis x line=bottom, axis y line=left,
  title={\small (a) The two tests at $p=\frac12$}, title style={yshift=-2pt},
  legend style={draw=none, fill=none, font=\footnotesize, at={(0.5,1.0)}, anchor=north, legend columns=2, column sep=4pt},
  legend cell align=left]
\addplot[black, thick, dashed] coordinates {(0,0) (0.5,0.1) (0.625,0) (1,0)};
\addlegendentry{Clearing test}
\addplot[black, thick, dotted] coordinates {(0,0) (0.444444,0) (0.5,0.05) (1,0)};
\addlegendentry{Confirming test}
\draw[black!55, thin] (axis cs:0.1,0) -- (axis cs:0.1,0.118);
\draw[black!55, thin] (axis cs:0.9,0) -- (axis cs:0.9,0.118);
\addplot[only marks, mark=*, mark size=1.7pt, forget plot] coordinates {(0.1,0.02) (0.9,0.01)};
\addplot[only marks, mark=o, mark size=1.7pt, forget plot] coordinates {(0.1,0) (0.9,0)};
\end{axis}
\begin{axis}[
  at={(ca.east)}, anchor=west, xshift=1.2cm,
  width=0.5\textwidth, height=0.38\textwidth,
  xmin=0, xmax=1, ymin=0, ymax=3.6,
  xtick={0,0.0769231,0.5,0.870968,1},
  xticklabels={$0$,$\frac1{13}$,$\frac12$,$\frac{27}{31}$,$1$},
  tick label style={font=\small}, label style={font=\small},
  xlabel={Belief $p$ that the file is malicious},
  axis x line=bottom, axis y line=none,
  title={\small (b) The test that is best in each block}, title style={yshift=-2pt},
  clip=false]
\fill[black!45] (axis cs:0,2.05) rectangle (axis cs:0.870968,2.5);
\fill[black!12] (axis cs:0.870968,2.05) rectangle (axis cs:1,2.5);
\fill[black!45] (axis cs:0,0.45) rectangle (axis cs:0.0769231,0.9);
\fill[black!12] (axis cs:0.0769231,0.45) rectangle (axis cs:1,0.9);
\draw (axis cs:0,2.05) rectangle (axis cs:1,2.5);
\draw (axis cs:0,0.45) rectangle (axis cs:1,0.9);
\node[anchor=south west, font=\footnotesize, inner sep=1.5pt] at (axis cs:0,2.52) {Block of the upper margin};
\node[anchor=south west, font=\footnotesize, inner sep=1.5pt] at (axis cs:0,0.92) {Block of the lower margin};
\draw[densely dotted] (axis cs:0.0769231,0) -- (axis cs:0.0769231,0.9);
\draw[densely dotted] (axis cs:0.0769231,1.45) -- (axis cs:0.0769231,2.5);
\draw[densely dotted] (axis cs:0.870968,0) -- (axis cs:0.870968,0.9);
\draw[densely dotted] (axis cs:0.870968,1.45) -- (axis cs:0.870968,2.5);
\draw[{Stealth[length=3pt]}-{Stealth[length=3pt]}, thin] (axis cs:0.0769231,1.62) -- (axis cs:0.870968,1.62);
\node[font=\footnotesize, fill=white, inner sep=1.5pt] at (axis cs:0.4739,1.62) {choice not settled};
\fill[black!45] (axis cs:0.02,3.22) rectangle (axis cs:0.08,3.42);
\draw (axis cs:0.02,3.22) rectangle (axis cs:0.08,3.42);
\node[anchor=west, font=\footnotesize] at (axis cs:0.09,3.32) {Confirming test};
\fill[black!12] (axis cs:0.56,3.22) rectangle (axis cs:0.62,3.42);
\draw (axis cs:0.56,3.22) rectangle (axis cs:0.62,3.42);
\node[anchor=west, font=\footnotesize] at (axis cs:0.63,3.32) {Clearing test};
\end{axis}
\end{tikzpicture}
\caption{The criterion in the file example. \footnotesize (a) The expected distance $x_E(t)$ by which each test carries the belief across $t$, at $p=1/2$. One test is more informative than the other in Blackwell's sense only if its curve is nowhere lower, and the two curves cross. The firm's decisions call for a comparison at the switching beliefs $t_1$ and $t_2$ alone, and a different test is higher at each. (b) The test that is best in each block at each belief $p$. The same test is best in both blocks, so that the firm's decisions settle the choice, when $p\leq1/13$ and when $p\geq27/31$.}
\label{fig:criterion}
\end{figure}
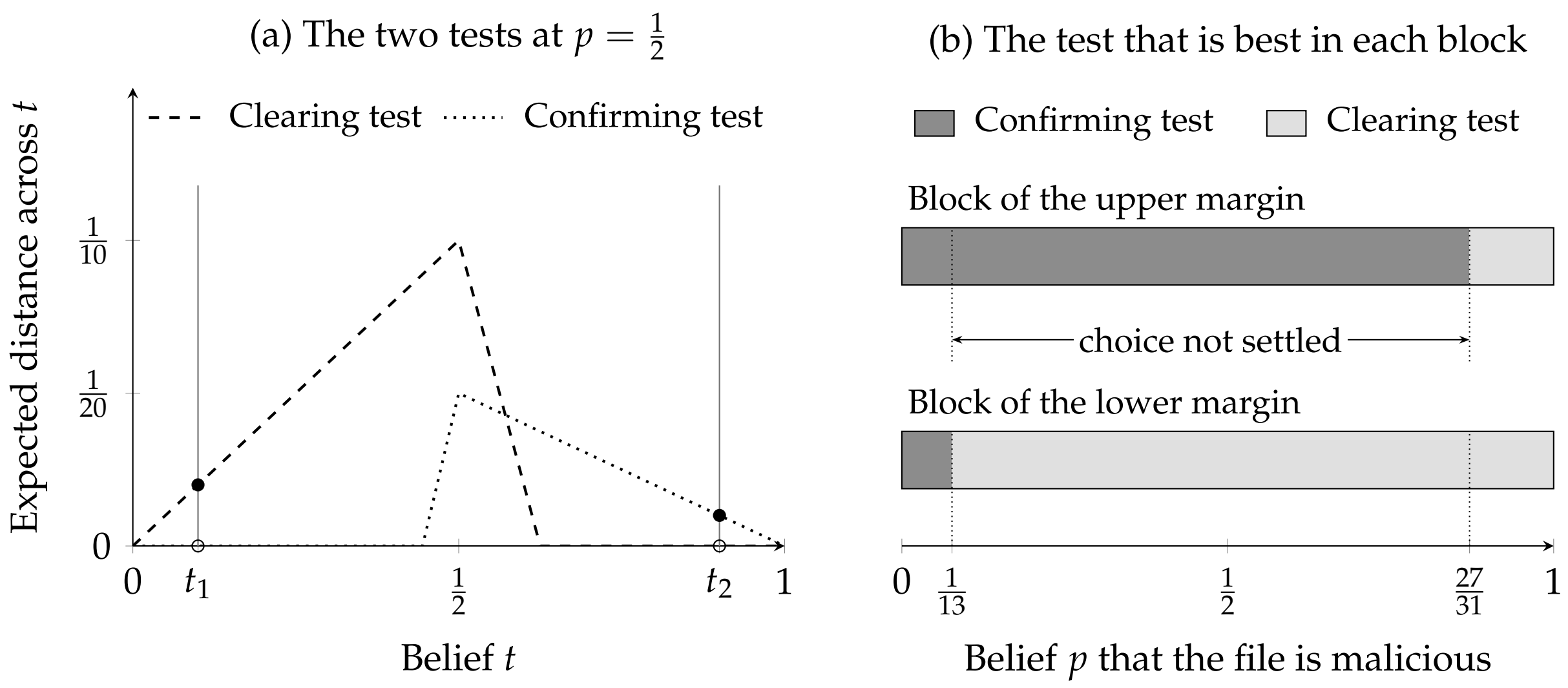

\paragraph{Evidence on one question.}\label{sec:onequestion} Suppose that the AI's evidence bears on a single question, as when a scanner bears only on whether a file is malicious, so that its belief moves along a segment. No three actions can tie on a segment, so every margin is a block, and every action in $A_Q$ other than those chosen at the ends of the segment separates the range, however many states there are. The principal's expected payoffs then have increasing differences in the action and the position of the belief, as in a monotone decision problem \citep{karlinrubin1956}, whatever its payoffs in each state (Supplemental Appendix~\ref{supp:fileexample}).\footnote{The limit is a continuum of ordered actions, such as a quantity or a reported probability. Every action then separates those below it from those above, the principal's marginal payoff at each action can be given any positive weight without changing a decision, and with two states and an optimal action that is interior and varies with the belief, the decisions settle a choice exactly when one experiment is at least as informative as each of the others in the sense of \citet{blackwell1953} (Supplemental Appendix~\ref{supp:fileexample}).} Let $p\in(0,1)$ be the position of the belief on the segment, which is $q(\omega_1)$ when there are two states, and let the principal switch at the beliefs $t_1<\dots<t_K$. As in \eqref{eq:kinks}, the value of an experiment $E$ in the block of the $i$th margin is $\phi_i(E,p)=d_i\,x_E(t_i)$, where, with $p_s$ the position of the posterior after signal $s$,
\[
x_E(t)=\sum_sP(s)\,(p_s-t)^+-(p-t)^+
\]
is the expected distance by which $E$ carries the belief across $t$. The sharpness $d_i$ multiplies the value of every experiment alike, so by Theorem~\ref{thm:test} the decisions settle the choice from a menu at $p$ if and only if some experiment on the menu has the largest $x_E(t_i)$ at every switching belief $t_i$. With two states, one experiment is more informative than another in the sense of \citet{blackwell1953} exactly when its $x_E(t)$ is at least as large at every $t\in[0,1]$. The decisions replace the comparison at every $t$ by one at $K$ of them (Figure~\ref{fig:criterion}a).

In the file example at $p=1/2$, the clearing test has $x_E(t_1)=1/50$ and $x_E(t_2)=0$, and the confirming test has $x_E(t_1)=0$ and $x_E(t_2)=1/100$, so neither is best at both switching beliefs. At other beliefs the same comparison shows that the firm's decisions settle the choice between the two tests when $p\leq1/13$, where the confirming test is best in both blocks, and when $p\geq27/31$, where the clearing test is, and at no belief in between (Figure~\ref{fig:criterion}b and Supplemental Appendix~\ref{supp:fileexample}). What the firm's decisions do not settle is thus the direction in which the AI looks for evidence, toward clearing the file or toward confirming that it is malicious.\footnote{\citet{chemierendorff2019} characterize how a decision maker who knows the stakes divides attention between sources of these two kinds.}

\paragraph{A cautious middle course.}\label{sec:middle} An action is \emph{cautious} if it is optimal for the principal at some belief in $Q$ but in no state that is known for certain, as declining to answer, choosing a reversible action and deferring to a person often are.\footnote{A record of right answers, the best action in each state, says nothing about a cautious action, and an AI that is rewarded only for right answers never takes one (Supplemental Appendix~\ref{supp:rules}). \citet{kalai2025} attribute the tendency of language models to guess rather than abstain to evaluations of this kind.} Suppose that three actions make up $A_Q$: two alternatives $a$ and $b$, and a middle course $c$, such as quarantining, that the principal takes when it is unsure which alternative is right. Then $c$ separates $a$ and $b$ if and only if there is $\alpha\in(0,1)$ such that
\[
V(c,q)\geq\alpha\,V(a,q)+(1-\alpha)\,V(b,q)\qquad\text{at every belief }q\in Q
\]
(Supplemental Appendix~\ref{supp:structure}). In other words, a middle course separates two alternatives exactly when it is at least as good as one lottery that mixes them at every belief that the AI can hold, which on the beliefs with full support means in every state. With more actions, the condition is that the other actions in $A_Q$ divide into two groups, one with each alternative, and that $c$ is at least as good as some lottery over each pair drawn one from each group. A separating action is thus never worse than both of the actions that it separates, the value restriction of \citet{sen1966} with beliefs in the place of voters. In the file example, quarantining costs $1$ whatever the file is, and tossing a coin between deleting and leaving costs $5$ whatever the file is. In the three-state example, the lottery that deletes with probability $2/3$ and otherwise leaves costs $10/3$ if the file is malicious, $20$ if it is critical and $10/3$ if it is routine, and every other lottery costs less than $10/3$ in some state. Quarantining therefore separates deleting and leaving on the beliefs with full support when $\kappa\leq10/3$, and it does not when $\kappa=5$. The principal that can cheaply defer, abstain or reverse is thus the one whose decisions reveal least.\footnote{When each state has a single right action, a cautious action whose payoff in every state falls short of the best by little enough separates the range. With payoffs drawn at random, several blocks are common on ranges of low dimension and rare when the range is the whole simplex over many states, and their frequency varies little with the number of actions, so on ranges of high dimension separating actions come from structure, such as evidence on one question or an option to be cautious, rather than from chance (Supplemental Appendix~\ref{supp:prevalence}).} But cheap caution also enlarges the set of beliefs at which the principal is cautious, and at such a belief an experiment has positive value in a block only if one of its signals leads to a belief at which another action of that block is better. What cheap caution leaves unrevealed thus matters only for sufficiently informative evidence.

\paragraph{Declining to answer.}\label{sec:answers} Consider an AI that answers a question or declines to. There are $n\geq3$ candidate answers, and $\omega_i$ is the state in which answer $i$ is correct. Giving answer $i$, the action $a_i$, is worth $r_i>0$ to the principal if the answer is correct and $-w_i<0$ if it is not, and declining, the action $c$, is worth zero. The principal prefers giving answer $i$ to declining when the probability that it is correct is at least $\tau_i=w_i/(r_i+w_i)$, its \emph{confidence threshold} for that answer. Let the range be the set of beliefs with full support, and let $\sum_i\tau_i>1$, so that the principal declines at some beliefs. Assumption~\ref{ass:acting} then holds, every block contains $c$, and the actions $a_i$ and $a_j$ belong to a common block if and only if $i$ and $j$ are joined by a chain of answers in which consecutive answers $k$ and $l$ have $\tau_k+\tau_l<1$ (Supplemental Appendix~\ref{supp:answers}). Intuitively, the principal gives answer $i$ only at beliefs that put probability at least $\tau_i$ on it. When $\tau_i+\tau_j\geq1$, no belief with full support does so for both answers, so the principal never switches directly between them. If every threshold exceeds one half, so that the AI is to answer only when it is more likely right than wrong, declining separates every two answers, and every answer forms a block of its own with $c$. The decisions then reveal each threshold $\tau_i$, which is the ratio of the cost of a wrong answer to the stakes $r_i+w_i$, and nothing about how the stakes compare across answers.

The comparison that is missing is the one that directs the AI's research. Suppose that the AI can verify one candidate answer, learning whether it is correct, at a belief at which the principal declines and would still decline if any one answer were ruled out. Verifying answer $i$ is worth $q(\omega_i)\,r_i$ to the principal and has value only in the block of $a_i$. Since $r_i=(1-\tau_i)(r_i+w_i)$, which answer to verify depends on the stakes, and by Theorem~\ref{thm:test} the decisions do not settle it when the answers lie in different blocks.

\section{Choices that reveal the relative scale}\label{sec:records}

When decisions do not settle a choice of evidence, the designer needs a record that restricts the relative scale. By Theorem~\ref{thm:main} a record does so only if some choice in it trades a gain in one block against a loss in another. Two common kinds of comparison are insufficient. A comparison in which one option is the principal's own choice at that belief adds nothing to its decisions because every aligned objective already ranks the principal's choice first. 

Reinforcement learning from human feedback fits a reward to comparisons of two responses \citep{christiano2017,ouyang2022}. When the responses are those of the AI's own current policy, the comparisons trade one block against another only where that policy errs, so they say less about the relative scale as the AI's decisions improve. A comparison of two actions in the same block adds nothing either, since decisions already fix the stakes within a block.\footnote{In general, every aligned objective ranks two actions as the principal does at a belief if and only if the principal's gains within the blocks that lie between the two actions are all nonnegative or all nonpositive. When the margin graph is a path, decisions thus fix how the principal ranks the actions on each side of its choice and nothing about comparisons across it (Supplemental Appendix~\ref{supp:records}).}

The comparisons that remain are between actions in different blocks, neither of which the principal would choose. We call two distinct blocks \emph{adjacent} if they share an action, and we call the principal's choice between two actions $a$ and $b$, at a belief at which it would take neither, a \emph{choice between mistakes}. In the file example, it is the firm's choice when quarantining is unavailable.

\begin{proposition}\label{prop:records}
Let Assumption~\ref{ass:acting} hold, let $A_Q$ have at least two actions, and let an objective $S$ be aligned on the range, with scales $\lambda_X$.
\begin{enumerate}[label=(\alph*)]
\item Let adjacent blocks $X$ and $Y$ share an action $c$, and let $\{a,c\}$ be a margin of $X$ and $\{c,b\}$ a margin of $Y$. At every belief $q\in Q$ at which $c$ is the principal's only optimal action, the principal strictly prefers $a$ to $b$, is indifferent between them, or strictly prefers $b$ to $a$ as $\lambda_Y/\lambda_X$ is less than, equal to or greater than
\[
\rho_S(q)=\frac{S(c,q)-S(b,q)}{S(c,q)-S(a,q)},
\]
and $\rho_S$ takes every positive value at such beliefs.
\item Take $K-1$ pairs of adjacent blocks that link all the blocks together. If, for each pair, $S$ is indifferent between the two actions $a$ and $b$ of part (a) at a belief in $Q$ at which the principal is indifferent between them, then $S$ is exact on the range. If a record consists of the principal's rankings of pairs of actions in $A_Q$ at finitely many beliefs in $Q$, and fewer than $K-1$ of these rankings are indifferences, then some objective that is aligned on the range and ranks every recorded pair as the principal does is not exact on the range.
\item Let $c\in A_Q$. For every sufficiently small $\Delta>0$, if the principal's payoff from $c$ is lowered by $\Delta$ in every state, and the objective obtained from $S$ by lowering its payoff from $c$ by some $\hat\Delta>0$ in every state is aligned on the range with the principal's new payoffs, then $\lambda_X=\hat\Delta/\Delta$ for every block $X$ that contains $c$.
\end{enumerate}
\end{proposition}

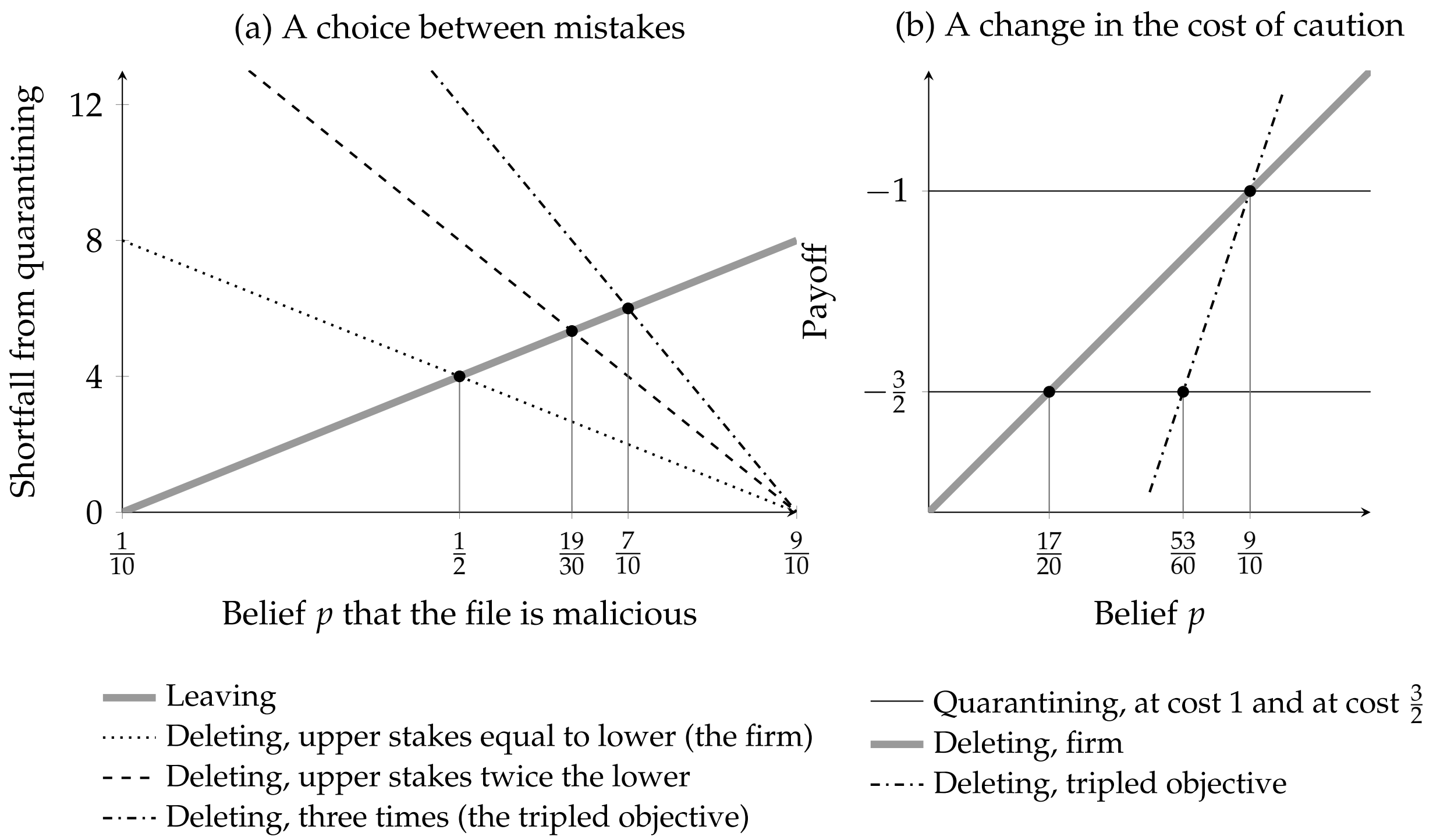
\begin{figure}[t]
\centering
\begin{tikzpicture}
\begin{axis}[
  name=qa, width=0.56\textwidth, height=0.40\textwidth,
  xmin=0.1, xmax=0.9, ymin=0, ymax=13,
  xtick={0.1,0.5,0.633333,0.7,0.9},
  xticklabels={$\frac1{10}$,$\frac12$,$\frac{19}{30}$,$\frac7{10}$,$\frac9{10}$},
  ytick={0,4,8,12},
  tick label style={font=\small}, label style={font=\small},
  xlabel={Belief $p$ that the file is malicious},
  ylabel={Shortfall from quarantining},
  axis x line=bottom, axis y line=left,
  title={\small (a) A choice between mistakes}, title style={yshift=-2pt},
  legend style={draw=none, font=\footnotesize, at={(0.5,-0.36)}, anchor=north, legend columns=1},
  legend cell align=left]
\addplot[black!40, line width=2.4pt, domain=0.1:0.9, samples=2] {10*x-1};
\addlegendentry{Leaving}
\addplot[black, thick, dotted, domain=0.1:0.9, samples=2] {9-10*x};
\addlegendentry{Deleting, upper stakes equal to lower (the firm)}
\addplot[black, thick, dashed, domain=0.25:0.9, samples=2] {18-20*x};
\addlegendentry{Deleting, upper stakes twice the lower}
\addplot[black, thick, dashdotted, domain=0.4667:0.9, samples=2] {27-30*x};
\addlegendentry{Deleting, three times (the tripled objective)}
\addplot[only marks, mark=*, mark size=1.7pt, forget plot] coordinates {(0.5,4) (0.633333,5.33333) (0.7,6)};
\draw[black!55, thin] (axis cs:0.5,0) -- (axis cs:0.5,4);
\draw[black!55, thin] (axis cs:0.633333,0) -- (axis cs:0.633333,5.33333);
\draw[black!55, thin] (axis cs:0.7,0) -- (axis cs:0.7,6);
\end{axis}
\begin{axis}[
  at={(qa.east)}, anchor=west, xshift=1.5cm,
  width=0.40\textwidth, height=0.40\textwidth,
  xmin=0.82, xmax=0.93, ymin=-1.8, ymax=-0.7,
  xtick={0.85,0.883333,0.9},
  xticklabels={$\frac{17}{20}$,$\frac{53}{60}$,$\frac9{10}$},
  ytick={-1.5,-1},
  yticklabels={$-\frac32$,$-1$},
  tick label style={font=\small}, label style={font=\small},
  xlabel={Belief $p$},
  ylabel={Payoff},
  axis x line=bottom, axis y line=left,
  title={\small (b) A change in the cost of caution}, title style={yshift=-2pt},
  legend style={draw=none, font=\footnotesize, at={(0.5,-0.36)}, anchor=north, legend columns=1},
  legend cell align=left]
\addplot[black, thin, domain=0.82:0.93, samples=2] {-1};
\addlegendentry{Quarantining, at cost $1$ and at cost $\frac32$}
\addplot[black, thin, domain=0.82:0.93, samples=2, forget plot] {-1.5};
\addplot[black!40, line width=2.4pt, domain=0.82:0.93, samples=2] {10*x-10};
\addlegendentry{Deleting, firm}
\addplot[black, thick, dashdotted, domain=0.875:0.908, samples=2] {30*x-28};
\addlegendentry{Deleting, tripled objective}
\addplot[only marks, mark=*, mark size=1.7pt, forget plot] coordinates {(0.9,-1) (0.85,-1.5) (0.883333,-1.5)};
\draw[black!55, thin] (axis cs:0.85,-1.8) -- (axis cs:0.85,-1.5);
\draw[black!55, thin] (axis cs:0.883333,-1.8) -- (axis cs:0.883333,-1.5);
\draw[black!55, thin] (axis cs:0.9,-1.8) -- (axis cs:0.9,-1);
\end{axis}
\end{tikzpicture}
\caption{Two records that reveal the relative scale in the file example. \footnotesize (a) Each line is the expected shortfall of a mistake relative to quarantining. A principal is indifferent between the two mistakes where its two lines cross, which is at $p=1/2$ if its stakes on the upper margin equal its stakes on the lower one, as the firm's do, at $19/30$ if they are twice as large, and at $7/10$ if they are three times as large, as under the tripled objective. A principal that prefers to delete at some belief is therefore indifferent at a lower one. (b) The upper switch is the belief at which deleting and quarantining are equally good. When quarantining costs $3/2$ rather than $1$, the firm's upper switch moves from $9/10$ to $17/20$ and the tripled objective's moves only to $53/60$, because each switch moves by the change in cost divided by the stakes there.}
\label{fig:question}
\end{figure}

\noindent Part (a) of Proposition~\ref{prop:records} says that a single question tests the relative scale of two adjacent blocks against a number of the designer's choosing. Intuitively, the designer can compute $\rho_S(q)$ from the objective it has, since $\rho_S(q)$ is the ratio of the two mistakes' shortfalls as that objective measures them. Because $\rho_S$ takes every positive value, the designer sets the threshold by choosing where to ask, and one answer shows which of two experiments the principal prefers. Let $E$ and $E'$ be experiments on a menu at $q$ whose block values differ only in $X$ and $Y$, with $E$ better in $X$ and $E'$ better in $Y$, and let
\[
\rho^*=\frac{\phi^S_Y(E',q)-\phi^S_Y(E,q)}{\phi^S_X(E,q)-\phi^S_X(E',q)}
\]
be the ratio of their advantages as the objective measures them. By part (c) of Theorem~\ref{thm:test}, the principal prefers $E$ to $E'$ exactly when $\lambda_Y/\lambda_X$ exceeds $\rho^*$, and so exactly when it prefers $b$ to $a$ at a belief at which $\rho_S$ equals $\rho^*$. In the file example at $p=1/2$, the tripled objective puts the clearing test's advantage in the block of the lower margin at $1/5$ and the confirming test's advantage in the block of the upper margin at $3/10$, so $\rho^*=3/2$, and its $\rho_S$ equals $3/2$ at $p=19/30$ (Figure~\ref{fig:question}a). The firm prefers to delete there, at an expected cost of $11/3$ against $19/3$ for leaving, so the clearing test is the right one. Choices between mistakes are second choices, the choices people would make if their first choice were unavailable, and like the second choices used in demand estimation \citep{blp2004,conlonmortimer2021}, they reveal substitution that first choices cannot.

Part (b) counts the choices that are needed. An answer at a belief at which the principal is indifferent between the two mistakes fixes the relative scale of two adjacent blocks, and an answer anywhere else only bounds it. The blocks are linked through the actions they share, so $K-1$ pairs of adjacent blocks link them all, and one indifference for each pair fixes every relative scale. No finite record of rankings of actions with fewer indifferences does. A designer need not find a belief of indifference at once, because each answer says on which side of the belief that was asked about the indifference lies. In the file example, bisecting on the belief narrows the relative scale so that the share of the value of costless evidence that is guaranteed after $k$ answers falls short of one by at most a multiple of $2^{-k}$ (Supplemental Appendix~\ref{supp:misdirection}).

Part (c) provides a record that needs no hypothetical question. Classical revealed preference recovers preferences from choices at different prices \citep{afriat1967,varian1982}, and decisions leave the relative scale free because beliefs vary while nothing is traded against a common unit. A change in the cost of a cautious action supplies such a unit. When the cost of quarantining rises by the same small amount whatever the file, each switch around quarantining moves by that amount divided by the stakes there, so the movements of the switches reveal how the stakes on the two sides compare (Figure~\ref{fig:question}b). Such variation arises without anyone asking for it, as when deferring to a person costs more at some times than at others, or when evaluations penalize abstention at different rates. What it requires is that the change in cost be known to be the same in every state, which plays the role of an exclusion restriction. Variation in the cost of one separating action thus fixes the relative scale of every block around it, and also their level when the size of the change is known.

\paragraph{Other routes to the relative scale.}\label{sec:robust} Two other routes are less reliable than they may seem. The first is noise. A model that explains how often each option is chosen, such as the logit model fitted to comparisons in reinforcement learning from human feedback \citep{bradleyterry1952,christiano2017}, recovers payoffs relative to the size of the noise \citep{swaitlouviere1993}. The frequencies of choices near the switches then reveal the relative scale if people are equally noisy in every block, and the principal's relative scale multiplied by the ratio of the precisions if they are not, and the record does not say which case holds (Supplemental Appendix~\ref{supp:records}).\footnote{A designer that knows how the stakes in different blocks are related can also recover the relative scale from decisions alone, but such knowledge is knowledge of the stakes, which is what the record was meant to supply. A fitting procedure that lacks it sets the relative scale by its own inductive bias. A procedure that fits decisions by maximizing the classification margin, for instance, returns a relative scale that reflects how densely the record covers each switch rather than what is at stake there (Supplemental Appendix~\ref{supp:maxmargin}).} The second is uncertainty on the AI's part. An AI that averages over the objectives that are consistent with its principal's decisions is itself aligned, since those objectives share an optimal action at every belief, but it ranks evidence by an average of their scales, which is the principal's only by coincidence.

\section{Capability}\label{sec:capability}

The blocks depend on the beliefs that the AI can hold, and so on its capability. A more capable AI gathers more informative evidence, so it can hold beliefs that a less capable AI cannot. Formally, let an AI's \emph{capability bound} be the most informative evidence that it could gather with every source available to it. The beliefs it can hold are those in the convex hull of the bound's posteriors, whose relative interior is its range, and a more informative bound gives a range at least as large (Supplemental Appendix~\ref{supp:range}).\footnote{This applies the splitting lemma of \citet{aumannmaschler1995} and \citet{kamenicagentzkow2011} to the garblings of a fixed experiment. If one bound is a garbling of another at a common prior with full support, its range lies in the other's \citep[Corollary~6.5.2]{rockafellar1970}. A range records which beliefs a bound makes attainable, not which splits of them it allows after the evidence already gathered, so whether a bound allows the experiment in part~(c) of Proposition~\ref{prop:capability} below depends on the bound. The bound in the example that follows allows it.} Greater capability is therefore a move from a range $Q$ to a larger range $Q'$. We say that a block $X$ of $Q$ \emph{lies within} a block $Y$ of $Q'$ if every margin of $X$ is a margin of $Y$.

\begin{proposition}\label{prop:capability}
Let Assumption~\ref{ass:acting} hold on ranges $Q\subseteq Q'$.
\begin{enumerate}[label=(\alph*)]
\item Every block of $Q$ lies within a block of $Q'$.
\item If an objective is aligned on $Q$, with scales $\lambda_X$ for the blocks of $Q$, and is also aligned on $Q'$, then $\lambda_X=\lambda_Y$ for any two blocks $X$ and $Y$ of $Q$ that lie within the same block of $Q'$.
\item Let an objective $S$ be aligned on $Q$, and suppose that at some belief in the closure of $Q'$ some action in $A_Q$ gives the principal a higher expected payoff than every action that is optimal for $S$. Then there are a belief $q_0\in Q$ and an experiment that keeps beliefs on the range $Q'$ at $q_0$ such that the principal's expected payoff is strictly lower when the AI runs the experiment and then acts than when the AI acts at $q_0$.
\end{enumerate}
\end{proposition}

\noindent Proposition~\ref{prop:capability} says that greater capability never splits a block but can merge blocks, that merged blocks must share a scale, and that an AI whose blocks do not share a scale can do worse for its principal with better evidence. Intuitively, every switch and every tie on the smaller range is still present on the larger one, so no chain of three-way ties is lost, and three-way ties at the new beliefs can join blocks that were separate, while actions that become optimal only at the new beliefs can form blocks of their own. Three implications follow. 

First, an objective that aligns a less capable AI can fail to align a more capable one, although the principal's preferences have not changed. If the objective gives different scales to two blocks that merge on the larger range, it is no longer aligned. At some belief that the less capable AI could not reach, an action that it prefers is incorrect for the principal. An error that was confined to the choice of evidence becomes an error in the action the AI takes.

Second, the error can outweigh the evidence. A principal that acts for itself is never worse off with more informative evidence \citep{blackwell1953}. With the AI acting for it, the principal gains the value of the additional evidence and loses whatever the AI's actions at the new beliefs forgo, and by part (c) the loss can exceed the gain.

Third, the principal's decisions at the more capable AI's beliefs reveal the relative scale of the blocks that merge. Records gathered where evidence is richer than it will be at deployment can therefore supply the relative scale for an AI that works with poorer evidence.

\begin{figure}[t]
\centering
\begin{tikzpicture}[scale=5.0, every node/.style={font=\footnotesize}]
\begin{scope}
\fill[black!10] (0,0) -- (1,0) -- (0.75,0.433013) -- (0.35,0.433013) -- (0.166667,0.288675) -- cycle;
\fill[black!26] (0.5,0.866025) -- (0.166667,0.288675) -- (0.35,0.433013) -- (0.583333,0.721688) -- cycle;
\fill[black!75] (0.1,0.173205) -- (0.166667,0.288675) -- (0.35,0.433013) -- cycle;
\draw (0.166667,0.288675) -- (0.35,0.433013) -- (0.75,0.433013);
\draw (0.35,0.433013) -- (0.583333,0.721688);
\draw[thick, dashed] (0.5,0.866025) -- (0.5,0);
\draw (0,0) -- (1,0) -- (0.5,0.866025) -- cycle;
\draw[densely dotted] (0.04,0.12) rectangle (0.44,0.50);
\node[below] at (0,0) {routine};
\node[below] at (1,0) {critical};
\node[above] at (0.5,0.866025) {malicious};
\node at (0.74,0.21) {leave};
\node[right] at (0.80,0.62) {quarantine};
\draw[-{Stealth[length=3pt]}, thin] (0.80,0.62) -- (0.62,0.535);
\node[left] at (0.20,0.66) {delete};
\draw[-{Stealth[length=3pt]}, thin] (0.20,0.66) -- (0.41,0.64);
\node[left] at (0.495,0.07) {scanner};
\node at (0.5,-0.22) {\small (a) The two ranges};
\end{scope}
\begin{scope}[shift={(1.22,-0.35)}, scale=2.5]
\begin{scope}
\clip (0.04,0.12) rectangle (0.44,0.50);
\fill[black!10] (0,0) -- (1,0) -- (0.75,0.433013) -- (0.35,0.433013) -- (0.166667,0.288675) -- cycle;
\fill[black!26] (0.5,0.866025) -- (0.166667,0.288675) -- (0.35,0.433013) -- (0.583333,0.721688) -- cycle;
\fill[black!75] (0.1,0.173205) -- (0.166667,0.288675) -- (0.35,0.433013) -- cycle;
\draw (0.166667,0.288675) -- (0.35,0.433013) -- (0.75,0.433013);
\draw (0.35,0.433013) -- (0.583333,0.721688);
\draw[very thick, dashed] (0.1,0.173205) -- (0.35,0.433013);
\draw (0,0) -- (0.5,0.866025);
\end{scope}
\draw[densely dotted] (0.04,0.12) rectangle (0.44,0.50);
\fill (0.35,0.433013) circle[radius=0.005];
\draw[fill=white] (0.25,0.346410) circle[radius=0.005];
\fill (0.1,0.173205) circle[radius=0.005];
\node at (0.35,0.20) {leave};
\node[anchor=east] at (0.185,0.455) {delete};
\draw[-{Stealth[length=3pt]}, thin] (0.185,0.455) -- (0.275,0.425);
\node[anchor=west] at (0.455,0.468) {quarantine};
\draw[-{Stealth[length=3pt]}, thin] (0.455,0.468) -- (0.415,0.468);
\node[anchor=west] at (0.455,0.395) {$(\tfrac12,\tfrac1{10},\tfrac25)$};
\draw[-{Stealth[length=3pt]}, thin] (0.455,0.395) -- (0.359,0.430);
\node[anchor=west] at (0.305,0.300) {$(\tfrac25,\tfrac1{20},\tfrac{11}{20})$};
\draw[-{Stealth[length=3pt]}, thin] (0.307,0.305) -- (0.257,0.341);
\node[anchor=west] at (0.118,0.158) {$(\tfrac15,0,\tfrac45)$};
\end{scope}
\node at (1.82,-0.22) {\small (b) Where the more capable AI errs};
\end{tikzpicture}
\caption{Capability merges blocks. \footnotesize The cost of quarantining is $\kappa=5$, and the shaded regions are the firm's choices, as in Figure~\ref{fig:regions}b. (a) The dashed segment is the range of a scanner that cannot tell critical files from routine ones, where half of the legitimate files are critical. On the segment the firm quarantines between leaving and deleting, so there are two blocks. On the triangle, the range of an AI that can tell them apart, the three actions tie and there is one. (b) The dotted rectangle in panel (a), enlarged. The tripled objective is aligned on the segment. On the triangle its boundary between deleting and leaving is the thick dashed line, which differs from the firm's, and in the dark region between the two the AI deletes files that the firm would leave. The open dot is the belief $(2/5,1/20,11/20)$.}
\label{fig:capability}
\end{figure}
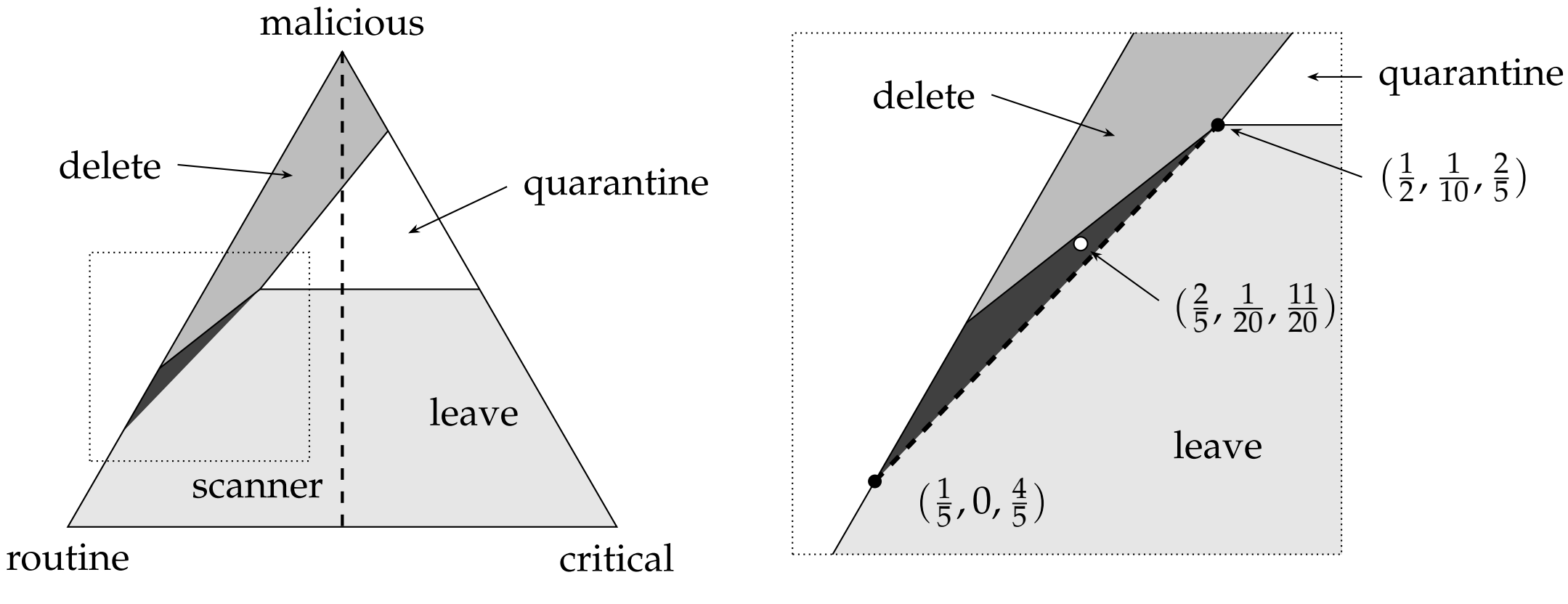

In the three-state example with $\kappa=5$, consider first a scanner that cannot tell critical files from routine ones, deployed where half of the legitimate files are critical. Its beliefs lie on the dashed segment in Figure~\ref{fig:capability} (Supplemental Appendix~\ref{supp:classes}), where the firm leaves a file when the probability that it is malicious is below $1/2$, deletes it above $5/7$ and quarantines it in between. Quarantining separates the segment, so the objective that triples the stakes between quarantining and deleting is aligned there.

An AI that can also inspect how a file is used can hold any belief in the triangle, where the three actions tie at $(1/2,1/10,2/5)$ and the two blocks merge. At the belief $(2/5,1/20,11/20)$ the firm leaves the file, at an expected cost of $4$ against $17/4$ for deleting, but the tripled objective prefers to delete, at a loss to the firm of $1/4$. The incorrect actions lie where the principal's own choice trades one of the old blocks against the other. In other words, the more capable AI makes as decisions some of the choices between mistakes that the less capable AI never faced. The loss at such a belief is at most $\lambda_{\max}/\lambda_{\min}-1$ times the size of the trade-off that the principal's choice makes there, a bound that is attained at $(1/5,0,4/5)$, where the loss is $2$ (Supplemental Appendix~\ref{supp:capbound} states the bound and its conditions).\footnote{Decisions on a range also say nothing about payoffs in directions in which beliefs on the range never move. An objective that puts the cost of deleting any legitimate file at the average over critical and routine files aligns a scanner that cannot tell them apart and can misalign an AI that can, and the same issue arises when an algorithm weighs the risks of different crimes by an average cost \citep{kleinberg2018}. This is the problem of extrapolating beyond the data, and the bound in the Supplemental Appendix has a term for it.}

The same example shows how the principal can be made worse off. Suppose that the scanner puts the probability that a file is malicious at $1/7$, so that the AI's belief over a malicious, a critical and a routine file is $(1/7,3/7,3/7)$. The firm would leave the file, and so does the AI, at an expected cost of $10/7$. Now let the AI also inspect whether the file is in use by a critical process, which is the case exactly when the file is critical. With probability $3/7$ it is, and the AI leaves the file, as the firm would. With probability $4/7$ it is not, the AI's belief becomes $(1/4,0,3/4)$, which lies in the dark region of Figure~\ref{fig:capability}b, and the firm would still leave the file, at an expected cost of $5/2$ against $15/4$ for deleting. The tripled objective deletes it.

The inspection is worth nothing to the firm at this belief, and it raises the firm's expected cost to $15/7$. The same happens at every scanner reading strictly between $1/9$ and $1/5$, and at readings below $1/9$ or of at least $1/5$ the more capable AI acts as the firm would (Supplemental Appendix~\ref{supp:capbound}). Intuitively, the firm quarantines a file because deleting a critical file is costly. Evidence that rules out a critical file removes that reason without showing whether the file is malicious, so deleting and leaving are compared directly, by a relative scale that the firm's decisions on the scanner's range never revealed. The harm arises at intermediate capability, since an AI that learned the state would again act as the firm would.

Capability has two other dimensions. An AI that can run more experiments needs the relative scale in more directions, and an added experiment can lower the principal's payoff although every final action remains right. At $p=1/2$ in the file example, adding the confirming test to a menu that contains only the clearing test lowers the value to the firm of the evidence that an AI with the tripled objective gathers from $1/5$ to $1/10$. An AI that can take more actions, including ways of gathering evidence that have consequences of their own, needs the stakes of the new actions, which no record of choices among the old ones contains.

\section{What designers can do}\label{sec:practice}

The results give a designer that trains an AI on records of a principal's choices a procedure in four steps, each of which uses only the objective that the AI has.

\paragraph{Find the blocks.} The margins are the pairs of actions between which the objective switches on the range, and an action separates two others when every path of margins from one to the other passes through it.\footnote{The blocks are the biconnected components of the margin graph, which standard algorithms find in time linear in the number of margins (Supplemental Appendix~\ref{supp:structure}).} A designer can often see the separating actions without computing anything, as in the three contexts of Section~\ref{sec:contexts}.

\paragraph{Check each choice of evidence.} For each choice of evidence that the AI will face, find the experiments that are best in each block, by valuing the experiments with the AI's objective when only the actions of the block are available, or by restricting the AI to those actions and observing which experiments it would run (Theorem~\ref{thm:test}). If one experiment is best in every block, nothing more is needed. If none is, the guaranteed share says how much the choice can lose, and the same ratio, with block values averaged over the beliefs at which the AI chooses evidence before the ratio is taken, bounds the share of the value of evidence that a whole deployment secures (Supplemental Appendix~\ref{supp:misdirection}).

\paragraph{Ask.} Where a choice is unsettled, obtain the relative scale on which it turns. Table~\ref{tab:records} lists records that are used to train AI systems and what each reveals. When two experiments differ in value only in two adjacent blocks, a choice between mistakes can be posed at the belief at which its answer shows which the principal prefers (Proposition~\ref{prop:records}). Otherwise the designer fixes the relative scales on which the choice turns, one pair of adjacent blocks at a time. Every answer can be kept for later decisions, and no more than $K-1$ answers at beliefs of indifference are ever needed.\footnote{Posing a comparison at a stated belief requires conveying the belief. One way is to group cases by the AI's calibrated belief and ask about the group, as in ``of files like these, two in three are malicious.'' Another is to record the principal's decisions on such cases when the separating action is unavailable and the other options are those it chooses somewhere on the range, which reveal the relative scale of all the blocks around it (Supplemental Appendix~\ref{supp:records}). An option that the principal never chooses can otherwise become its fallback and hide the comparison, which a choice between two named mistakes avoids.}

\paragraph{Scope the task and manage upgrades.} Where the relative scale cannot be obtained, a designer can restrict the choice, by giving the AI experiments that are best in every block, by letting it run its experiments together, or by fixing which evidence it gathers where the guaranteed share is low. An AI that reaches new beliefs needs decisions at those beliefs, without which it can leave its principal worse off than the AI it replaces, and collecting them where blocks merge also supplies the old relative scale (Proposition~\ref{prop:capability}). An AI that gains experiments faces choices of evidence that the old decisions need not settle. And an AI that loses a separating action, as when no one is available to defer to, chooses among the remaining actions by the relative scale and can err although its objective is aligned.

\begin{table}[t]
\centering
\footnotesize
\setstretch{1.05}
\begin{tabularx}{\textwidth}{@{}>{\raggedright\arraybackslash}p{0.30\textwidth}>{\raggedright\arraybackslash}X>{\raggedright\arraybackslash}X@{}}
\toprule
Record & What it reveals & What it leaves open\\
\midrule
Correct answers, as in rewards for verified answers & Decisions when the state is known & Everything about cautious actions, which an AI rewarded only for right answers never takes\\\addlinespace[4pt]
Demonstrations, or decisions at known beliefs & The stakes within each block & The relative scale of the blocks, and the level of the stakes relative to the cost of evidence\\\addlinespace[4pt]
Comparisons in which one option is the principal's choice, including comparisons of the AI's own responses once its decisions are right & Nothing beyond decisions & The same\\\addlinespace[4pt]
Choices between two mistakes on opposite sides of a separating action & The relative scale of the two adjacent blocks & Other blocks, until $K-1$ answers at beliefs of indifference are in hand, and the level of the stakes relative to the cost of evidence\\\addlinespace[4pt]
Decisions at slightly different costs of a separating action & The relative scale of every block around that action, and the level when the size of the change is known & Nothing for those blocks, provided the change in cost is the same in every state and of known size\\\addlinespace[4pt]
Frequencies of noisy choices, fitted by a logit model & The relative scale, if people are equally noisy in every block & Otherwise, how much of the fitted scale is stakes and how much is precision\\\addlinespace[4pt]
The principal's own choices of evidence & The relative scale of blocks on which the experiments bear differently & Whether the choices reflect the stakes or people's mistakes in choosing evidence\\
\bottomrule
\end{tabularx}
\caption{What records of a principal's choices reveal about its stakes. \footnotesize A strict ranking bounds a relative scale, and an indifference fixes it.}
\label{tab:records}
\end{table}

The results also yield (potentially testable) implications for AI systems that maximize their objectives. First, objectives that share an AI's switches but scale its rewards differently across blocks should produce the same actions on given evidence and, from menus that decisions do not settle, can produce different choices of evidence. Second, restricting such an AI to the actions of one block should make its choice among costless experiments independent of how its rewards are scaled. Third, an AI with an incorrect relative scale can gather too much evidence at some beliefs and too little at others, sorted by belief rather than by any error in its predictions. Fourth, an additional source of evidence can lower the principal's payoff even when every final action remains correct, and more informative evidence can lower it through wrong actions. Fifth, the cheaper deferral or abstention is in the training data, the less those data reveal about the relative costs of the mistakes that they guard against. Sixth, with a separating action removed from a menu of the actions that the principal chooses somewhere on the range, objectives that differ only in such scaling can produce different actions.

\section{Conclusion}\label{sec:conclusion}

Are revealed preferences all that an AI needs? The principal's decisions are all it needs for acting, and they are all it needs for a choice among costless experiments when the experiment that is best in one block is best in every other. The blocks arise because a separating action, of which a cheap cautious action is the leading case, stands between them. Where the blocks disagree about the evidence, the AI acts as its principal would and gathers evidence by a relative scale that its training left arbitrary, and it can take wrong actions once it is capable enough to reach beliefs at which the blocks merge. What decisions miss for such a choice is $K-1$ numbers, which choices between two mistakes or small changes in the cost of caution reveal, and costly evidence also needs the level of the stakes. In other words, revealed preferences are all that an AI needs provided that they include the principal's choices among options it would never choose.

These results qualify the proposal of \citet{russell2019} that an AI learn its principal's preferences from behavior. Behavior reveals what the task in which it is observed requires. Demonstrations of decisions at given beliefs, from which inverse reinforcement learning infers a reward \citep{ngrussell2000}, reveal what acting requires, and comparisons of an AI's own responses reveal the relative scale only while the AI still errs. The obstacle is identification rather than the size of the sample, so more such records do not remove it.

Our results also bear on the foundations of incomplete contracts. \citet{maskintirole1999} argue that parties who can foresee their payoffs can contract on them and elicit the realized state by a revelation mechanism, a step that is not robust to small perturbations of common knowledge \citep{afhkt2012}. With an AI whose evidence no one else shares, the elicitation step has no counterpart \citep{kunimoto2008}, and having the AI report its belief under a proper scoring rule does not help, because the rule directs the gathering of evidence as the principal would, on every menu, only if it weights the switching beliefs alone and in the principal's proportions \citep{osband1989,li2022}. The objective must carry the stakes itself. A complete contingent contract can be incomplete over beliefs, and one that is complete over beliefs can be incomplete over experiments, a source of incompleteness distinct from the rigidity and discretion of \citet{battigallimaggi2002}.

Our assumptions favor learning from records, and apart from noise that is the same in every block, which reveals the relative scale, relaxing them adds to the problems identified here rather than removing them. Two directions seem to us the most useful. The implications listed in Section~\ref{sec:practice} can be tested on AI agents with rewards scaled differently across blocks, and it remains to ask when conveying the stakes of caution is worth its cost if the AI can be corrected after deployment.

\clearpage

\small\setstretch{1.15}
\appendix
\phantomsection\label{pg:appstart}
\section{Proofs}\label{app:proofs}

This appendix proves Theorems~\ref{thm:main} and~\ref{thm:test} and Propositions~\ref{prop:records} and~\ref{prop:capability}, and the Supplemental Appendix proves the remaining claims in the text. Throughout, $L=\aff Q$, and (R1) and (R2) are the two parts of the regularity of $V$ in Assumption~\ref{ass:acting}(i): every action in $A_Q$ is optimal for the principal at every belief in some nonempty subset of $Q$ that is open relative to $L$, and no two actions in $A_Q$ have equal expected payoffs at every belief in $L$.

\paragraph{Notation.} For a payoff function $W$, an action $a$ and $\xi\in\R^\Omega$, we write $W(a,\xi)=\sum_\omega\xi(\omega)W(a,\omega)$, which extends the expected payoff to vectors that need not be beliefs, so that $V(a,\xi)-V(b,\xi)=g_{ab}\cdot\xi$ and $S(a,\xi)-S(b,\xi)=e_{ab}\cdot\xi$. Stakes satisfy $g_{ba}=-g_{ab}$ and $g_{ac}=g_{ab}+g_{bc}$, and likewise for $e$. The set $\mathcal R_a=\{q\in\Delta(\Omega):a\in\BR_V(q)\}$ of beliefs at which $a$ is optimal for the principal is closed and convex, being the intersection of $\Delta(\Omega)$ with the closed half-spaces $\{q:g_{ab}\cdot q\geq0\}$, $b\in A$, and $L_0=\{\xi-\xi':\xi,\xi'\in L\}$ is the direction space of $L$. The three actions of a three-way tie form a \emph{triple}. By Assumption~\ref{ass:acting}(ii), each pair of actions in a triple is a margin, and these three margins form the \emph{triangle} of the triple. Two margins are \emph{linked} if they are two of the three margins of a triple. The proofs construct the blocks from three-way ties, and Lemma~\ref{lem:blocks} shows that the construction agrees with the definition in Section~\ref{sec:decisions}. Until that lemma, a block is a maximal set of margins that are joined to one another by chains of linked margins, together with the actions of those margins, so that a margin that belongs to no triple is a block on its own and every margin belongs to exactly one block. An objective is \emph{block-proportional} with scales $\lambda_X>0$ if no action outside $A_Q$ is optimal for it at any belief in $\cl Q$ and
\begin{equation}\label{eq:blocks}
S(a,q)-S(b,q)=\lambda_X\bigl[V(a,q)-V(b,q)\bigr]\quad\text{for all actions $a$ and $b$ of each block $X$ and every }q\in Q.
\end{equation}
By Fact~3 below, \eqref{eq:blocks} holds if and only if \eqref{eq:rep} does, so the block-proportional objectives are those that satisfy the two conditions of Theorem~\ref{thm:main}(a). A payoff function $W$ \emph{makes the principal's decisions} on a set of beliefs if $\BR_W(q)=\BR_V(q)$ at every belief in the set, and it is aligned on the set if $\BR_W(q)\subseteq\BR_V(q)$ there, so that an objective is aligned on the range when it is aligned on $\cl Q$.

\paragraph{Six facts.} \emph{Fact 1. For every $q\in L$, $\spn Q=L_0\oplus\R q$.} Indeed $L=q+L_0$, the coordinates of every element of $L_0$ sum to zero and those of $q$ sum to one.

\emph{Fact 2. If $f$ and $h$ are linear functionals on a finite-dimensional vector space, $h\neq0$, and $f$ vanishes wherever $h$ does, then $f=\lambda h$ for some $\lambda\in\R$.}

\emph{Fact 3. For $e,g\in\R^\Omega$ and $\lambda\in\R$, $e\cdot q=\lambda\,g\cdot q$ at every $q\in Q$ if and only if $e\cdot\xi=\lambda\,g\cdot\xi$ for every $\xi\in\spn Q$}, since both sides are linear in $\xi$. In particular, \eqref{eq:blocks} holds for actions $a$ and $b$ with scale $\lambda_X$ if and only if $e_{ab}\cdot\xi=\lambda_X\,g_{ab}\cdot\xi$ for every $\xi\in\spn Q$, and since $\cl Q\subseteq L\subseteq\spn Q$, it then holds at every belief in $\cl Q$.

\emph{Fact 4. A nonempty subset of $L$ that is open relative to $L$ is not contained in any finite union of affine subspaces of $L$ of smaller dimension}, since such subspaces have Lebesgue measure zero in $L$.

\emph{Fact 5. Under Assumption~\ref{ass:acting}, every action $b\in A_Q$ is the only optimal action at every belief of a nonempty set $U_b\subseteq Q$ that is open relative to $L$.} By (R1), $b$ is optimal at every belief of a nonempty set $O_b\subseteq Q$ that is open relative to $L$. By (R2), the points of $L$ at which $b$ ties with another action of $A_Q$ lie in finitely many affine subspaces of $L$ of smaller dimension, which are closed, and actions outside $A_Q$ are optimal at no belief in $Q$. By Fact~4, removing these subspaces from $O_b$ leaves the required set $U_b$.

\emph{Fact 6. For every payoff function $W$ and belief $q$, some open ball $B$ around $q$ has $\BR_W(q')\subseteq\BR_W(q)$ for every belief $q'\in B$}, since each action outside $\BR_W(q)$ falls short of $W^*(q)$ at $q$ and expected payoffs are continuous.

\subsection*{Margins and three-way ties}

\begin{applemma}\label{lem:local}
Let $Q$ be a range, let $q_1\in Q$ with $\BR_V(q_1)=\{a,b\}$, and suppose that $g_{ab}\cdot z\neq0$ for some $z\in L_0$. Suppose that the AI is aligned at every belief in a subset of $Q$ that is open relative to $L$ and contains $q_1$. Then there is $\lambda>0$ with $e_{ab}\cdot\xi=\lambda\,g_{ab}\cdot\xi$ for every $\xi\in\spn Q$.
\end{applemma}

\begin{proof}
Write $g=g_{ab}$ and $e=e_{ab}$, and rescale $z$ to obtain $w\in L_0$ with $g\cdot w=1$. By the hypothesis and Fact~6, there is a set $O\subseteq Q$ that is open relative to $L$ and contains $q_1$, at each belief of which the AI is aligned and only $a$ and $b$ can be optimal for the principal.

\emph{Step 1: $e$ vanishes where $g$ does, on $O$.} Let $q\in O$ with $g\cdot q=0$. For small $\varepsilon>0$, the beliefs $q\pm\varepsilon w$ lie in $O$ and $g\cdot(q\pm\varepsilon w)=\pm\varepsilon$, so the principal's only optimal action is $a$ at $q+\varepsilon w$ and $b$ at $q-\varepsilon w$. By alignment the same holds for the AI, so $e\cdot(q+\varepsilon w)>0>e\cdot(q-\varepsilon w)$, and letting $\varepsilon\to0$ gives $e\cdot q=0$.

\emph{Step 2: proportionality on $\spn Q$.} Let $\xi\in\spn Q$ have $g\cdot\xi=0$, and write $\xi=z'+tq_1$ with $z'\in L_0$ and $t\in\R$ (Fact~1). Since $g\cdot q_1=0$, also $g\cdot z'=0$, so for small $|s|$ the belief $q_1+sz'$ lies in $O$ and has $g\cdot(q_1+sz')=0$. Step~1 then gives $e\cdot q_1+s\,e\cdot z'=0$ for all small $|s|$, so $e\cdot q_1=e\cdot z'=0$ and $e\cdot\xi=0$. As linear functionals on $\spn Q$, $e$ therefore vanishes wherever $g$ does, and $g\neq0$ there because $g\cdot w=1$, so Fact~2 gives $\lambda\in\R$ with $e\cdot\xi=\lambda\,g\cdot\xi$ for every $\xi\in\spn Q$. Finally, Step~1 at $q_1$ gives $0<e\cdot(q_1+\varepsilon w)=\lambda\varepsilon$ for small $\varepsilon>0$, so $\lambda>0$.
\end{proof}

\noindent The lemma uses no property of $V$ beyond those stated, so it applies to any payoff function in place of $V$.

The next lemma shows that the stakes among any three actions of $A_Q$ are independent, so that three-way ties occupy a set of dimension two less than the range. The one after it shows that alignment near a belief at which exactly three actions tie forces a common scale on their stakes.

\begin{applemma}\label{lem:triple}\label{lem:independent}
Let $V$ be regular on a range $Q$.
\begin{enumerate}[label=(\alph*)]
\item For any three distinct actions $a$, $b$ and $c$ in $A_Q$, the maps $\xi\mapsto g_{ab}\cdot\xi$ and $\xi\mapsto g_{bc}\cdot\xi$ are linearly independent on $\spn Q$.
\item If, in addition, $a$, $b$ and $c$ have equal expected payoffs at some $q_0\in L$, the maps $z\mapsto g_{ab}\cdot z$ and $z\mapsto g_{bc}\cdot z$ are linearly independent on $L_0$.
\item The beliefs in $Q$ at which three or more actions are optimal for the principal lie in finitely many affine subspaces of $L$, each of dimension $\dim L-2$, and there are no such beliefs if $\dim L\leq1$.
\end{enumerate}
\end{applemma}

\begin{proof}
\emph{Part (a).} Suppose that $\alpha_1\,g_{ab}\cdot\xi+\alpha_2\,g_{bc}\cdot\xi=0$ for every $\xi\in\spn Q$, with $(\alpha_1,\alpha_2)\neq(0,0)$. If $\alpha_2=0$, then $g_{ab}$ vanishes on $L\subseteq\spn Q$, contradicting (R2). Otherwise let $\alpha=-\alpha_1/\alpha_2$, so that $g_{bc}\cdot\xi=\alpha\,g_{ab}\cdot\xi$ and $g_{ac}\cdot\xi=(1+\alpha)\,g_{ab}\cdot\xi$ on $\spn Q$. If $\alpha=0$ or $\alpha=-1$, then $g_{bc}$ or $g_{ac}$ vanishes on $L$, contradicting (R2). In every other case, one of the three actions has an expected payoff that is a combination of those of the other two with positive weights that sum to one. Writing $V(x)$ for the function $\xi\mapsto V(x,\xi)$ on $\spn Q$, the identity $V(b)-V(c)=\alpha[V(a)-V(b)]$ rearranges to
\[
\begin{aligned}
V(b)&=\tfrac{1}{1+\alpha}V(c)+\tfrac{\alpha}{1+\alpha}V(a)&&\text{if }\alpha>0,\\
V(c)&=(1+\alpha)V(b)+(-\alpha)V(a)&&\text{if }-1<\alpha<0,\\
V(a)&=\tfrac{1+\alpha}{\alpha}V(b)+\bigl(-\tfrac1\alpha\bigr)V(c)&&\text{if }\alpha<-1.
\end{aligned}
\]
Call the action on the left $x$ and the other two $y$ and $z$. At a belief at which $x$ is optimal, its expected payoff is at least those of $y$ and $z$ and is a combination of them with positive weights, so all three are equal. Hence $x$ is optimal only at beliefs at which $g_{yz}$ vanishes, which by (R2) lie in an affine subspace of $L$ of smaller dimension, and by Fact~4 this contradicts (R1) for $x$.

\emph{Part (b).} Both maps vanish at $q_0$, so a linear combination of them that vanished on $L_0$ would vanish on $L_0\oplus\R q_0=\spn Q$ (Fact~1), contrary to part (a).

\emph{Part (c).} Let three or more actions be optimal at $q\in Q$, and choose three of them, $a$, $b$ and $c$, which belong to $A_Q$. Then $q$ lies in $\{\xi\in L:g_{ab}\cdot\xi=g_{bc}\cdot\xi=0\}$, an affine subspace of $L$ whose direction space is the kernel of the map $z\mapsto(g_{ab}\cdot z,g_{bc}\cdot z)$ on $L_0$. By part (b) this map has rank two, so the subspace has dimension $\dim L-2$, and there are finitely many triples of actions. If $\dim L\leq1$, then $L_0$ cannot carry two linearly independent functionals, so by part (b) no belief in $Q$ has three optimal actions.
\end{proof}

\begin{applemma}\label{lem:localtriple}
Let $Q$ be a range with affine hull $L$, let $q_0\in Q$ with $\BR_V(q_0)=\{a,b,c\}$, and suppose that the maps $z\mapsto g_{ab}\cdot z$ and $z\mapsto g_{bc}\cdot z$ are linearly independent on $L_0$. Suppose that the AI is aligned at every belief in a subset of $Q$ that is open relative to $L$ and contains $q_0$. Then there is $\lambda>0$ with $e_{xy}\cdot\xi=\lambda\,g_{xy}\cdot\xi$ for all $x,y\in\{a,b,c\}$ and every $\xi\in\spn Q$.
\end{applemma}

\begin{proof}
By the linear independence, the map $z\mapsto(g_{ab}\cdot z,\,g_{bc}\cdot z)$ from $L_0$ to $\R^2$ is onto, so there are $z_{ab},z_{bc},z_{ac}\in L_0$ at which it equals $(0,1)$, $(-1,0)$ and $(1,-1)$. By the hypothesis and Fact~6, some set $O\subseteq Q$ that is open relative to $L$ and contains $q_0$ has, at each of its beliefs, the AI aligned and no action outside $\{a,b,c\}$ optimal for the principal. For small $\varepsilon>0$, the beliefs $q_0+\varepsilon z_{ab}$, $q_0+\varepsilon z_{bc}$ and $q_0+\varepsilon z_{ac}$ lie in $O$, and since $g_{ab}\cdot q_0=g_{bc}\cdot q_0=0$, exactly $a$ and $b$ are optimal at the first, exactly $b$ and $c$ at the second, and exactly $a$ and $c$ at the third. Each of $g_{ab}$, $g_{bc}$ and $g_{ac}=g_{ab}+g_{bc}$ is nonzero somewhere on $L_0$, since $g_{ab}\cdot z_{bc}=-1$ and $g_{bc}\cdot z_{ab}=g_{ac}\cdot z_{ab}=1$. Lemma~\ref{lem:local}, applied at each of the three beliefs, therefore gives positive scales with $e_{xy}\cdot\xi=\lambda_{xy}\,g_{xy}\cdot\xi$ for every $\xi\in\spn Q$ and $xy\in\{ab,bc,ac\}$. Because $e_{ac}=e_{ab}+e_{bc}$ and $g_{ac}=g_{ab}+g_{bc}$, it follows that $(\lambda_{ab}-\lambda_{ac})\,g_{ab}\cdot\xi+(\lambda_{bc}-\lambda_{ac})\,g_{bc}\cdot\xi=0$ for every $\xi\in L_0$, and the linear independence gives $\lambda_{ab}=\lambda_{bc}=\lambda_{ac}$. The remaining pairs follow from $e_{yx}=-e_{xy}$ and $g_{yx}=-g_{xy}$, and both sides vanish when $x=y$.
\end{proof}

\subsection*{The tree of blocks}

Let $\mathcal G_Q$ be the graph whose vertices are the actions in $A_Q$, with an edge between two actions if they form a margin, and let $\mathcal H$ be the graph whose vertices are the actions in $A_Q$ and the blocks, with an edge between an action and a block if the action belongs to the block.

\begin{applemma}\label{lem:structure}
Let Assumption~\ref{ass:acting} hold.
\begin{enumerate}[label=(\alph*)]
\item The graph $\mathcal G_Q$ is connected.
\item If $Z$ is a set of margins in which every action belongs to an even number of margins, then $Z$ is the symmetric difference of the triangles of finitely many triples.
\item The margins of each block form a connected subgraph of $\mathcal G_Q$, and $\mathcal H$ is a tree.
\item Let $q\in\cl Q$ and $v\in\BR_V(q)$, let $X$ be a block, and let $x$ be the action adjacent to $X$ on the path in $\mathcal H$ from $X$ to $v$, with $x=v$ if $v$ belongs to $X$. Then $V(x,q)\geq V(a,q)$ for every action $a$ in $X$.
\end{enumerate}
\end{applemma}

\begin{proof}[Proof of Lemma~\ref{lem:structure}(a)]
Suppose that $A_Q$ were the union of disjoint nonempty sets $C_1$ and $C_2$ with no margin between them. The sets $Q_j=Q\cap\bigcup_{c\in C_j}\mathcal R_c$ are nonempty, since every action in $A_Q$ is optimal somewhere in $Q$, they are closed relative to $Q$, and they cover $Q$, since every belief in $Q$ has an optimal action in $A_Q$. As $Q$ is connected, they intersect, so some belief in $Q$ has optimal actions $c\in C_1$ and $d\in C_2$, and by Assumption~\ref{ass:acting}(ii) $\{c,d\}$ is a margin, a contradiction.
\end{proof}

\begin{proof}[Proof of Lemma~\ref{lem:structure}(b)]
Let $Z$ be nonempty, since otherwise there is nothing to prove. For each margin $\{a,b\}\in Z$ choose $p_{ab}\in Q$ at which exactly $a$ and $b$ are optimal, and let $C\subseteq Q$ be the convex hull of these beliefs. The polytopes $C^a=C\cap\mathcal R_a$, $a\in A_Q$, cover $C$, and every nonempty intersection of them is compact and convex, and hence an absolute retract. By the nerve theorem \citep{borsuk1948}, the nerve $N$ of this cover, the simplicial complex whose simplices are the sets of actions whose sets $C^a$ have a point in common, is homotopy equivalent to the contractible set $C$, so its first homology group with coefficients in $\mathbb Z/2$ is zero. By Assumption~\ref{ass:acting}(ii), every edge $\{a,b\}$ of $N$ is a margin and every triangle $\{a,b,c\}$ of $N$ is a triple, whose triangle consists of its three edges. Each margin $\{a,b\}\in Z$ is an edge of $N$, because $a$ and $b$ are both optimal at $p_{ab}\in C$. As a $1$-chain of $N$ with coefficients in $\mathbb Z/2$, $Z$ has zero boundary, because each action belongs to an even number of its margins, so it is the boundary of a $2$-chain, that is, the symmetric difference of the boundaries of finitely many triangles of $N$, each of which is the triangle of a triple.
\end{proof}

\begin{proof}[Proof of Lemma~\ref{lem:structure}(c)]
Two linked margins share an action, so by induction along chains of links, the margins of a block form a connected subgraph of $\mathcal G_Q$. The graph $\mathcal H$ is connected, since by part (a) any two actions are joined by a path in $\mathcal G_Q$, each margin $\{b,b'\}$ of which gives the path $b,Y,b'$ in $\mathcal H$ through its block $Y$, and every block contains an action. Suppose, for contradiction, that $\mathcal H$ contained a cycle. It alternates between actions and blocks, so it has the form $a_1,X_1,a_2,\dots,a_m,X_m,a_1$ with $m\geq2$, distinct actions and distinct blocks, where $a_i$ and $a_{i+1}$ belong to $X_i$ and $a_{m+1}=a_1$. For each $i$, choose a path $P_i$ in $\mathcal G_Q$ from $a_i$ to $a_{i+1}$ that uses only margins of $X_i$ and visits no action twice, and let $Z$ be the symmetric difference of the sets of margins of $P_1,\dots,P_m$. Each $a_i$ is an endpoint of exactly two of the paths, and every other action of a path belongs to two of its margins, so every action belongs to an even number of margins of $Z$, and by part (b) $Z$ is the symmetric difference of the triangles of finitely many triples. The three margins of a triple belong to one block, and distinct blocks share no margin, so the margins of $Z$ that belong to $X_1$, which are the margins of $P_1$, are the symmetric difference of the triangles of those triples whose margins belong to $X_1$. In that set every action belongs to an even number of margins, whereas $a_1$ belongs to exactly one margin of $P_1$, a contradiction.
\end{proof}

\noindent Parts (b) and (c) are the local-to-global step described in Section~\ref{sec:general}. Consistency around three-way ties is the local condition, and the convexity of the range makes it global.

\begin{proof}[Proof of Lemma~\ref{lem:structure}(d)]
\emph{Step 1: reduction.} By Assumption~\ref{ass:acting}(iii), $v\in A_Q$, so the set $U_v$ of Fact~5 is defined. Fix $\tilde q\in U_v$. For $s\in(0,1]$, the belief $q^s=(1-s)q+s\tilde q$ lies in $Q$, because $Q$ is convex and open relative to its affine hull and $q$ lies in its closure \citep[Theorem~6.1]{rockafellar1970}, and $v$ is the only optimal action there, because $V(v,\cdot)-V(b,\cdot)$ is nonnegative at $q$ and positive at $\tilde q$ for every action $b\neq v$. The action $x$ depends only on $v$ and $X$, $q^s\to q$ as $s\to0$, and the inequality to be proved is preserved under limits, so it suffices to prove part (d) at beliefs in $Q$ at which $v$ is the only optimal action, which we assume from now on.

\emph{Step 2: the position of $x$.} If $v$ belongs to $X$, then $x=v$ and the claim holds because $v$ is optimal at $q$. Suppose that $v$ does not belong to $X$ and, for contradiction, that $V(a,q)>V(x,q)$ for some action $a$ of $X$, so that $a\neq x$. The path in $\mathcal H$ from $X$ to $v$ begins $X,x$ and does not contain $a$, since otherwise the edge between $X$ and $a$ and the part of the path from $X$ to $a$ would form a cycle. So the path from $a$ to $v$ is $a$ followed by the path from $X$ to $v$, and it passes through $x$. In a tree every walk between two vertices passes through every vertex of the path between them, so every walk in $\mathcal H$ from $v$ to $a$ passes through $x$.

\emph{Step 3: a walk of optimal actions.} By Lemma~\ref{lem:triple}(c), the beliefs in $Q$ at which three or more actions are optimal lie in finitely many affine subspaces $M_1,\dots,M_N$ of $L$ of dimension $\dim L-2$, and there are none if $\dim L\leq1$. If $(1-t)q+tp\in M_j$ for some $t\in(0,1]$, then $p$ lies in the affine hull of $M_j\cup\{q\}$, which has dimension at most $\dim L-1$. By Fact~4, some $q_a\in U_a$ lies outside these $N$ hulls, so that no belief on the segment from $q$ to $q_a$ other than $q$ has three or more optimal actions. For $t\in[0,1]$ let $q_t=(1-t)q+tq_a$, which lies in $Q$, and for each action $b$ let $\mathcal I_b$ be the set of $t$ at which $b$ is optimal at $q_t$, a closed interval, possibly empty, since it is defined by finitely many affine inequalities in $t$. These intervals cover $[0,1]$, $0$ belongs to $\mathcal I_v$ only, $1$ belongs to $\mathcal I_a$ only, and $\max\mathcal I_v>0$ by Fact~6. Let $b_0=v$. Given $b_j$ with $\tau_j=\max\mathcal I_{b_j}<1$, some action $b_{j+1}$ is optimal at $q_t$ for a sequence of values of $t>\tau_j$ that decreases to $\tau_j$, so $\tau_j\in\mathcal I_{b_{j+1}}$ and $\max\mathcal I_{b_{j+1}}>\tau_j$. Both $b_j$ and $b_{j+1}$ are optimal at $q_{\tau_j}$, which has at most two optimal actions because $\tau_j>0$, so they form a margin. The numbers $\max\mathcal I_{b_j}$ increase strictly, so no action appears twice, and the construction ends with an action whose interval contains $1$, which is $a$. Hence $v=b_0,b_1,\dots,b_m=a$ is a walk in $\mathcal G_Q$ in which each $b_j$ is optimal at some belief $q_t$.

\emph{Step 4: the contradiction.} Replacing each margin $\{b_{j-1},b_j\}$ of this walk by the path $b_{j-1},Y_j,b_j$ in $\mathcal H$, where $Y_j$ is the block that contains the margin, gives a walk in $\mathcal H$ from $v$ to $a$. By Step~2 it passes through $x$, so $x=b_j$ for some $j$, and $V(x,q_t)\geq V(a,q_t)$ for some $t\in[0,1]$. But $t\mapsto V(a,q_t)-V(x,q_t)$ is affine, positive at $t=0$ by assumption and positive at $t=1$ because $a$ is the only optimal action at $q_a$, so it is positive on $[0,1]$, a contradiction.
\end{proof}

Root the tree $\mathcal H$ at an action $r\in A_Q$. For each block $X$, let $r_X$ be its parent, the action adjacent to $X$ on the path from $X$ to $r$, and let $h_X(q)=V^*_X(q)-V(r_X,q)$. For an action $v\in A_Q$, let $\mathcal P(v)$ be the set of blocks on the path from $r$ to $v$ and, for $X\in\mathcal P(v)$, let $x^v_X$ be the action that follows $X$ on that path.

\begin{applemma}\label{lem:decomp}
Let Assumption~\ref{ass:acting} hold.
\begin{enumerate}[label=(\alph*)]
\item If $q\in\cl Q$ and $v\in\BR_V(q)$, then $V(x^v_X,q)=V^*_X(q)$ for every $X\in\mathcal P(v)$, and $V(r_X,q)=V^*_X(q)$ for every block $X\notin\mathcal P(v)$.
\item Let $\lambda_X>0$ for each block $X$, and let $S$ satisfy $e_{ab}\cdot\xi=\lambda_X\,g_{ab}\cdot\xi$ for all actions $a$ and $b$ in $X$, every block $X$ and every $\xi\in\spn Q$. For every $v\in A_Q$ and every $q$ in $\cl Q$,
\[
S(v,q)-S(r,q)=\sum_{X\in\mathcal P(v)}\lambda_X\bigl[V(x^v_X,q)-V(r_X,q)\bigr]\leq\sum_X\lambda_Xh_X(q),
\]
with equality if and only if $V(x^v_X,q)=V^*_X(q)$ for every $X\in\mathcal P(v)$ and $V(r_X,q)=V^*_X(q)$ for every block $X\notin\mathcal P(v)$.
\item If, in addition, no action outside $A_Q$ is optimal for the AI at any belief in $Q$, then $S^*(q)=S(r,q)+\sum_X\lambda_Xh_X(q)$ at every $q\in\cl Q$, and every experiment $E$ whose posteriors at a belief $q\in Q$ lie in $\cl Q$ has $I_S(E,q)=\sum_X\lambda_X\phi_X(E,q)$, with $\phi_X(E,q)\geq0$. In particular, taking $S=V$ and every $\lambda_X=1$, $V^*(q)=V(r,q)+\sum_Xh_X(q)$ on $\cl Q$ and $I_V(E,q)=\sum_X\phi_X(E,q)$.
\item If $S$ satisfies the hypothesis of part (b) with the same scale $\lambda$ on every block, then $S(v,q)=\lambda V(v,q)+\beta\cdot q$ for every $v\in A_Q$ and every $q\in\cl Q$, where $\beta=S(r,\cdot)-\lambda V(r,\cdot)$. Conversely, if $S(v,q)=\lambda V(v,q)+\beta\cdot q$ for every $v\in A_Q$ and $q\in Q$, for some $\lambda>0$ and $\beta:\Omega\to\R$, then $S$ satisfies the hypothesis of part (b) with the scale $\lambda$ on every block, and these scales are the only ones with which it does.
\item $\sum_XV(r_X,\xi)-V(r,\xi)=\sum_{c\in A_Q}(n_c-1)V(c,\xi)$ for every $\xi\in\R^\Omega$.
\end{enumerate}
\end{applemma}

\begin{proof}
\emph{Part (a).} In $\mathcal H$ rooted at $r$, $r_X$ is the parent of $X$, the other actions of $X$ are its children, and $\mathcal P(v)$ is the set of blocks that are ancestors of $v$. The path from $X$ to $v$ therefore begins with the child $x^v_X$ if $X\in\mathcal P(v)$, and with the parent $r_X$ otherwise. In either case Lemma~\ref{lem:structure}(d) shows that this action, which belongs to $X$, is best in $X$ at $q$.

\emph{Part (b).} Write the path from $r$ to $v$ as $r=y_0,X_1,y_1,\dots,X_m,y_m=v$, so that $\mathcal P(v)=\{X_1,\dots,X_m\}$, $y_{j-1}=r_{X_j}$ and $y_j=x^v_{X_j}$. Because $\cl Q\subseteq\spn Q$, the hypothesis gives $S(y_j,q)-S(y_{j-1},q)=\lambda_{X_j}[V(y_j,q)-V(y_{j-1},q)]$, and summing over $j$ gives the equality, with an empty sum if $v=r$. Each term with $X\in\mathcal P(v)$ is at most $\lambda_Xh_X(q)$, because $V(x^v_X,q)\leq V^*_X(q)$, and each $\lambda_Xh_X(q)$ with $X\notin\mathcal P(v)$ is nonnegative, because $r_X$ belongs to $X$. Since the scales are positive, equality holds exactly under the stated conditions.

\emph{Part (c).} Fix $q\in Q$ and $v\in\BR_V(q)$. By part (a), $v$ meets the condition for equality in part (b), so $S(v,q)-S(r,q)=\sum_X\lambda_Xh_X(q)\geq S(v',q)-S(r,q)$ for every $v'\in A_Q$. Since no action outside $A_Q$ is optimal for the AI, $S^*(q)=S(r,q)+\sum_X\lambda_Xh_X(q)$, and since both sides are continuous, the equality extends to $\cl Q$. Let $E$ have signals $s$ with probabilities $P(s)>0$ and posteriors $q_s\in\cl Q$ at $q$. The posteriors average to $q$, so $\sum_sP(s)W(c,q_s)=W(c,q)$ for every payoff function $W$ and action $c$, and since $h_X$ and $V^*_X$ differ by a linear function,
\[
I_S(E,q)=\sum_X\lambda_X\Bigl[\sum_sP(s)h_X(q_s)-h_X(q)\Bigr]=\sum_X\lambda_X\phi_X(E,q).
\]
Each $\phi_X(E,q)$ is nonnegative by Jensen's inequality, because $V^*_X$ is convex. The hypotheses hold for $S=V$ with every $\lambda_X=1$, since then $e_{ab}=g_{ab}$ and no action outside $A_Q$ is optimal for the principal at any belief in $Q$.

\emph{Part (d).} Part (b), applied to $S$ with every scale equal to $\lambda$ and to $V$ with every scale equal to one, gives $S(v,q)-S(r,q)=\lambda[V(v,q)-V(r,q)]$ for every $v\in A_Q$ and $q\in\cl Q$, which rearranges to $S(v,q)=\lambda V(v,q)+\beta\cdot q$. Conversely, subtracting the identities for $a$ and $b$ gives $e_{ab}\cdot q=\lambda\,g_{ab}\cdot q$ at every $q\in Q$, which Fact~3 extends to $\spn Q$. If $S$ also satisfied the hypothesis of part (b) with a scale $\lambda'_X$ on a block $X$, a margin $\{a,b\}$ of $X$ would give $(\lambda'_X-\lambda)\,g_{ab}\cdot\xi=0$ for every $\xi\in\spn Q$, and since $g_{ab}$ does not vanish on $L$ by (R2), $\lambda'_X=\lambda$.

\emph{Part (e).} Every block has exactly one parent. An action $c\neq r$ is the parent of $n_c-1$ of the $n_c$ blocks that contain it, the remaining one being its own parent, and $r$ is the parent of all $n_r$ blocks that contain it. Hence $\sum_XV(r_X,\xi)=\sum_{c\neq r}(n_c-1)V(c,\xi)+n_rV(r,\xi)$, and subtracting $V(r,\xi)$ gives the claim.
\end{proof}

The next lemma splits the principal's gain from one action over another into gains within blocks.

\begin{applemma}\label{lem:gains}
Let Assumption~\ref{ass:acting} hold. For actions $a,b\in A_Q$, let $a=z_0,X_1,z_1,\dots,X_m,z_m=b$ be the path from $a$ to $b$ in the tree $\mathcal H$, and for $\xi\in\R^\Omega$ let the principal's \emph{gain from $a$ over $b$ within block $X$} be $\delta_X(a,b;\xi)=V(z_{j-1},\xi)-V(z_j,\xi)$ if $X=X_j$ and $\delta_X(a,b;\xi)=0$ if $X$ is not on the path. Write $\delta(a,b;\xi)\in\R^K$ for the vector of these gains.
\begin{enumerate}[label=(\alph*)]
\item For every $\xi\in\R^\Omega$, $V(a,\xi)-V(b,\xi)=\sum_X\delta_X(a,b;\xi)$ and $\delta(b,a;\xi)=-\delta(a,b;\xi)$.
\item If $S$ is block-proportional with scales $\lambda_X$, then $S(a,\xi)-S(b,\xi)=\sum_X\lambda_X\,\delta_X(a,b;\xi)$ for every $\xi\in\spn Q$.
\item If $q\in\cl Q$ and $a\in\BR_V(q)$, then $\delta_X(a,b;q)\geq0$ for every block $X$.
\end{enumerate}
\end{applemma}

\begin{proof}
A path in a tree visits no vertex twice, so the gains are well defined. Part (a) holds because the gains telescope, and reversing the path reverses the sign of each gain. For part (b), $z_{j-1}$ and $z_j$ belong to $X_j$, so by Fact~3 $S(z_{j-1},\xi)-S(z_j,\xi)=\lambda_{X_j}[V(z_{j-1},\xi)-V(z_j,\xi)]$ for every $\xi\in\spn Q$, and summing over $j$ gives the claim. For part (c), the action adjacent to $X_j$ on the path from $X_j$ to $a$ is $z_{j-1}$, so Lemma~\ref{lem:structure}(d), applied with $v=a$, gives $V(z_{j-1},q)\geq V(z_j,q)$, and the gains within blocks off the path are zero.
\end{proof}

\subsection*{Separating actions}

The next lemma relates the blocks constructed from three-way ties to the actions that separate others, and shows that they are the blocks of Section~\ref{sec:decisions}.

\begin{applemma}\label{lem:blocks}
Let Assumption~\ref{ass:acting} hold, with the blocks constructed from three-way ties.
\begin{enumerate}[label=(\alph*)]
\item For every block $X$ and every action $x$ of $X$, any two actions of $X$ other than $x$ are joined by a path of margins of $X$ that does not visit $x$.
\item For distinct actions $a$, $b$ and $c$ in $A_Q$, the following are equivalent: (i) $c$ separates $a$ and $b$, (ii) $c$ lies on the path from $a$ to $b$ in the tree $\mathcal H$, and (iii) every path from $a$ to $b$ in $\mathcal G_Q$ visits $c$.
\item For an action $c\in A_Q$, the following are equivalent: (i) $c$ separates some pair of actions in $A_Q$, (ii) $c$ belongs to at least two blocks, and (iii) the beliefs in $Q$ at which $c$ is not optimal for the principal form a disconnected set.
\item Two distinct actions in $A_Q$ belong to a common block if and only if no action separates them. A set of at least two actions in $A_Q$ is the set of actions of a block if and only if no two of its actions are separated by another action and it is contained in no larger set with this property. A margin whose two actions belong to a block $X$ is a margin of $X$.
\end{enumerate}
\end{applemma}

\noindent By part (d), the construction yields the blocks of Section~\ref{sec:decisions}, with the margins between the actions of a block as its margins, so any two margins of a block are joined by a chain of linked margins.

\begin{proof}
Parts (c) and (d) follow from part (b) because $\mathcal H$ is a tree (Lemma~\ref{lem:structure}(c)). We use two facts throughout. Two distinct blocks share at most one action, since blocks $X\neq X'$ that both contained distinct actions $a$ and $b$ would form the cycle $a,X,b,X',a$ in $\mathcal H$. And for actions $v\neq c$ in $A_Q$, the set $D^c_v$ of beliefs in $Q$ at which $v$ is optimal for the principal and $c$ is not equals $\{q\in Q:g_{vb}\cdot q\geq0\text{ for every }b\in A,\ g_{vc}\cdot q>0\}$, so it is convex, and hence connected, and it is nonempty because it contains $U_v$.

\emph{Part (a).} Let $x$ be an action of a block $X$, and suppose, for contradiction, that the actions of $X$ other than $x$ were the union of disjoint nonempty sets $T_1$ and $T_2$ with no margin of $X$ between them. Every margin of $X$ contains an action other than $x$, and its actions other than $x$ all lie in one of the two sets, which we call its type. If two margins of $X$ are linked, the three actions of their triple are pairwise joined by margins of $X$, so at most one of them is $x$ and the others all lie in one set $T_j$. Each of the three margins contains two of the three actions, and so an action of $T_j$, so linked margins have the same type. Chains of links connect the margins of $X$, so all of them have one type, say $T_1$. But an action of $T_2$ belongs to a margin of $X$, whose type is $T_2$.

\emph{Part (b), (ii) implies (i).} Let $c$ lie on the path from $a$ to $b$ in $\mathcal H$. Then $a$ and $b$ lie in different components of the forest obtained by removing $c$ from $\mathcal H$. Let $C_1$ be the set of actions in the component that contains $a$, and $C_2$ the set of the other actions of $A_Q\setminus\{c\}$, which contains $b$. No margin joins $C_1$ and $C_2$, because a margin $\{x,y\}$ with $x,y\neq c$ gives the path $x,Z,y$ through its block $Z$, which avoids $c$. For $j=1,2$, let $W_j$ be the set of beliefs in $Q\setminus\mathcal R_c$ at which some action of $C_j$ is optimal. Every belief in $Q\setminus\mathcal R_c$ has an optimal action in $A_Q\setminus\{c\}$, so the two sets cover $Q\setminus\mathcal R_c$. They are disjoint, since by Assumption~\ref{ass:acting}(ii) actions $x\in C_1$ and $y\in C_2$ that were both optimal at a belief in $Q$ would form a margin. Each is closed relative to $Q\setminus\mathcal R_c$, being its intersection with $\bigcup_{x\in C_j}\mathcal R_x$, and so also open relative to it, as the complement of the other. A connected subset of $Q\setminus\mathcal R_c$ therefore lies in $W_1$ or in $W_2$, whereas the beliefs at which $a$ is optimal lie in $W_1$ and those at which $b$ is optimal lie in $W_2$. Hence $c$ separates $a$ and $b$.

\emph{Part (b), (i) implies (iii).} Let $a=v_0,v_1,\dots,v_m=b$ be a path in $\mathcal G_Q$ that does not visit $c$. For each $i\geq1$, a belief $q_i\in Q$ with $\BR_V(q_i)=\{v_{i-1},v_i\}$ lies in $D^c_{v_{i-1}}\cap D^c_{v_i}$. The sets $D^c_{v_0},\dots,D^c_{v_m}$ are connected and consecutive ones intersect, so their union is a connected set of beliefs in $Q$ at which $c$ is not optimal. It contains $q_1$, at which $a$ is optimal, and $q_m$, at which $b$ is, so $c$ does not separate $a$ and $b$.

\emph{Part (b), (iii) implies (ii).} Let $a=z_0,X_1,z_1,\dots,X_m,z_m=b$ be the path from $a$ to $b$ in $\mathcal H$, and suppose that no $z_j$ is $c$. For each $j$, the actions $z_{j-1}$ and $z_j$ are joined by a path of margins of $X_j$ that does not visit $c$, by part (a) if $c$ is an action of $X_j$, and by Lemma~\ref{lem:structure}(c) if it is not. Following these paths in turn gives a walk in $\mathcal G_Q$ from $a$ to $b$ that does not visit $c$, and the walk contains a path.

\emph{Part (c).} If $c$ separates $a$ and $b$, then by part (b) it lies on the path from $a$ to $b$ in $\mathcal H$, between two distinct blocks that contain it. Conversely, if $c$ belongs to distinct blocks $X$ and $Y$, choose actions $a\neq c$ of $X$ and $b\neq c$ of $Y$, which are distinct because $X$ and $Y$ share at most one action. Then $a,X,c,Y,b$ is the path from $a$ to $b$ in $\mathcal H$, and $c$ separates $a$ and $b$ by part (b). So (i) and (ii) are equivalent. If $c$ separates $a$ and $b$, the nonempty sets $D^c_a$ and $D^c_b$ lie in $Q\setminus\mathcal R_c$, which is therefore disconnected, so (i) implies (iii). Conversely, let $Q\setminus\mathcal R_c$ be the union of disjoint nonempty sets $W_1$ and $W_2$ that are open relative to it. For $j=1,2$, choose $q_j\in W_j$ and an action $a_j\neq c$ that is optimal at $q_j$. The connected set $D^c_{a_j}$ contains $q_j$ and so lies in $W_j$. Every belief in $Q\setminus\mathcal R_c$ at which $a_j$ is optimal therefore lies in $W_j$, so $a_1\neq a_2$, and no connected subset of $Q\setminus\mathcal R_c$ contains a belief at which $a_1$ is optimal and one at which $a_2$ is. Hence $c$ separates $a_1$ and $a_2$, and (iii) implies (i). This step uses the convexity of the sets $D^c_v$ and not Assumption~\ref{ass:acting}.

\emph{Part (d).} An action outside $A_Q$ separates no two actions in $A_Q$, because it is optimal at no belief in the connected set $Q$. By part (b), no action separates distinct actions $a$ and $b$ in $A_Q$ if and only if the path between them in $\mathcal H$ contains no other action, that is, if and only if it is $a,X,b$ for some block $X$, which proves the first statement. Call a set of at least two actions in $A_Q$ \emph{unseparated} if no two of its actions are separated by another action, that is, if any two of them belong to a common block. The actions of a block form an unseparated set. Every unseparated set $P$ is contained in the set of actions of a single block. Indeed, fix distinct $u,v\in P$ and the block $X$ that contains both, let $x\in P$ differ from them, and let $X_u$ and $X_v$ be the blocks that contain $x$ together with $u$ and with $v$. If $x$ were not an action of $X$, then $X_u$ and $X_v$ would differ from $X$. They could not coincide, since that block would share $u$ and $v$ with $X$, and if they differed, $u,X,v,X_v,x,X_u,u$ would be a cycle in $\mathcal H$. Consequently the actions of a block form a maximal unseparated set, since an unseparated set that contains them lies in the actions of a block that shares at least two actions with the first and so is the same block, and every maximal unseparated set, which lies in the actions of a block, equals them. Finally, a margin whose actions belong to a block $X$ belongs to a block that contains both actions, which is $X$.
\end{proof}

\subsection*{Proof of Theorem~\ref{thm:main}}

\begin{proof}
By Lemma~\ref{lem:blocks}(d), the blocks of Section~\ref{sec:decisions} and their margins are those constructed from three-way ties. By Fact~3, an objective satisfies the two conditions of part (a) with scales $\lambda_X$ if and only if it is block-proportional with those scales, that is, if and only if no action outside $A_Q$ is optimal for it on $\cl Q$ and $e_{ab}\cdot\xi=\lambda_X\,g_{ab}\cdot\xi$ for all actions $a$ and $b$ of each block $X$ and every $\xi\in\spn Q$.

\emph{Part (a): alignment implies block-proportionality.} Let the AI be aligned on $\cl Q$, and let $\{a,b\}$ be a margin, with $\BR_V(q_1)=\{a,b\}$ at some $q_1\in Q$. The stake $g_{ab}$ is not zero on $L_0$, since otherwise $a$ and $b$ would have equal expected payoffs throughout $L$, contrary to (R2), so Lemma~\ref{lem:local}, with $Q$ as the set on which the AI is aligned, gives $\lambda_{ab}>0$ with $e_{ab}\cdot\xi=\lambda_{ab}\,g_{ab}\cdot\xi$ for every $\xi\in\spn Q$. Now let $\{a,b,c\}$ be a triple, with $\BR_V(q_0)=\{a,b,c\}$ at some $q_0\in Q$. Lemma~\ref{lem:triple}(b) and Lemma~\ref{lem:localtriple} give $\lambda>0$ with $e_{xy}\cdot\xi=\lambda\,g_{xy}\cdot\xi$ for all $x,y\in\{a,b,c\}$ and every $\xi\in\spn Q$. By the linear independence, each of $g_{ab}$, $g_{bc}$ and $g_{ac}=g_{ab}+g_{bc}$ is nonzero somewhere on $L_0$, so $\lambda_{ab}=\lambda_{bc}=\lambda_{ac}=\lambda$. Linked margins thus have the same scale, and so do all the margins of a block, which chains of links connect. Calling this scale $\lambda_X$ and summing along a path of margins of $X$ between two of its actions (Lemma~\ref{lem:structure}(c)) gives $e_{ab}\cdot\xi=\lambda_X\,g_{ab}\cdot\xi$ for all actions $a$ and $b$ of $X$ and every $\xi\in\spn Q$. Finally, an action outside $A_Q$ that was optimal for the AI at a belief in $\cl Q$ would not be optimal for the principal there, by Assumption~\ref{ass:acting}(iii), and the AI would be misaligned.

\emph{Part (a): block-proportionality implies the same decisions.} Let $S$ be block-proportional and let $q\in\cl Q$. The actions that are optimal at $q$ for the AI or for the principal belong to $A_Q$, by assumption and by Assumption~\ref{ass:acting}(iii). By Lemma~\ref{lem:decomp}(c), applied to $S$ and to $V$, an action $v\in A_Q$ is optimal for the AI at $q$ if and only if $S(v,q)-S(r,q)=\sum_X\lambda_Xh_X(q)$, and optimal for the principal if and only if $V(v,q)-V(r,q)=\sum_Xh_X(q)$. By Lemma~\ref{lem:decomp}(b), each of these holds if and only if the condition for equality there holds, and that condition does not involve the scales. Hence $\BR_S(q)=\BR_V(q)$, and in particular the AI is aligned on the range.

\emph{Part (a): the scales are unique, and every vector of scales arises.} If \eqref{eq:blocks} holds for a block $X$ with scales $\lambda_X$ and $\lambda'_X$, a margin $\{a,b\}$ of $X$ gives $(\lambda_X-\lambda'_X)\,g_{ab}\cdot\xi=0$ for every $\xi\in\spn Q$, and $g_{ab}$ does not vanish on $L$ by (R2), so $\lambda_X=\lambda'_X$. Given $\lambda_X>0$ for each block $X$, let $S(v,\cdot)=V(r,\cdot)+\sum_{X\in\mathcal P(v)}\lambda_X[V(x^v_X,\cdot)-V(r_X,\cdot)]$ for $v\in A_Q$. If $v$ is a child of the block $X$, then $\mathcal P(v)=\mathcal P(r_X)\cup\{X\}$, $x^v_X=v$ and $x^v_Y=x^{r_X}_Y$ for every $Y\in\mathcal P(r_X)$, so $S(v,\cdot)=S(r_X,\cdot)+\lambda_X[V(v,\cdot)-V(r_X,\cdot)]$ for every action $v$ of $X$, including $r_X$. Hence $e_{ab}=\lambda_X\,g_{ab}$ for all actions $a$ and $b$ of $X$. Setting $S(c,\omega)=\underline s-1$ for every action $c\notin A_Q$ and every $\omega$, where $\underline s$ is the smallest value of $S(v,\omega)$ over $v\in A_Q$ and $\omega\in\Omega$, ensures that no such action is optimal for the AI at any belief, so $S$ is block-proportional with the given scales.

\emph{Part (b).} By part (a), $S$ is block-proportional, and Lemma~\ref{lem:decomp}(c) gives the two decompositions and $\phi_X(E,q)\geq0$. The same lemma with $S=V$, together with Lemma~\ref{lem:decomp}(e), gives $V^*(q)=V(r,q)+\sum_Xh_X(q)=\sum_XV^*_X(q)-\sum_{c\in A_Q}(n_c-1)V(c,q)$ on $\cl Q$, which is the formula for $V^*$ in the text.

\emph{Part (c).} An action that separates two actions in $A_Q$ belongs to $A_Q$, since an action outside it is optimal at no belief in the connected set $Q$, and by Lemma~\ref{lem:blocks}(c) it belongs to at least two blocks, so $K\geq2$. Conversely, if $X$ and $Y$ are distinct blocks, the path from $X$ to $Y$ in $\mathcal H$ begins $X,c,X'$ for an action $c$ and a block $X'\neq X$, both of which contain $c$, so $c$ separates the range by Lemma~\ref{lem:blocks}(c). For the last statement, suppose first that no action separates the range, so that $K\leq1$, and let $S$ be aligned on the range, and hence block-proportional. If $K=1$, Lemma~\ref{lem:decomp}(d) shows that $S$ is exact on the range. If $K=0$, then $\mathcal G_Q$ has no edges and is connected (Lemma~\ref{lem:structure}(a)), so $A_Q$ is a single action $a$, and $S(a,q)=\lambda V(a,q)+\beta\cdot q$ with any $\lambda>0$ and $\beta=S(a,\cdot)-\lambda V(a,\cdot)$. Suppose next that some action separates the range, so that $K\geq2$. By part (a), some objective is aligned on the range with scales that are not all equal, and by the converse in Lemma~\ref{lem:decomp}(d) it is not exact on the range.
\end{proof}

\subsection*{Proof of Theorem~\ref{thm:test}}

\begin{proof}
By Theorem~\ref{thm:main}(a) and~(b), an objective $W$ that is aligned on the range is block-proportional with some scales $w_X>0$ and has $I_W(E,q)=\sum_Xw_X\phi_X(E,q)$ for every experiment $E$ on the menu, and every vector of positive scales arises from some such objective, $V$ itself having every scale equal to one. Write $y(E)=\phi(E,q)\in\R^K_+$ for the block values of an experiment $E$ on the menu, and for a block $Y$ and $\varepsilon>0$ let $W^{Y,\varepsilon}$ be an objective that is aligned on the range with scale one on $Y$ and $\varepsilon$ on every other block, so that $I_{W^{Y,\varepsilon}}(E,q)=y_Y(E)+\varepsilon\sum_{X\neq Y}y_X(E)$.

\emph{Part (a).} If $E^*\in M$ is best in every block, then for every objective $W$ that is aligned on the range and every $E\in M$,
\[
I_W(E^*,q)-I_W(E,q)=\sum_Xw_X\bigl[m_X-y_X(E)\bigr]\geq0,
\]
with equality if and only if $y_X(E)=m_X$ for every block $X$. So the decisions settle the choice, an experiment on $M$ is optimal for $W$ if and only if it is best in every block, and taking $W=V$ shows that every such experiment is optimal for the principal. Conversely, if some $E^*\in M$ is optimal for every objective that is aligned on the range, then $I_{W^{Y,\varepsilon}}(E^*,q)\geq I_{W^{Y,\varepsilon}}(E,q)$ for every $E\in M$, every block $Y$ and every $\varepsilon>0$, and letting $\varepsilon\to0$ gives $y_Y(E^*)\geq y_Y(E)$.

\emph{Part (b).} If no experiment on the menu has positive value, both sides of the inequality are zero. Otherwise $y_X(E)\geq G(E)\,m_X$ for every block $X$, by the definition of $G(E)$ when $m_X>0$ and trivially when $m_X=0$, and $m_X\geq y_X(E')$ for every $E'\in M$, so
\[
I_W(E,q)=\sum_Xw_Xy_X(E)\geq G(E)\sum_Xw_Xm_X\geq G(E)\,I_W(E',q).
\]
To show that no larger number has this property, let $g>G(E)$, choose a block $X^*$ with $m_{X^*}>0$ and $y_{X^*}(E)=G(E)\,m_{X^*}$, and choose $E^*\in M$ with $y_{X^*}(E^*)=m_{X^*}$. As $\varepsilon\to0$, $I_{W^{X^*,\varepsilon}}(E,q)$ tends to $G(E)\,m_{X^*}<g\,m_{X^*}$, whereas $I_{W^{X^*,\varepsilon}}(E^*,q)\geq m_{X^*}$ for every $\varepsilon$, so $I_{W^{X^*,\varepsilon}}(E,q)<g\max_{E'\in M}I_{W^{X^*,\varepsilon}}(E',q)$ once $\varepsilon$ is small enough.

\emph{Part (c).} By Theorem~\ref{thm:main}(a), $S$ is block-proportional and makes the principal's decisions on $\cl Q$, so it has the principal's margins, separating actions and blocks. Fix a block $X$ and an action $r_X$ of $X$. By \eqref{eq:blocks} and Fact~3, $S^*_X(q')=S(r_X,q')+\lambda_X[V^*_X(q')-V(r_X,q')]$ at every $q'\in\cl Q$. The posteriors of an experiment $E\in M$ lie in $\cl Q$ and average to $q$, so the linear terms cancel in its value, and $\phi^S_X(E,q)=\lambda_X\,\phi_X(E,q)$. Because $\lambda_X>0$, the experiments that are best in $X$, and the ratios that define the guaranteed share, are the same under $S$ as under $V$.
\end{proof}

\subsection*{Proof of Proposition~\ref{prop:records}}

\begin{proof}
By Theorem~\ref{thm:main}(a), the objectives that are aligned on the range are the block-proportional ones, they make the principal's decisions on $\cl Q$, and every vector of positive scales arises. For $a,b\in A_Q$ and $q\in\cl Q$, write $\delta=\delta(a,b;q)\in\R^K$ for the vector of gains of Lemma~\ref{lem:gains}, so that an aligned objective with scales $\lambda$ has $S(a,q)-S(b,q)=\lambda\cdot\delta$, whereas $V(a,q)-V(b,q)=\one\cdot\delta$.

\emph{Part (a).} The actions $a$, $b$ and $c$ are distinct, since a margin consists of two distinct actions and $a=b$ would put the margin $\{a,c\}$ in both $X$ and $Y$. Moreover $a\notin Y$, since otherwise $X$ and $Y$ would share the two actions $a$ and $c$, and likewise $b\notin X$. So $a,X,c,Y,b$ is the path from $a$ to $b$ in $\mathcal H$, $\delta_X(a,b;q)=V(a,q)-V(c,q)$, $\delta_Y(a,b;q)=V(c,q)-V(b,q)$, and every other gain is zero.

Let $q\in Q$ have $\BR_V(q)=\{c\}$, so that $\ell_a=V(c,q)-V(a,q)$ and $\ell_b=V(c,q)-V(b,q)$ are positive, and let $\rho_V(q)=\ell_b/\ell_a$. The principal strictly prefers $a$ to $b$, is indifferent, or strictly prefers $b$ as $\rho_V(q)$ is greater than, equal to or less than one. By \eqref{eq:blocks} and Fact~3, $S(c,q)-S(a,q)=\lambda_X\ell_a$ and $S(c,q)-S(b,q)=\lambda_Y\ell_b$, so $\rho_S(q)=(\lambda_Y/\lambda_X)\,\rho_V(q)$, which gives the first claim.

For the values of $\rho_S$, let $q_1,q_2\in Q$ have $\BR_V(q_1)=\{a,c\}$ and $\BR_V(q_2)=\{c,b\}$, and let $q(t)=(1-t)q_1+tq_2$, which lies in $Q$. For every action $d\neq c$ and $t\in(0,1)$, $V(d,q(t))-V(c,q(t))$ combines with positive weights $V(d,q_1)-V(c,q_1)$ and $V(d,q_2)-V(c,q_2)$, which are nonpositive and not both zero, since $a$ is not optimal at $q_2$, $b$ is not optimal at $q_1$, and any other action is optimal at neither. So $c$ is the only optimal action at $q(t)$. Linearity gives $\ell_a(q(t))=t\,\ell_a(q_2)$ and $\ell_b(q(t))=(1-t)\,\ell_b(q_1)$, where both factors are positive, so
\[
\rho_V(q(t))=\frac{1-t}{t}\cdot\frac{\ell_b(q_1)}{\ell_a(q_2)}
\]
is continuous on $(0,1)$ and takes every positive value, and so does $\rho_S=(\lambda_Y/\lambda_X)\,\rho_V$.

\emph{Part (b).} Let $X$, $Y$, $a$, $b$ and $c$ be as in part (a). We first show that $V(c,q)>V(a,q)$ at every $q\in Q$ at which $V(a,q)=V(b,q)$. Take $c$ as the root of $\mathcal H$, which is possible because Lemma~\ref{lem:decomp} holds for every choice of root in $A_Q$, so that $r_X=r_Y=c$. For $q\in Q$ and $v\in\BR_V(q)$, the path from $c$ to $v$ in $\mathcal H$ visits $c$ only at its start, so it contains at most one of $X$ and $Y$, since a second block on it that contained $c$ would close a cycle through $c$. By Lemma~\ref{lem:decomp}(a), $V(c,q)=V^*_X(q)\geq V(a,q)$ if $X\notin\mathcal P(v)$ and $V(c,q)=V^*_Y(q)\geq V(b,q)$ if $Y\notin\mathcal P(v)$, so $V(c,q)\geq\min\{V(a,q),V(b,q)\}$ at every $q\in Q$. If $V(a,q)=V(b,q)=V(c,q)$ at some $q\in Q$, Lemma~\ref{lem:triple}(b), applied to $a$, $c$ and $b$, gives $z\in L_0$ with $g_{ac}\cdot z=1$ and $g_{cb}\cdot z=-1$. For small $\varepsilon>0$ the belief $q+\varepsilon z$ lies in $Q$, and there $V(a,\cdot)-V(c,\cdot)=V(b,\cdot)-V(c,\cdot)=\varepsilon$, a contradiction.

Now let $q\in Q$ be a belief at which the principal is indifferent between $a$ and $b$, which exists by part (a). There $\delta_X(a,b;q)=V(a,q)-V(c,q)<0$ and $\delta_Y(a,b;q)=-\delta_X(a,b;q)$, so by Lemma~\ref{lem:gains}(b) $S(a,q)-S(b,q)=(\lambda_X-\lambda_Y)\,\delta_X(a,b;q)$, which is zero if and only if $\lambda_X=\lambda_Y$. The graph whose vertices are the blocks, with an edge between adjacent blocks, is connected, since consecutive blocks on a path in $\mathcal H$ share the action between them, and $K-1$ pairs of adjacent blocks that link all the blocks are the edges of a spanning tree of it. For each pair, the shared action belongs to a margin of each block, which supplies $a$ and $b$. If, for each pair, the objective is indifferent between $a$ and $b$ at a belief in $Q$ at which the principal is indifferent between them, then the scales are equal along the spanning tree, and by Lemma~\ref{lem:decomp}(d) the objective is exact on the range.

For the second claim, let the record consist of the principal's rankings of pairs $(a_i,b_i)$ of actions in $A_Q$ at beliefs $q_i\in Q$, $i=1,\dots,N$, let $\delta^i=\delta(a_i,b_i;q_i)$, and let $I_0$ be the set of $i$ at which the principal is indifferent, so that $\one\cdot\delta^i=0$ exactly for $i\in I_0$. If $|I_0|<K-1$, the subspace $\{z\in\R^K:z\cdot\delta^i=0\text{ for every }i\in I_0\}$ has dimension at least two, so it contains a vector $z$ that is not a multiple of $\one$. For small $\varepsilon>0$, the vector $\lambda=\one+\varepsilon z$ is positive, has $\lambda\cdot\delta^i=0$ for $i\in I_0$, and gives $\lambda\cdot\delta^i$ the sign of $\one\cdot\delta^i$ for each of the finitely many other $i$. An objective that is aligned on the range with these scales ranks every recorded pair as the principal does, by Lemma~\ref{lem:gains}, and its scales are not all equal, so by Lemma~\ref{lem:decomp}(d) it is not exact on the range.

\emph{Part (c).} Write $V^\Delta$ and $S^{\hat\Delta}$ for $V$ and $S$ with the payoff from $c$ lowered in every state by $\Delta$ and by $\hat\Delta$. For each block $X$ that contains $c$, fix a margin $\{a_X,c\}$ of $X$, which exists because every action of a block belongs to one of its margins, and a belief $q_X\in Q$ with $\BR_V(q_X)=\{a_X,c\}$. Write $g_X=g_{a_Xc}$, which is not zero on $L_0$, and choose $z_X\in L_0$ with $g_X\cdot z_X=-1$. Under $V^\Delta$ the stakes between $a_X$ and $c$ are $g_X+\Delta\one$, and $\one\cdot q=1$ for every $q\in L$, so at $q_X^\Delta=q_X+\Delta z_X$
\[
V^\Delta(a_X,q_X^\Delta)-V^\Delta(c,q_X^\Delta)=g_X\cdot q_X+\Delta\,g_X\cdot z_X+\Delta=0-\Delta+\Delta=0.
\]
Every other action $d$ has $V^\Delta(d,\cdot)=V(d,\cdot)$ and $V(a_X,q_X)>V(d,q_X)$, and $Q$ is open relative to $L$, so by continuity there is $\bar\Delta>0$ such that, whenever $0<\Delta<\bar\Delta$, every such belief $q_X^\Delta$ lies in $Q$ and has $\BR_{V^\Delta}(q_X^\Delta)=\{a_X,c\}$. Let $0<\Delta<\bar\Delta$, and let $S^{\hat\Delta}$ be aligned with $V^\Delta$ on the range. The stakes of $S^{\hat\Delta}$ between $a_X$ and $c$ are $e_{a_Xc}+\hat\Delta\one$, and $(g_X+\Delta\one)\cdot z_X=-1$, so Lemma~\ref{lem:local}, applied to $V^\Delta$ and $S^{\hat\Delta}$ at $q_X^\Delta$, gives $\lambda'_X>0$ with $(e_{a_Xc}+\hat\Delta\one)\cdot\xi=\lambda'_X(g_X+\Delta\one)\cdot\xi$ for every $\xi\in\spn Q$. Since also $e_{a_Xc}\cdot\xi=\lambda_X\,g_X\cdot\xi$ there, by \eqref{eq:blocks} and Fact~3, subtracting gives
\[
(\lambda_X-\lambda'_X)\,g_X\cdot\xi+(\hat\Delta-\lambda'_X\Delta)\,\one\cdot\xi=0\qquad\text{for every }\xi\in\spn Q.
\]
The two maps are linearly independent on $\spn Q$, since at $q_X$ the first is zero and the second is one, and at $z_X$ the first is $-1$ and the second is zero. Hence $\lambda'_X=\lambda_X$ and $\hat\Delta=\lambda_X\Delta$, that is, $\lambda_X=\hat\Delta/\Delta$.
\end{proof}

\subsection*{Proof of Proposition~\ref{prop:capability}}

\begin{proof}
\emph{Part (a).} Since $Q\subseteq Q'$, every margin and every three-way tie on $Q$ is one on $Q'$. Margins that are linked on $Q$ are therefore linked on $Q'$, so the margins of a block of $Q$, which chains of links connect, all belong to one block of $Q'$.

\emph{Part (b).} By Theorem~\ref{thm:main}(a), $S$ is block-proportional on $Q$ with the unique scales $\lambda_X$. The part of the proof of Theorem~\ref{thm:main}(a) that derives \eqref{eq:blocks} from alignment uses only alignment at the beliefs of the range and Assumption~\ref{ass:acting} on it. Applied on $Q'$, it gives, for each block $Y$ of $Q'$, a scale $\lambda'_Y>0$ with $e_{ab}\cdot\xi=\lambda'_Y\,g_{ab}\cdot\xi$ for every margin $\{a,b\}$ of $Y$ and every $\xi\in\spn Q'$. Let a block $X$ of $Q$ lie within $Y$, and let $\{a,b\}$ be a margin of $X$, which is a margin of $Y$. Since $\spn Q\subseteq\spn Q'$ and $e_{ab}\cdot\xi=\lambda_X\,g_{ab}\cdot\xi$ on $\spn Q$, we have $(\lambda'_Y-\lambda_X)\,g_{ab}\cdot\xi=0$ for every $\xi\in\spn Q$, and by (R2) on $Q$, $\lambda_X=\lambda'_Y$.

\emph{Part (c).} Let $q_1\in\cl Q'$ and $a\in A_Q$ have $V(x,q_1)<V(a,q_1)$ for every $x\in\BR_S(q_1)$. By Fact~5 on $Q$, $a$ is the principal's only optimal action at some belief $q_0\in Q$, and by alignment it is also the AI's. For $\varepsilon\in(0,1)$, the vector $q_2^\varepsilon=(q_0-\varepsilon q_1)/(1-\varepsilon)$ lies in $\aff Q'$ and tends to $q_0$ as $\varepsilon\to0$. Because $Q'$ is open relative to $\aff Q'$, and by Fact~6 applied to $S$, there is $\varepsilon\in(0,1)$ for which $q_2=q_2^\varepsilon$ lies in $Q'$ and $\BR_S(q_2)=\{a\}$. Then $q_0=\varepsilon q_1+(1-\varepsilon)q_2$, so by the splitting lemma \citep{aumannmaschler1995,kamenicagentzkow2011} some experiment $E$ has two signals that, at $q_0$, have probabilities $\varepsilon$ and $1-\varepsilon$ and lead to the posteriors $q_1$ and $q_2$, both of which lie in $\cl Q'$. If the AI acts at $q_0$, it takes $a$, and the principal's expected payoff is $V(a,q_0)=\varepsilon\,V(a,q_1)+(1-\varepsilon)\,V(a,q_2)$. If the AI runs $E$ first, it takes $a$ after the second signal and some $x\in\BR_S(q_1)$ after the first, which lowers the principal's expected payoff by $\varepsilon\,[V(a,q_1)-V(x,q_1)]>0$.
\end{proof}

\clearpage
\phantomsection\label{pg:refsstart}
\begingroup
\setstretch{1}\setlength{\bibsep}{0.12em plus 0.05em}\def\sloppy{\tolerance 9999 \emergencystretch 3em \hfuzz 1pt \vfuzz\hfuzz}\sloppy
\bibliographystyle{aerab}
\bibliography{paper2A-v18-references}
\endgroup

\clearpage
\begin{bibunit}[aerab]
\makeatletter 
\let\oa@natanchorstart\hyper@natanchorstart\def\hyper@natanchorstart#1{\oa@natanchorstart{oa-#1}}
\let\oa@natlinkstart\hyper@natlinkstart\def\hyper@natlinkstart#1{\oa@natlinkstart{oa-#1}}
\let\oa@natlinkbreak\hyper@natlinkbreak\def\hyper@natlinkbreak#1#2{\oa@natlinkbreak{#1}{oa-#2}}
\makeatother
\pagenumbering{arabic}
\renewcommand{\thepage}{S\arabic{page}}
\setcounter{page}{1}
\changefontsizes{11pt}
\singlespacing
\setcounter{section}{0}
\renewcommand{\thesection}{S\arabic{section}}
\renewcommand{\theHsection}{supp.\arabic{section}}
\setcounter{equation}{0}
\renewcommand{\theequation}{S.\arabic{equation}}
\renewcommand{\theHequation}{supp.\arabic{equation}}
\setcounter{footnote}{0}
\setcounter{table}{0}
\renewcommand{\thetable}{S\arabic{table}}
\renewcommand{\theHtable}{supp.\arabic{table}}
\phantomsection\label{pg:suppstart}
\begin{center}
{\large\bfseries Supplemental Appendix to\\[0.2em] ``Are Revealed Preferences All You Need for AI Alignment?''}\\[0.8em]
Joshua S. Gans and Richard Holden\\[0.3em]
{\itshape For online publication}
\end{center}

\medskip

This Supplemental Appendix proves the claims in the text that Appendix~\ref{app:proofs} does not, in the order of the sections to which they belong, except that Section~\ref{supp:penalty} concerns a remark in Section~\ref{sec:model}, and reports the computations behind some of the numerical illustrations. Numbered results, equations and sections without the prefix S are those of the paper. Table~\ref{tab:guide} lists the results that follow and says why each is here.

{\footnotesize\renewcommand{\tablename}{\normalsize Table}
\setlength{\LTcapwidth}{\textwidth}\setlength{\LTleft}{0pt}\setlength{\LTright}{0pt plus 1fil}\setlength{\LTpre}{\bigskipamount}\setlength{\LTpost}{\bigskipamount}
\begin{longtable}{@{}>{\raggedright\arraybackslash}p{0.13\textwidth}>{\raggedright\arraybackslash}p{0.20\textwidth}>{\raggedright\arraybackslash}p{\dimexpr0.67\textwidth-4\tabcolsep\relax}@{}}
\caption{The results of this Supplemental Appendix, with the assumption of the paper that each supports or relaxes (Panel A) or the statement of the paper that it proves (Panel B).}\label{tab:guide}\\
\toprule
Result & In the paper & In this Supplemental Appendix\\
\midrule
\endfirsthead
\multicolumn{3}{@{}l@{}}{Table~\ref{tab:guide}, continued}\\\addlinespace[3pt]
\toprule
Result & In the paper & In this Supplemental Appendix\\
\midrule
\endhead
\bottomrule
\endfoot
\bottomrule
\endlastfoot
\multicolumn{3}{@{}l@{}}{\emph{Panel A. Assumptions of the paper that the results support or relax}}\\*\addlinespace[3pt]
Claim~\ref{claim:generic} & Assumption~\ref{ass:acting} holds (Section~\ref{sec:decisions}). & The assumption, which rules out ties that arise only because payoffs satisfy particular equalities, fails only for a set of payoff functions of measure zero when the range is the relative interior of a polytope, as the set of beliefs with full support is.\\\addlinespace[3pt]
Claim~\ref{claim:separable} & Assumption~\ref{ass:acting} holds (Section~\ref{sec:decisions}). & The assumption can fail when an AI makes several decisions whose payoffs add up. Each decision's payoffs, or each block of a decision that has several, can then be scaled by its own positive number without changing a choice, and Theorem~\ref{thm:test}(a) and~(b) and the bounds on the loss hold with the blocks of the separate decisions in place of blocks.\\\addlinespace[3pt]
Lemma~\ref{lem:overvalue}, Claim~\ref{claim:avgloss} & The experiments on a menu are costless (Theorem~\ref{thm:test}). & With costly experiments, the principal's loss from the AI's choice is at most the difference between the amounts by which the objective overvalues the two experiments. The expected loss is of second order in the error in the scales when the menu is the same at every belief and the AI's belief has a bounded density.\\\addlinespace[3pt]
Claim~\ref{claim:sequential} & The AI runs one experiment before it acts (Section~\ref{sec:test}). & The decomposition of Theorem~\ref{thm:main}(b) holds for every plan that gathers evidence in sequence and stops at beliefs on the range, and with it Theorem~\ref{thm:test}(a) and~(b) and the bounds on the loss.\\\addlinespace[3pt]
Claim~\ref{claim:continuum} & The set of actions is finite (Section~\ref{sec:model}). & With a continuum of ordered actions and strictly concave payoffs, every action separates those below it from those above, and the principal's marginal payoff at each action can be given any positive weight without changing a decision. With two states and an optimal action that is interior at every belief and varies with it, decisions settle a choice exactly when one experiment is at least as informative as each of the others \citep{blackwell1953}.\\\addlinespace[3pt]
Claim~\ref{claim:noise} & Records are free of noise (Section~\ref{sec:records}). & A logit model fitted to noisy choices returns the principal's relative scale if the noise is the same in every comparison, and, when each block's comparisons have a temperature of their own and no comparison spans two blocks, that relative scale multiplied by the ratio of the precisions.\\\addlinespace[3pt]
Claim~\ref{claim:penalty} & The AI maximizes its objective (Section~\ref{sec:model}). & With two states, when training penalizes divergence from a reference policy, the AI takes each action with a probability that rises with its payoff under the objective. Each switch then moves by an amount proportional to the weight on the penalty divided by the scale of its margin, and the principal's expected loss from the AI's actions is of second order in that weight and inversely proportional to the square of each scale. The scales that decisions leave free therefore matter for acting.\\\addlinespace[7pt]
\multicolumn{3}{@{}l@{}}{\emph{Panel B. Statements of the paper that the results prove}}\\*\addlinespace[3pt]
Lemma~\ref{lem:cycle}, Claims~\ref{claim:cut} and~\ref{claim:biconnected} & Sections~\ref{sec:theorem} and~\ref{sec:practice}: finding the blocks from the margin graph. & An action separates the range exactly when removing it disconnects the margin graph, and the blocks are the biconnected components of that graph.\\\addlinespace[3pt]
Lemma~\ref{lem:strict}, Claim~\ref{claim:between} & Sections~\ref{sec:theorem} and~\ref{sec:contexts}: separation by comparisons of payoffs. & An action separates two others exactly when the remaining actions divide into two groups, one with each of them, such that it is at least as good as some lottery over each pair drawn one from each group, at every belief on the range. A separating action beats one of the two actions that it separates at every belief, which is the value restriction of \citet{sen1966}.\\\addlinespace[3pt]
Claim~\ref{claim:peaked} & Section~\ref{sec:theorem}: single-peaked payoffs. & No three actions tie exactly when the margin graph is a tree, when every margin is a block, and when the principal's expected payoffs are single-peaked on a tree. The margin graph is a path exactly when the expected payoffs are concave in a suitably spaced action at every belief on the range.\\\addlinespace[3pt]
Claim~\ref{claim:twoq} & Section~\ref{sec:theorem}: decisions that interact. & For two yes-or-no decisions, a slight complementarity or substitutability at a belief at which both are marginal puts all four actions in one block. The freedom of scales across decisions whose payoffs add up therefore depends on the payoffs adding up exactly.\\\addlinespace[3pt]
Claims~\ref{claim:tasks} and~\ref{claim:psi} & Section~\ref{sec:test} and Table~\ref{tab:records}: ranking evidence and paying for it. & An aligned objective ranks experiments as the principal does exactly when, at each belief, they have the same average scale, and it pays for them as the principal does exactly when that average is one. The principal's own choices among experiments thus reveal the relative scale of blocks on which the experiments bear differently.\\\addlinespace[3pt]
Claim~\ref{claim:choose} & Section~\ref{sec:test}: choices that decisions settle. & Decisions settle the choice when one experiment on the menu is at least as valuable as the others in every problem of a class that contains each block's problem. This is so when every other experiment is a garbling of it and, when the principal's problem is single crossing, when it is more accurate than the others in the sense of \citet{lehmann1988}. Decisions also settle the choice when the AI can run its experiments together and when the experiments have positive value in at most one block.\\\addlinespace[3pt]
Claim~\ref{claim:ratio} & Section~\ref{sec:test}: how much an unsettled choice can lose. & The experiment that the AI prefers is worth at least $\lambda_{\min}/\lambda_{\max}$ of the principal's choice, and the bound is attained whenever there are two blocks.\\\addlinespace[3pt]
Claims~\ref{claim:lottery}, \ref{claim:finiterecords} and~\ref{claim:deployment} & Sections~\ref{sec:test}, \ref{sec:records} and~\ref{sec:practice}: lotteries over experiments, answers that only bound the relative scale, and the share that a deployment secures. & A lottery over the experiments that are best in each block guarantees the share $1/K'$, and no lottery guarantees more when each experiment has value in a single block. The largest share that a lottery guarantees is the value of a zero-sum game in which the designer chooses an experiment and an opponent chooses a block, and at most $K'$ experiments are needed. The share that an experiment guarantees after a finite record of rankings is found by linear programming, and the share of the value of evidence that a deployment secures is the smallest ratio of averaged block values. All are computed from the AI's objective.\\\addlinespace[3pt]
Section~\ref{supp:misdirection}, the guarantee from bisection & Section~\ref{sec:records}: bisection. & In the file example, $k$ answers found by bisection leave a guaranteed share that falls short of one by at most a multiple of $2^{-k}$.\\\addlinespace[3pt]
Section~\ref{supp:fileexample}, first part & Sections~\ref{sec:example} and~\ref{sec:contexts}: evidence on one question. & When beliefs move along a segment, every margin is a block, every action in $A_Q$ other than those chosen at the ends of the segment separates the range, and the principal's expected payoffs have increasing differences in the action and the position of the belief. In the file example, the decisions settle the choice between the two tests exactly when $p\leq1/13$ or $p\geq27/31$.\\\addlinespace[3pt]
Claim~\ref{claim:rules} & Section~\ref{sec:contexts}, Table~\ref{tab:records}: right answers. & An AI that is rewarded only for right answers never takes a cautious action.\\\addlinespace[3pt]
Claim~\ref{claim:middle} & Section~\ref{sec:contexts}: a cautious middle course. & When a middle course and two alternatives are the only actions, the middle course separates the alternatives exactly when it is better than both wherever they tie, and exactly when it is at least as good as one lottery over them at every belief on the range. In the three-state example this is so when $\kappa\leq10/3$.\\\addlinespace[3pt]
Claim~\ref{claim:cheapcaution}, Table~\ref{tab:prevalence} & Section~\ref{sec:contexts}: cheap caution, and how common several blocks are. & When each state has a single right action, a cautious action that falls short of it by little enough in every state separates the range. With payoffs drawn at random, several blocks are common on ranges of low dimension and rare on ranges of high dimension.\\\addlinespace[3pt]
Claim~\ref{claim:answers} & Section~\ref{sec:contexts}: declining to answer. & Every block contains the option to decline, and two answers share a block exactly when they are joined by a chain of answers in which consecutive confidence thresholds sum to less than one.\\\addlinespace[3pt]
Claim~\ref{claim:informative} & Sections~\ref{sec:theorem} and~\ref{sec:records}: comparisons that add nothing to decisions. & Every aligned objective ranks two actions at a belief as the principal does exactly when the principal's gains within blocks are all nonnegative or all nonpositive. Two actions thus share a block exactly when every aligned objective ranks them as the principal does at every belief.\\\addlinespace[3pt]
Claim~\ref{claim:sides} & Section~\ref{sec:records}: rankings that decisions fix. & Decisions fix how the principal ranks the actions along every path of blocks that leads away from its choice. When the margin graph is a path, they fix its ranking of two actions on the same side of its choice and leave open every comparison across it.\\\addlinespace[3pt]
Claim~\ref{claim:unavailable} & Sections~\ref{sec:records} and~\ref{sec:practice}: decisions with a separating action unavailable. & An aligned objective makes the principal's decisions when an action is unavailable and the other actions are those chosen somewhere on the range exactly when the blocks that contain that action have equal scales. Those decisions therefore reveal the relative scale of the blocks around a separating action, an aligned AI whose scales differ there can err when the action is unavailable, and no separating action can be left out if such decisions are to make the objective exact.\\\addlinespace[3pt]
Claim~\ref{claim:maxmargin} & Section~\ref{sec:records}: fitting procedures. & With two states, fitting a finite record of decisions by maximizing the classification margin gives a relative scale that reflects how densely the record covers each switching belief, not what is at stake there.\\\addlinespace[3pt]
Claim~\ref{claim:range} & Section~\ref{sec:capability}: the beliefs that an AI can hold. & The beliefs that arise under some garbling of an AI's capability bound are exactly those in the convex hull of the bound's posteriors.\\\addlinespace[3pt]
Claim~\ref{claim:classes} & Section~\ref{sec:capability}: states that the AI cannot tell apart. & When the AI's evidence has the same distribution in every state of a class, the AI is aligned exactly when it is aligned in the problem whose states are the classes, with payoffs averaged within each class.\\\addlinespace[3pt]
Claim~\ref{claim:capbound} & Section~\ref{sec:capability}: the loss at beliefs outside the range. & When the actions chosen at a belief outside the range are in $A_Q$, the principal's loss there is at most $\lambda_{\max}/\lambda_{\min}-1$ times the size of the trade-off between blocks that its choice makes, plus a term for errors in directions in which beliefs on the range never move.\\
\end{longtable}
}

The notation is that of Appendix~\ref{app:proofs}, including (R1) and (R2), the margin graph $\mathcal G_Q$, the blocks as constructed from three-way ties, which by Lemma~\ref{lem:blocks} are those of the text, the tree $\mathcal H$ of blocks and actions, the gains within blocks $\delta_X(a,b;q)$ of Lemma~\ref{lem:gains} and the term block-proportional. We use the following consequences of Theorem~\ref{thm:main} throughout, often without mention. The objectives that are aligned on the range are exactly the block-proportional ones, they make the principal's decisions at every belief in $\cl Q$, and every vector of positive scales is the vector of scales of one of them. For an experiment $E$ that keeps beliefs on the range at a belief $q\in Q$, we write $y(E)=\phi(E,q)\in\R^K_+$ for its block values, so that an objective $S$ that is aligned on the range with scales $\lambda=(\lambda_X)$ has $I_S(E,q)=\lambda\cdot y(E)$, whereas $I_V(E,q)=\one\cdot y(E)$.

\section{Genericity}\label{supp:generic}

We show that Assumption~\ref{ass:acting} holds for almost every payoff function when the range is the relative interior of a polytope, as the set of beliefs with full support is.

\begin{suppclaim}\label{claim:generic}
Let $P\subseteq\Delta(\Omega)$ be a polytope and $Q$ its relative interior. The set of payoff functions $V\in\R^{A\times\Omega}$ for which Assumption~\ref{ass:acting} fails on $Q$ has Lebesgue measure zero.
\end{suppclaim}

\begin{proof}
The proof has three steps. We first state a general-position condition and show that it fails only on a null set, and then show that it implies each condition of Assumption~\ref{ass:acting}. Recall that $P$ has finitely many faces \citep[Theorem~19.1]{rockafellar1970}, that $P$ is itself a face, that every point of $P$ lies in the relative interior of exactly one nonempty face \citep[Theorem~18.2]{rockafellar1970}, and that $Q$ has closure $P$ and affine hull $\aff P$ \citep[Theorems~6.2 and~6.3]{rockafellar1970}. For a nonempty face $F$ of dimension $f$, let $L_F=\aff F$ and let $L_{F,0}$ be its direction space.

\emph{Step 1: a general-position condition.} For a face $F$, a set $J\subseteq A$ of $m\geq2$ actions and an action $a_0\in J$, consider the affine map $\Lambda_{F,J}:L_F\to\R^{m-1}$ given by $\xi\mapsto(g_{ba_0}\cdot\xi)_{b\in J\setminus\{a_0\}}$. We say that $V$ is in \emph{general position} if, for every face $F$ and every such $J$ and $a_0$: (i) if $m-1\leq f$, the linear functionals $z\mapsto g_{ba_0}\cdot z$, $b\in J\setminus\{a_0\}$, are linearly independent on $L_{F,0}$, and (ii) if $m-1>f$, $\Lambda_{F,J}$ has no zero in $L_F$.

\emph{Step 2: general position fails only on a null set.} Fix $F$, $J$ and $a_0$. Choose $\xi_0\in L_F$ and a basis $z_1,\dots,z_f$ of $L_{F,0}$. The vectors $z_1,\dots,z_f,\xi_0$ are linearly independent in $\R^\Omega$, because every $z\in L_{F,0}$ has $z\cdot\one=0$ whereas $\xi_0\cdot\one=1$. Let $M(V)$ be the $(m-1)\times(f+1)$ matrix whose row for $b\in J\setminus\{a_0\}$ is $(g_{ba_0}\cdot z_1,\dots,g_{ba_0}\cdot z_f,\,g_{ba_0}\cdot\xi_0)$, and let $M_1(V)$ be its first $f$ columns. Its entries are linear in $V$, and the linear map $V\mapsto M(V)$ is onto, because setting $V(a_0,\cdot)=0$ and choosing each $V(b,\cdot)$ prescribes the inner products of $g_{ba_0}=V(b,\cdot)$ with the $f+1$ independent vectors arbitrarily. Condition (i) says that $M_1(V)$ has rank $m-1$. If $m-1\leq f$, it fails exactly where the sum of the squares of the $(m-1)\times(m-1)$ minors of $M_1(V)$ vanishes. This is a polynomial in $V$, and it is not identically zero, because $M_1$ is onto the $(m-1)\times f$ matrices, among which some have rank $m-1$. If $m-1>f$, condition (ii) fails only if the system $M_1(V)y=-M(V)_{\cdot,f+1}$, which says that $\Lambda_{F,J}$ vanishes at $\xi_0+\sum_ky_kz_k$, has a solution $y\in\R^f$, which requires $M(V)$ to have rank at most $f$. That happens exactly where the sum of the squares of the $(f+1)\times(f+1)$ minors of $M(V)$ vanishes, a polynomial that is not identically zero because $M$ is onto and $m-1\geq f+1$. The zero set of a polynomial that is not identically zero has Lebesgue measure zero, and there are finitely many faces, sets $J$ and actions $a_0$. So general position fails only on a null set.

\emph{Step 3: general position implies Assumption~\ref{ass:acting}.} Let $V$ be in general position. Consider first a belief $q_0\in Q$, which lies in the relative interior of the face $P$, of dimension $d=\dim L$. Let $J=\BR_V(q_0)$ and $m=|J|$, and fix $a_0\in J$. If $m\geq2$, then $\Lambda_{P,J}(q_0)=0$, so $m-1\leq d$ by (ii), and by (i) the map $z\mapsto(g_{ba_0}\cdot z)_{b\in J\setminus\{a_0\}}$ from $L_0$ to $\R^{m-1}$ is onto. Hence, for any $J'\subseteq J$ that contains $a_0$, there is $z\in L_0$ with $g_{ba_0}\cdot z=0$ for $b\in J'\setminus\{a_0\}$ and $g_{ba_0}\cdot z=-1$ for $b\in J\setminus J'$. For small $\varepsilon>0$, the belief $q_0+\varepsilon z$ lies in $Q$, which is open relative to $L$, the actions in $J'$ tie there, those in $J\setminus J'$ fall short of them by $\varepsilon$, and actions outside $J$ remain strictly worse by continuity. So $\BR_V(q_0+\varepsilon z)=J'$. When $m=1$, the only such $J'$ is $J=\{a_0\}$, and $z=0$ gives the same conclusion. Taking $J'=\{a_0\}$ for an action $a_0\in A_Q$ optimal at $q_0$ shows that $a_0$ is the only optimal action at a belief of $Q$, and so, by continuity, at every belief of a nonempty subset of $Q$ that is open relative to $L$, which is (R1). Taking $J'$ to be any two or three actions that are optimal at $q_0$ gives Assumption~\ref{ass:acting}(ii). For (R2), let $a\neq b$ belong to $A_Q$. If $d\geq1$, condition (i) with $F=P$ and $J=\{a,b\}$ shows that $g_{ab}$ is not constant on $L$, and if $d=0$, condition (ii) shows that $g_{ab}\cdot\xi\neq0$ at the single point of $L$. Either way $a$ and $b$ do not have equal expected payoffs throughout $L$.

For Assumption~\ref{ass:acting}(iii), suppose, toward a contradiction, that an action $c\notin A_Q$ is optimal at a belief $q_0$ in the closure of $Q$, which is $P$. Then $q_0\notin Q$, so $q_0$ lies in the relative interior of a proper face $F$ of $P$. Let $J=\BR_V(q_0)$, which contains $c$, and $m=|J|$. If $m\geq2$, then $\Lambda_{F,J}(q_0)=0$ with $a_0=c$, so $m-1\leq f$ by (ii), and by (i) there is $z\in L_{F,0}$ with $g_{bc}\cdot z=-1$ for every $b\in J\setminus\{c\}$. For small $\varepsilon>0$, the belief $q_1=q_0+\varepsilon z$ lies in the relative interior of $F$, and $c$ is the only optimal action there. If $m=1$, let $q_1=q_0$. By continuity, $c$ is the only optimal action at every belief in some ball around $q_1$, and because $q_1$ lies in $P$, the closure of $Q$, that ball contains a belief of $Q$. Then $c\in A_Q$, a contradiction.
\end{proof}

\section{Separating actions and the margin graph}\label{supp:structure}

This section shows that an action separates the range exactly when removing it disconnects the margin graph $\mathcal G_Q$, and that the blocks are the biconnected components of that graph, as stated in Section~\ref{sec:theorem}. The second result uses a lemma on cycles of margins. The section then characterizes separation by comparisons of payoffs and relates the blocks to single-peaked payoffs.

\begin{supplemma}\label{lem:cycle}
Let Assumption~\ref{ass:acting} hold. The margins of every simple cycle of $\mathcal G_Q$ belong to a single block.
\end{supplemma}

\begin{proof}
Group the margins of a simple cycle of $\mathcal G_Q$, taken in order around the cycle, into maximal runs of consecutive margins that belong to the same block. Suppose, for contradiction, that there are $k\geq2$ runs. Let $a_1,\dots,a_k$ be the actions at which one run ends and the next begins, which are distinct because the cycle is simple, and let $Y_i$ be the block of the run from $a_i$ to $a_{i+1}$, with $a_{k+1}=a_1$ and $Y_0=Y_k$. Each $a_i$ belongs to $Y_{i-1}$ and to $Y_i$, so $a_1,Y_1,a_2,\dots,a_k,Y_k,a_1$ is a closed walk in the tree $\mathcal H$ (Lemma~\ref{lem:structure}(c)). Consecutive blocks on the walk differ, because the runs are maximal, and consecutive actions differ, so the walk never returns at once along the edge by which it arrived. But a closed walk in a tree does so at any of its vertices that is farthest from a fixed vertex of the tree, since both of that vertex's neighbors on the walk are then its neighbor on the path to the fixed vertex. Hence there is a single run, and all the margins of the cycle belong to one block.
\end{proof}

We say that an action $c\in A_Q$ is a \emph{cut vertex} of $\mathcal G_Q$ if $A_Q\setminus\{c\}$ is the union of two disjoint nonempty sets with no margin between them.

\begin{suppclaim}\label{claim:cut}
Let Assumption~\ref{ass:acting} hold and let $c\in A_Q$. Then $c$ is a cut vertex of $\mathcal G_Q$ if and only if it belongs to at least two blocks, and so, by Lemma~\ref{lem:blocks}(c), if and only if it separates the range.
\end{suppclaim}

\begin{proof}
Let $c$ be a cut vertex, so that $A_Q\setminus\{c\}=C_1\cup C_2$ with $C_1$ and $C_2$ disjoint, nonempty and joined by no margin, and choose $a\in C_1$ and $b\in C_2$. Every path from $a$ to $b$ in $\mathcal G_Q$ visits $c$, so by Lemma~\ref{lem:blocks}(b) $c$ separates $a$ and $b$, and by Lemma~\ref{lem:blocks}(c) it belongs to at least two blocks. Conversely, let $c$ belong to at least two blocks. By Lemma~\ref{lem:blocks}(c) it separates some actions $a$ and $b$ in $A_Q$, and by Lemma~\ref{lem:blocks}(b) every path from $a$ to $b$ in $\mathcal G_Q$ visits $c$. Let $C_1$ be the set of actions that are joined to $a$ by a path in $\mathcal G_Q$ that does not visit $c$, including $a$ itself, and let $C_2$ be the set of the other actions of $A_Q\setminus\{c\}$, which contains $b$. A margin between an action of $C_1$ and an action $y\neq c$ extends such a path to $y$, so $y\in C_1$. No margin therefore joins $C_1$ and $C_2$, and $c$ is a cut vertex.
\end{proof}

\noindent If $c$ is not a cut vertex and $\{a,c\}$ and $\{c,b\}$ are margins with $a\neq b$, a path in $\mathcal G_Q$ from $a$ to $b$ that avoids $c$ closes a simple cycle through both margins, which by Lemma~\ref{lem:cycle} belong to one block. So when no action separates the range, any two margins that share an action lie on a cycle of margins and belong to one block.

A \emph{biconnected component} of $\mathcal G_Q$ is a maximal connected subgraph without a cut vertex, where a subgraph with two vertices joined by an edge has no cut vertex. Regard each block as the subgraph formed by its margins and their actions.

\begin{suppclaim}\label{claim:biconnected}
Let Assumption~\ref{ass:acting} hold and let $A_Q$ contain at least two actions. The blocks are the biconnected components of $\mathcal G_Q$.
\end{suppclaim}

\begin{proof}
Let $X$ be a block. It is connected by Lemma~\ref{lem:structure}(c), and by Lemma~\ref{lem:blocks}(a) it has no cut vertex.

For maximality, let $Z\supsetneq X$ be a connected subgraph of $\mathcal G_Q$, and suppose, toward a contradiction, that $Z$ has no cut vertex. Because $Z$ is connected and contains a margin outside $X$, some margin $f=\{v,w\}$ of $Z$ outside $X$ has an action $v$ in $X$. Let $u\neq v$ be another action of $X$. Because $v$ is not a cut vertex of $Z$, some path in $Z$ from $w$ to $u$ avoids $v$. Let $x$ be the first action of $X$ on this path, which differs from $v$, and let $P$ be the part of the path from $w$ to $x$, whose actions other than $x$ lie outside $X$. Let $P'$ be a path from $x$ to $v$ that uses only margins of $X$, which exists because $X$ is connected. Following $f$ from $v$ to $w$, then $P$ to $x$, then $P'$ back to $v$ gives a closed walk in which no action repeats, since $P$ and $P'$ share only $x$. If $w=x$, then $P$ is trivial and $P'$ has at least two margins, because $f$ is the only margin between $v$ and $w$ and $f$ is not in $X$. Either way the walk is a simple cycle that contains $f$ and a margin of $X$, and by Lemma~\ref{lem:cycle} these belong to one block, so $f$ is in $X$, a contradiction. Hence $X$ is a maximal connected subgraph without a cut vertex, that is, a biconnected component. Conversely, let $B$ be a biconnected component. Because $\mathcal G_Q$ is connected and has at least two vertices, $B$ contains a margin, which belongs to some block $X$. Removing any one action from the subgraph $B\cup X$ leaves what remains of $B$ and of $X$ connected, since neither has a cut vertex, and the two still share an action of that margin, so $B\cup X$ is connected and has no cut vertex. By the maximality of $B$ and of $X$, $B=B\cup X=X$.
\end{proof}

\subsection*{Separation as betweenness, and single-peaked payoffs}

The rest of this section proves the claims in Sections~\ref{sec:theorem} and~\ref{sec:middle} that describe separation through comparisons of payoffs. We say that an action $x$ \emph{beats} an action $y$ at a belief $q$ if $V(x,q)>V(y,q)$, and that two actions $x$ and $y$ in $A_Q\setminus\{c\}$ \emph{beat $c$ together} if both beat $c$ at some belief in $Q$. For $c\in A_Q$, the \emph{sides} of $c$ are the vertex sets of the connected components of the graph obtained from $\mathcal G_Q$ by deleting $c$ and its margins. By Lemma~\ref{lem:blocks}(b), $c$ separates two actions if and only if they lie on different sides of $c$.

\begin{supplemma}\label{lem:strict}
Let Assumption~\ref{ass:acting} hold, and let $x$, $y$ and $c$ be distinct actions in $A_Q$. If $V(x,q)\geq V(c,q)$ and $V(y,q)\geq V(c,q)$ at some belief $q\in Q$, then $x$ and $y$ beat $c$ together.
\end{supplemma}

\begin{proof}
For distinct $u,w\in A_Q$, the stake $g_{uw}$ is not zero on $L_0$. Otherwise $g_{uw}\cdot\xi$ would be the same at every $\xi\in L$, whereas by Fact~5 $u$ beats $w$ at some belief in $Q$ and $w$ beats $u$ at another. If both inequalities of the lemma are strict at $q$, there is nothing to prove. Suppose that $V(x,q)=V(c,q)$ and $V(y,q)>V(c,q)$, and choose $z\in L_0$ with $g_{xc}\cdot z>0$. For small $\varepsilon>0$ the belief $q+\varepsilon z$ lies in $Q$, which is open relative to $L$, and there $g_{xc}\cdot(q+\varepsilon z)=\varepsilon\,g_{xc}\cdot z>0$, while $g_{yc}\cdot(q+\varepsilon z)>0$ by continuity. The case in which only the second inequality is an equality is the same with $x$ and $y$ exchanged. Suppose finally that $V(x,q)=V(y,q)=V(c,q)$. By Lemma~\ref{lem:triple}(b), applied to $x$, $c$ and $y$, the map $z\mapsto(g_{xc}\cdot z,\,g_{cy}\cdot z)$ from $L_0$ to $\R^2$ is onto, so some $z\in L_0$ has $g_{xc}\cdot z=1$ and $g_{yc}\cdot z=-g_{cy}\cdot z=1$, and at $q+\varepsilon z$, for small $\varepsilon>0$, both $x$ and $y$ beat $c$.
\end{proof}

\begin{suppclaim}\label{claim:between}
Let Assumption~\ref{ass:acting} hold and let $c\in A_Q$.
\begin{enumerate}[label=(\alph*)]
\item Two actions of $A_Q\setminus\{c\}$ lie on the same side of $c$ if and only if they are joined by a chain of actions in $A_Q\setminus\{c\}$ in which each action and the next beat $c$ together.
\item For distinct actions $a$ and $b$ in $A_Q\setminus\{c\}$, the following are equivalent.
\begin{enumerate}[label=(\roman*)]
\item The action $c$ separates $a$ and $b$.
\item The set $A_Q\setminus\{c\}$ is the union of disjoint sets $A_1$, which contains $a$, and $A_2$, which contains $b$, such that at every belief on the range $c$ is at least as good for the principal as every action of $A_1$ or at least as good as every action of $A_2$.
\item There are such sets $A_1$ and $A_2$ with the following property. For every $x\in A_1$ and $y\in A_2$ there is $\alpha\in(0,1)$ such that $V(c,q)\geq\alpha\,V(x,q)+(1-\alpha)\,V(y,q)$ at every belief $q$ on the range.
\end{enumerate}
In (ii) and (iii), $A_1$ can be taken to be the side of $c$ that contains $a$.
\item If $c$ separates $a$ and $b$, then at every belief in $Q$ the action $c$ beats $a$ or beats $b$.
\end{enumerate}
\end{suppclaim}

\begin{proof}
\emph{Part (a).} We show that two actions other than $c$ that form a margin beat $c$ together, and that two actions that beat $c$ together lie on the same side of $c$. Part (a) follows, since a path in $\mathcal G_Q$ that does not visit $c$ is then a chain of the kind described, and such a chain stays on one side.

Let $\{x,y\}$ be a margin with $x,y\neq c$, and let $q\in Q$ have $\BR_V(q)=\{x,y\}$. Then $c$ is not optimal at $q$, so $V(x,q)=V(y,q)>V(c,q)$, and $x$ and $y$ beat $c$ together. Conversely, let $x$ and $y$ both beat $c$ at $q_0\in Q$. The sets $D_x=\{q\in Q:V(x,q)>V(c,q)\}$ and $D_y=\{q\in Q:V(y,q)>V(c,q)\}$ are convex, being intersections of the convex set $Q$ with open half-spaces, and both contain $q_0$, so $D_x\cup D_y$ is connected. The action $c$ is optimal at no belief of $D_x\cup D_y$. By Fact~5, $x$ is the only optimal action at some belief of $Q$, which lies in $D_x$, and $y$ is the only optimal action at some belief of $Q$, which lies in $D_y$. So $D_x\cup D_y$ is a connected set of beliefs in $Q$ at which $c$ is not optimal, and it contains a belief at which $x$ is optimal and one at which $y$ is. Hence $c$ does not separate $x$ and $y$, and by Lemma~\ref{lem:blocks}(b) some path from $x$ to $y$ in $\mathcal G_Q$ does not visit $c$, so $x$ and $y$ lie on the same side.

\emph{Part (b), (i) implies (ii).} Let $c$ separate $a$ and $b$, let $A_1$ be the side of $c$ that contains $a$, and let $A_2$ consist of the other actions of $A_Q\setminus\{c\}$, which include $b$. By part (a), no action of $A_1$ beats $c$ together with an action of $A_2$, so at no belief in $Q$ do an action of $A_1$ and an action of $A_2$ both beat $c$. The same is true at every belief $q\in\cl Q$, because two strict inequalities that held at $q$ would hold at nearby beliefs of $Q$. Hence at every belief on the range, no action of $A_1$ beats $c$ or no action of $A_2$ does.

\emph{(ii) implies (iii).} Fix $x\in A_1$ and $y\in A_2$. The set $D=\{(g_{xc}\cdot q,\,g_{yc}\cdot q):q\in\cl Q\}$ is a compact convex subset of $\R^2$, and by (ii) it does not meet the open convex set $P=\{(u,v):u>0,\ v>0\}$. By the separating hyperplane theorem \citep[Theorem~11.3]{rockafellar1970}, there are $(\alpha_1,\alpha_2)\neq(0,0)$ and $\kappa\in\R$ with $\alpha_1u+\alpha_2v\leq\kappa$ on $D$ and $\alpha_1u+\alpha_2v\geq\kappa$ on $P$. The second inequality requires $\alpha_1\geq0$ and $\alpha_2\geq0$, since otherwise its left-hand side is unbounded below on $P$, and letting $(u,v)$ tend to the origin within $P$ gives $\kappa\leq0$. Dividing by $\alpha_1+\alpha_2>0$ and writing $\alpha$ for the first weight, we obtain $\alpha\,g_{xc}\cdot q+(1-\alpha)\,g_{yc}\cdot q\leq0$ at every $q\in\cl Q$, which is the inequality in (iii) with $\alpha\in[0,1]$. If $\alpha=1$, then $V(c,q)\geq V(x,q)$ at every belief in $Q$, which is false at a belief at which $x$ is the only optimal action (Fact~5). Likewise $\alpha\neq0$.

\emph{(iii) implies (i).} Suppose that $c$ does not separate $a$ and $b$. They then lie on the same side of $c$, and by part (a) they are joined by a chain in which each action and the next beat $c$ together. The chain starts in $A_1$ and ends in $A_2$, so some action $x\in A_1$ of the chain beats $c$ together with the next action, $y\in A_2$. At a belief at which both beat $c$, every lottery over $x$ and $y$ is strictly better than $c$, which contradicts (iii).

\emph{Part (c).} Let $c$ separate $a$ and $b$. If $V(a,q)\geq V(c,q)$ and $V(b,q)\geq V(c,q)$ at some $q\in Q$, then by Lemma~\ref{lem:strict} $a$ and $b$ beat $c$ together, and by part (a) they lie on the same side of $c$, a contradiction.
\end{proof}

\noindent A separating action is thus one that is never beaten from two sides at once. Condition (iii) is the form that a designer can check from payoffs. The inequality is linear in the belief, so when the range is the set of beliefs with full support it holds at every belief on the range if and only if it holds in every state. Part (c) says that, with beliefs in the place of voters who rank the actions by expected payoff, the three actions $a$, $c$ and $b$ satisfy the value restriction of \citet{sen1966}: no belief ranks $c$ last among them. With three actions in $A_Q$ this is also sufficient for $c$ to separate $a$ and $b$ (Claim~\ref{claim:middle}). With more it is not, because the principal's choice can pass from $a$ to $b$ around $c$, as it does between two answers whose confidence thresholds sum to at least one when the threshold of a third answer sums to less than one with each of them (Claim~\ref{claim:answers}).

The next claim uses the notion of single-peaked payoffs on a tree \citep{black1948,demange1982}. For a tree $T$ whose vertices are the actions in $A_Q$, we say that the principal's expected payoffs are \emph{single-peaked on $T$} if $V(c,q)\geq\min\{V(x,q),V(y,q)\}$ at every belief $q\in Q$, for all distinct actions $x$, $c$ and $y$ such that $c$ lies on the path from $x$ to $y$ in $T$.

\begin{suppclaim}\label{claim:peaked}
Let Assumption~\ref{ass:acting} hold, and let $A_Q$ have $n\geq2$ actions.
\begin{enumerate}[label=(\alph*)]
\item There are at most $n-1$ blocks, and the following are equivalent.
\begin{enumerate}[label=(\roman*)]
\item No belief in $Q$ has three or more optimal actions for the principal.
\item The margin graph $\mathcal G_Q$ is a tree.
\item $K=n-1$, that is, every margin is a block.
\item The principal's expected payoffs are single-peaked on some tree whose vertices are the actions in $A_Q$.
\end{enumerate}
The tree in (iv) is then the margin graph.
\item Let $n\geq3$, and let $a_1,\dots,a_n$ list the actions in $A_Q$. The margin graph is the path that joins each $a_i$ to $a_{i+1}$ if and only if there are numbers $h_1<\dots<h_n$ such that, at every belief $q$ on the range and for $j=2,\dots,n-1$,
\begin{equation}\tag{C}\label{eq:respaced}
\frac{V(a_j,q)-V(a_{j-1},q)}{h_j-h_{j-1}}\;\geq\;\frac{V(a_{j+1},q)-V(a_j,q)}{h_{j+1}-h_j}.
\end{equation}
\end{enumerate}
\end{suppclaim}

\begin{proof}
\emph{Part (a).} The tree $\mathcal H$ has $n+K$ vertices and so $n+K-1$ edges. Its edges join each block to its actions, and each block has at least two actions, so $n+K-1\geq2K$ and $K\leq n-1$, with equality if and only if every block has exactly two actions. A block with two actions has a single margin, and every margin belongs to exactly one block, so $K=n-1$ if and only if every margin is a block.

\emph{(i) and (iii).} If three or more actions are optimal at a belief in $Q$, any three of them form a three-way tie by Assumption~\ref{ass:acting}(ii), the margins of the triple are linked, and their block has at least three actions, so $K<n-1$. Conversely, a block with at least three actions has at least two margins, which are joined by a chain of linked margins, so there is a three-way tie.

\emph{(ii) and (iii).} If $K=n-1$, the margin graph has $n-1$ edges, one for each block, and it is connected (Lemma~\ref{lem:structure}(a)), so it is a tree. If the margin graph is a tree, it has no cycle. A three-way tie would give a triangle of margins, so there is none, and (iii) holds by the previous step.

\emph{(ii) implies (iv).} Take $T=\mathcal G_Q$. If $c$ lies on the path from $x$ to $y$ in $T$, every path from $x$ to $y$ in $\mathcal G_Q$ visits $c$, so $c$ separates $x$ and $y$ (Lemma~\ref{lem:blocks}(b)) and beats one of them at every belief in $Q$ (Claim~\ref{claim:between}(c)).

\emph{(iv) implies (ii).} Let the payoffs be single-peaked on $T$, and let $\{x,y\}$ be a margin, with $\BR_V(q)=\{x,y\}$ at $q\in Q$. If $x$ and $y$ were not adjacent in $T$, some action $c$ would lie between them on the path in $T$, and $x$ and $y$ would both beat it at $q$, contrary to (iv). So every margin is an edge of $T$. The margin graph is connected and has the $n$ vertices of $T$, so it has at least $n-1$ edges, which are all the edges of $T$. Hence $\mathcal G_Q=T$.

\emph{Part (b).} Let the margin graph be the path. For $j=2,\dots,n-1$, every path from $a_{j-1}$ to $a_{j+1}$ in $\mathcal G_Q$ visits $a_j$, so $a_j$ separates $a_{j-1}$ and $a_{j+1}$ (Lemma~\ref{lem:blocks}(b)), and Claim~\ref{claim:between}(b) gives $\alpha_j\in(0,1)$ with $V(a_j,q)\geq\alpha_jV(a_{j-1},q)+(1-\alpha_j)V(a_{j+1},q)$ at every belief $q$ on the range, that is,
\[
\alpha_j\bigl[V(a_j,q)-V(a_{j-1},q)\bigr]\geq(1-\alpha_j)\bigl[V(a_{j+1},q)-V(a_j,q)\bigr].
\]
Let $\gamma_1=1$ and $\gamma_j=\gamma_{j-1}\,\alpha_j/(1-\alpha_j)$ for $j=2,\dots,n-1$, and let $h_1=0$ and $h_{j+1}=h_j+\gamma_j$ for $j=1,\dots,n-1$. Dividing the displayed inequality by $\alpha_j\gamma_{j-1}=(1-\alpha_j)\gamma_j>0$ gives \eqref{eq:respaced}.

Conversely, let \eqref{eq:respaced} hold. Fix $q\in Q$ and let $s_m=[V(a_{m+1},q)-V(a_m,q)]/(h_{m+1}-h_m)$, which by \eqref{eq:respaced} is nonincreasing in $m$. Let $i<j<k$ and suppose that $V(a_j,q)<V(a_i,q)$. Then $\sum_{m=i}^{j-1}s_m(h_{m+1}-h_m)<0$, so $s_m<0$ for some $m<j$ and hence for every $m\geq j$, and $V(a_k,q)-V(a_j,q)=\sum_{m=j}^{k-1}s_m(h_{m+1}-h_m)<0$. So $V(a_j,q)\geq\min\{V(a_i,q),V(a_k,q)\}$, the payoffs are single-peaked on the path that joins each $a_i$ to $a_{i+1}$, and by part (a) that path is the margin graph.
\end{proof}

\noindent Part (a) says that decisions leave every margin a scale of its own exactly when the principal's expected payoffs are single-peaked on a tree. With the actions on a path and beliefs anywhere in the simplex, this is the single-agent counterpart of Proposition~7 of \citet{morrisui2004}, under which the differences between the payoffs of adjacent actions can be weighted independently exactly when payoffs are unimodal. Evidence on one question gives a path (Section~\ref{supp:fileexample}), a middle course that separates two alternatives gives a path of three actions, and declining to answer with thresholds above one half gives a star whose center is the action of declining. Condition \eqref{eq:respaced} says that the expected payoff is a concave function of the position $h_j$ of the action at every belief on the range, and on the beliefs with full support it holds if and only if the payoff is concave in the position in every state, because it is linear in the belief. With equally spaced positions it says that each action is at least as good as the equal lottery over its two neighbors.

\section{Separable decisions}\label{supp:separable}

We state precisely and prove the claims in Section~\ref{sec:theorem} on decisions whose payoffs add up and on decisions that interact. Let $A=A_1\times A_2$, with actions $a=(a_1,a_2)$, and let $V(a,\omega)=V_1(a_1,\omega)+V_2(a_2,\omega)$. For an experiment $E$ and a belief $q$, let $I_{V_k}(E,q)$ be the value of $E$ at $q$ in the decision problem with actions $A_k$ and payoff function $V_k$, for $k=1,2$.

\begin{suppclaim}\label{claim:separable}
Let $\alpha_1,\alpha_2>0$ and $\beta:\Omega\to\R$, and let $S(a,\omega)=\alpha_1V_1(a_1,\omega)+\alpha_2V_2(a_2,\omega)+\beta(\omega)$.
\begin{enumerate}[label=(\alph*)]
\item At every belief $q$, $\BR_S(q)=\BR_V(q)$, and every experiment $E$ has $I_V(E,q)=I_{V_1}(E,q)+I_{V_2}(E,q)$ and $I_S(E,q)=\alpha_1I_{V_1}(E,q)+\alpha_2I_{V_2}(E,q)$, where $I_{V_1}(E,q)$ and $I_{V_2}(E,q)$ are nonnegative.
\item Let $(x_1,x_2)$ be the principal's only optimal action at a belief $q$, and let $y_1\neq x_1$ and $y_2\neq x_2$. The principal is indifferent at $q$ between the mistakes $(y_1,x_2)$ and $(x_1,y_2)$ if and only if $V_1(x_1,q)-V_1(y_1,q)=V_2(x_2,q)-V_2(y_2,q)$. At such a belief the objective is indifferent between them if $\alpha_1=\alpha_2$, strictly prefers $(y_1,x_2)$ if $\alpha_1<\alpha_2$, and strictly prefers $(x_1,y_2)$ if $\alpha_1>\alpha_2$.
\end{enumerate}
\end{suppclaim}

\begin{proof}
\emph{Part (a).} At every belief $q$, $S(a,q)=\alpha_1V_1(a_1,q)+\alpha_2V_2(a_2,q)+\beta\cdot q$. The two terms depend on different components of $a$, and $\alpha_1$ and $\alpha_2$ are positive, so $a$ maximizes $S(\cdot,q)$ exactly when $a_k$ maximizes $V_k(\cdot,q)$ over $A_k$ for $k=1,2$, and the same holds for $V$. Hence $\BR_S(q)=\BR_V(q)$, $S^*(q)=\alpha_1V_1^*(q)+\alpha_2V_2^*(q)+\beta\cdot q$ and $V^*(q)=V_1^*(q)+V_2^*(q)$, where $V_k^*(q)=\max_{a_k\in A_k}V_k(a_k,q)$. The posteriors of $E$ average to $q$, so $\beta\cdot q$ cancels in \eqref{eq:value}, which gives the expressions for $I_V$ and $I_S$, and each $I_{V_k}(E,q)$ is nonnegative because $V_k^*$ is convex.

\emph{Part (b).} The principal's losses from $(y_1,x_2)$ and $(x_1,y_2)$ are $d_1=V_1(x_1,q)-V_1(y_1,q)$ and $d_2=V_2(x_2,q)-V_2(y_2,q)$, both positive because $(x_1,x_2)$ is its only optimal action, so it is indifferent exactly when $d_1=d_2$. The objective's losses from the two mistakes are $\alpha_1d_1$ and $\alpha_2d_2$, so when $d_1=d_2$ it prefers $(y_1,x_2)$ if $\alpha_1<\alpha_2$, prefers $(x_1,y_2)$ if $\alpha_1>\alpha_2$, and is indifferent if $\alpha_1=\alpha_2$.
\end{proof}

By part (a), the decomposition of Theorem~\ref{thm:main}(b) holds for such objectives with the two decisions in place of blocks and $\alpha_1$ and $\alpha_2$ as their scales. The proof of Theorem~\ref{thm:test}(a) and~(b) and the proofs of the claims in Sections~\ref{supp:corollary} to~\ref{supp:misdirection}, which use only that decomposition, therefore apply to them, with every statement about the objectives that are aligned on the range read as a statement about the objectives of this form. When a decision satisfies Assumption~\ref{ass:acting} and has several blocks of its own, the same argument applies to objectives that scale each block of each decision by its own positive number and add the decisions' payoffs, with the blocks of the separate decisions in place of the decisions. The blocks of the combined problem cannot take their place, because that problem can fail Assumption~\ref{ass:acting} and have a single block while every scale of each decision remains free. Part (b) plays the role of Proposition~\ref{prop:records}(a) and~(b). With two actions in each decision, if the beliefs at which the two decisions switch cross inside the range, all four actions are optimal where they cross and no belief has exactly three of them optimal, so Assumption~\ref{ass:acting} fails. No action separates the range either. An action $(x_1,x_2)$ is not optimal exactly where the other action of the first decision beats $x_1$ or the other action of the second decision beats $x_2$. Each of these sets of beliefs in $Q$ is convex, and they intersect near the belief at which the switches cross, so their union is connected.

\paragraph{Decisions that interact.} The freedom of the scales $\alpha_1$ and $\alpha_2$ depends on the payoffs adding up. Let the AI make two yes-or-no decisions, so that its actions are the pairs $(x,y)$ with $x,y\in\{0,1\}$, and write the principal's payoffs as
\[
V\bigl((x,y),\omega\bigr)=x\,r_1(\omega)+y\,r_2(\omega)+x\,y\,k(\omega)
\]
for vectors $r_1,r_2,k\in\R^\Omega$. Every payoff function on these four actions has this form once the payoffs of $(0,0)$ are subtracted, which changes no choice and no value of an experiment. The two decisions are complements at a belief $q$ if $k\cdot q>0$ and substitutes if $k\cdot q<0$.

\begin{suppclaim}\label{claim:twoq}
Let the principal's payoffs over two yes-or-no decisions be as above.
\begin{enumerate}[label=(\alph*)]
\item If $(0,0)$ and $(1,1)$ are the principal's only optimal actions at a belief $q$, then $k\cdot q>0$. If $(1,0)$ and $(0,1)$ are, then $k\cdot q<0$.
\item Let $Q$ be a range, and suppose that $r_1\cdot\hat q=r_2\cdot\hat q=0$ at some $\hat q\in Q$, that the maps $z\mapsto r_1\cdot z$ and $z\mapsto r_2\cdot z$ are linearly independent on $L_0$, and that $k\cdot\hat q\neq0$. For $\varepsilon\neq0$, let $V^\varepsilon$ be the payoff function with $\varepsilon k$ in place of $k$. There is $\bar\varepsilon>0$ such that, whenever $0<|\varepsilon|<\bar\varepsilon$, each of two triples of actions is the set of optimal actions under $V^\varepsilon$ at some belief in $Q$: the triples $\{(0,0),(1,0),(1,1)\}$ and $\{(0,0),(0,1),(1,1)\}$ if $\varepsilon\,k\cdot\hat q>0$, and the triples $\{(0,0),(1,0),(0,1)\}$ and $\{(1,0),(0,1),(1,1)\}$ if $\varepsilon\,k\cdot\hat q<0$. If $V^\varepsilon$ also satisfies Assumption~\ref{ass:acting} on $Q$, there is a single block, and every objective that is aligned with $V^\varepsilon$ on the range is exact on the range.
\end{enumerate}
\end{suppclaim}

\begin{proof}
Write $u=r_1\cdot q$, $v=r_2\cdot q$ and $\kappa=k\cdot q$, so that the actions $(0,0)$, $(1,0)$, $(0,1)$ and $(1,1)$ have expected payoffs $0$, $u$, $v$ and $u+v+\kappa$.

\emph{Part (a).} If exactly $(0,0)$ and $(1,1)$ are optimal, then $u+v+\kappa=0$, $u<0$ and $v<0$, so $\kappa=-(u+v)>0$. If exactly $(1,0)$ and $(0,1)$ are optimal, then $u=v>0$ and $u>u+v+\kappa$, so $\kappa<-v<0$.

\emph{Part (b).} Choose $z_1,z_2\in L_0$ with $(r_1\cdot z_1,r_2\cdot z_1)=(1,0)$ and $(r_1\cdot z_2,r_2\cdot z_2)=(0,1)$, and consider the beliefs $q=\hat q+s\,z_1+t\,z_2$, which lie in $Q$ when $s$ and $t$ are small. At such a belief $u=s$, $v=t$ and, for the payoffs $V^\varepsilon$, $\kappa=\varepsilon\,(k\cdot\hat q+s\,k\cdot z_1+t\,k\cdot z_2)$. Let $\varepsilon\,k\cdot\hat q>0$. The belief with $s=0$ and $t=-\varepsilon\,k\cdot\hat q/(1+\varepsilon\,k\cdot z_2)$ has $u=0$ and $v+\kappa=t\,(1+\varepsilon\,k\cdot z_2)+\varepsilon\,k\cdot\hat q=0$, and $t<0$ when $|\varepsilon|$ is small. The payoffs of $(0,0)$, $(1,0)$ and $(1,1)$ are then $0$ and the payoff of $(0,1)$ is $t<0$, so exactly those three actions are optimal. By symmetry, the belief with $t=0$ and $s=-\varepsilon\,k\cdot\hat q/(1+\varepsilon\,k\cdot z_1)$ has exactly $(0,0)$, $(0,1)$ and $(1,1)$ optimal. Let $\varepsilon\,k\cdot\hat q<0$. At $\hat q$ itself $u=v=0$ and $\kappa<0$, so exactly $(0,0)$, $(1,0)$ and $(0,1)$ are optimal. The belief with $s=t=-\varepsilon\,k\cdot\hat q/(1+\varepsilon\,k\cdot z_1+\varepsilon\,k\cdot z_2)$ has $s>0$ when $|\varepsilon|$ is small and $\kappa=\varepsilon\,k\cdot\hat q+s\,(\varepsilon\,k\cdot z_1+\varepsilon\,k\cdot z_2)=-s$, so the payoffs of $(1,0)$, $(0,1)$ and $(1,1)$ are all $s$ and the payoff of $(0,0)$ is $0<s$. All these beliefs tend to $\hat q$ as $\varepsilon$ tends to zero, so they lie in $Q$ when $|\varepsilon|$ is small. Under Assumption~\ref{ass:acting} each of the two triples is a three-way tie, the three margins of a triple are linked, and the two triples share a margin, so their five margins lie in one block, which contains the four actions. Any other margin joins two actions of this block and so belongs to it (Lemma~\ref{lem:blocks}(d)). Hence $K=1$, and by Theorem~\ref{thm:main}(c) every objective that is aligned on the range is exact on the range.
\end{proof}

\noindent With additive payoffs the principal never switches directly between two actions that differ in both decisions, by part (a), and each decision keeps a scale of its own, by Claim~\ref{claim:separable}. With a slight interaction where both decisions are marginal it does switch directly, between taking neither action and both under complements and between the two single actions under substitutes, and the belief at which it does so depends on how the stakes of the two decisions compare, which ties their scales. In the terms of \citet{topkis1978} and \citet{milgromshannon1994}, the payoff is supermodular in $(x,y)$ under complements and submodular under substitutes, and the free scales belong to the modular case between them. The identification is weak when the interaction is small. The objective with payoffs $x\,r_1(\omega)+2y\,r_2(\omega)+\varepsilon\,x\,y\,k(\omega)$, which doubles the stakes of the second decision, errs only on a set of beliefs whose measure is of order $|\varepsilon|$, at a loss of order $|\varepsilon|$, yet it values evidence about the second decision about twice as highly as the principal does, so it misranks experiments by amounts of order one.

\section{Ranking evidence and paying for it}\label{supp:corollary}

This section shows when a given aligned objective ranks evidence, and pays for it, as the principal does (Section~\ref{sec:test}), and derives what the principal's own choices of evidence reveal. For an objective that is aligned on the range with scales $\lambda_X$ and an experiment $E$ with $I_V(E,q)>0$, let $\bar\lambda(E,q)=\sum_X\lambda_X\,\phi_X(E,q)/\sum_X\phi_X(E,q)$ be the \emph{average scale} of $E$ at $q$, the average of the scales weighted by the shares of the experiment's value that come from each block, so that $I_S(E,q)=\bar\lambda(E,q)\,I_V(E,q)$. For each belief $q\in Q$, let $T(q)$ be a finite set of experiments that keep beliefs on the range at $q$, the experiments available to the AI at that belief.\footnote{Every experiment keeps beliefs on the range when $Q$ is the set of beliefs with full support. When the range comes from a capability bound (Section~\ref{sec:capability}), an experiment keeps beliefs on the range at $q$ if, combined with the evidence that led to $q$, it is a garbling of the bound (Section~\ref{supp:range}). The set $T(q)$ depends on the belief because the evidence that the AI can still gather depends on the evidence that it already has.} The \emph{dilution} $E^\alpha$ of an experiment $E$, for $\alpha\in[0,1]$, reports the signal of $E$ with probability $\alpha$ and an uninformative signal otherwise. Dilutions play the role here that lotteries play in expected utility, since indifference between one experiment and a dilution of another reveals the ratio of their values. Let $S$ be aligned on the range. We say that $S$ \emph{ranks evidence as the principal does} if, at every belief $q\in Q$ and on every menu of two costless experiments, each a dilution of an experiment in $T(q)$, every experiment that is optimal for the AI is optimal for the principal. We say that $S$ \emph{pays for evidence as the principal does} if, at every belief $q\in Q$, for every experiment $E\in T(q)$ and every cost $c>0$, which the objective counts as the principal does, every choice between running $E$ at that cost and not running it that is optimal for the AI is optimal for the principal.

\begin{suppclaim}\label{claim:tasks}
Let Assumption~\ref{ass:acting} hold, and let $S$ be aligned on the range with scales $\lambda_X$.
\begin{enumerate}[label=(\alph*)]
\item $S$ ranks evidence as the principal does if and only if, at each belief $q\in Q$, all experiments in $T(q)$ with positive value to the principal have the same average scale.
\item $S$ pays for evidence as the principal does if and only if, at every belief $q\in Q$, every experiment in $T(q)$ with positive value to the principal has an average scale of one.
\end{enumerate}
\end{suppclaim}

\begin{proof}
Fix $q\in Q$, and let $\lambda=(\lambda_X)$.

\emph{Part (a).} A dilution $E^\alpha$ leads to the posterior $q$ with probability $1-\alpha$ and otherwise to the posteriors of $E$, with their probabilities multiplied by $\alpha$, so it also keeps beliefs on the range at $q$, and
\[
\phi_X(E^\alpha,q)=(1-\alpha)V^*_X(q)+\alpha\sum_sP(s)V^*_X(q_s)-V^*_X(q)=\alpha\,\phi_X(E,q).
\]
The principal values $E^\alpha$ at $\alpha\,\one\cdot y(E)$ and the AI at $\alpha\,\lambda\cdot y(E)$. Suppose first that all experiments in $T(q)$ with positive value have the same average scale $\nu>0$. Then $\lambda\cdot y(E)=\nu\,\one\cdot y(E)$ for every $E\in T(q)$, with both sides zero when $y(E)=0$, so the AI's value of every dilution of every experiment in $T(q)$ is $\nu$ times the principal's, and on every menu of such dilutions the AI and the principal have the same optimal experiments.

Suppose next that $S$ ranks evidence as the principal does. Let $E,E'\in T(q)$ have $y=y(E)\neq0$ and $y'=y(E')\neq0$. The numbers $\one\cdot y$, $\one\cdot y'$, $\lambda\cdot y$ and $\lambda\cdot y'$ are positive, because $y$ and $y'$ are nonnegative and nonzero and every $\lambda_X$ is positive. Suppose that $\lambda\cdot y\leq\lambda\cdot y'$, and let $\alpha=(\lambda\cdot y)/(\lambda\cdot y')\in(0,1]$. The AI values $E$ and $E'^\alpha$ equally, at $\lambda\cdot y$, so both are among its optimal choices from the menu $\{E,E'^\alpha\}$, and both must therefore be optimal for the principal, which gives $\one\cdot y=\alpha\,\one\cdot y'$. Hence
\[
\frac{\lambda\cdot y}{\one\cdot y}=\frac{\alpha\,\lambda\cdot y'}{\alpha\,\one\cdot y'}=\frac{\lambda\cdot y'}{\one\cdot y'}.
\]
If instead $\lambda\cdot y>\lambda\cdot y'$, the same argument with $E$ and $E'$ exchanged gives the same equality. So every experiment in $T(q)$ with positive value has the same average scale.

\emph{Part (b).} Fix $E\in T(q)$, and write $I_V=\one\cdot y(E)$ and $I_S=\lambda\cdot y(E)$. If $I_V=0$, then $y(E)=0$, so $I_S=0$, and at every cost $c>0$ both the AI and the principal strictly prefer not to run $E$. If $I_V>0$, the principal strictly prefers to run $E$ when $c<I_V$ and strictly prefers not to when $c>I_V$, while the AI strictly prefers to run $E$ when $c<I_S$ and strictly prefers not to when $c>I_S$. If $I_S\neq I_V$, then at every cost strictly between the two numbers one of them strictly prefers to run $E$ and the other strictly prefers not to, so the AI does not choose as the principal would. If $I_S=I_V$, the AI's net value $I_S-c$ equals the principal's at every cost, so they make the same choices. Hence the AI chooses as the principal would at every cost if and only if $I_S=I_V$, that is, if and only if $E$ has an average scale of one whenever $I_V>0$.
\end{proof}

\noindent In the proof of part (a), when two experiments have different average scales, the AI is indifferent between one of them and a dilution of the other, whereas the principal is not. Diluting one of the two slightly further then makes the AI strictly prefer the option that the principal strictly rejects, which is the statement in Section~\ref{sec:test}. In the file example, the clearing and confirming tests bear on different blocks, so only an objective with $\lambda_1=\lambda_2$ ranks every pair of their dilutions as the firm does.

The next claim characterizes the scales that these two properties allow. For an experiment $E\in T(q)$ with $I_V(E,q)>0$, let $u(E,q)=\phi(E,q)/I_V(E,q)$ be its \emph{value shares}, the fractions of its value to the principal that come from each block, and let $\Psi$ be the span of the differences $u(E,q)-u(E',q)$ over all beliefs $q\in Q$ and all pairs of experiments in $T(q)$ with positive value. Every such difference has components that sum to zero, so $\Psi$ lies in the set $\one^\perp$ of vectors whose components sum to zero.

\begin{suppclaim}\label{claim:psi}
Let Assumption~\ref{ass:acting} hold.
\begin{enumerate}[label=(\alph*)]
\item At a belief $q\in Q$, some objective that is aligned on the range gives two experiments in $T(q)$ with positive value different average scales if and only if two experiments in $T(q)$ have linearly independent block values at $q$.
\item An objective that is aligned on the range with scales $\lambda=(\lambda_X)$ ranks evidence as the principal does if and only if $\lambda$ is orthogonal to $\Psi$. Every aligned objective that ranks evidence as the principal does is exact on the range if and only if $\Psi=\one^\perp$.
\item Every aligned objective that pays for evidence as the principal does has $\lambda_X=1$ for every block $X$ if and only if the block values $\phi(E,q)$, $E\in T(q)$, $q\in Q$, span $\R^K$.
\end{enumerate}
\end{suppclaim}

\noindent Claim~\ref{claim:psi} says what the principal's own choices of evidence reveal (Table~\ref{tab:records}). Choices among experiments reveal more than decisions as soon as two experiments bear on the blocks in different proportions, and each further direction in which the experiments' value shares vary fixes one more relative scale. Choices of whether to run an experiment at a cost reveal more still, because they fix the level of the scales. With the sandbox of Section~\ref{sec:test} as the only experiment, choices among experiments restrict nothing, and at any one belief the decision whether to sandbox fixes only the sandbox's average scale. But the sandbox's block values $(\min\{9p,1-p\},\min\{p,9(1-p)\})$ at different beliefs span $\R^2$, so by part (c) the decisions whether to sandbox at every belief fix both scales at one.

\begin{proof}
The conditions of Claim~\ref{claim:tasks} depend on an objective that is aligned on the range only through its scales. Say that the \emph{ranking condition} holds at $q$ if all experiments in $T(q)$ with positive value have the same average scale. It holds exactly when there is $\nu(q)>0$ with $\sum_X(\lambda_X-\nu(q))\phi_X(E,q)=0$ for every $E\in T(q)$, since experiments with zero value satisfy this for every $\nu(q)$. The condition of Claim~\ref{claim:tasks}(b) says that $\sum_X(\lambda_X-1)\phi_X(E,q)=0$ for every $q\in Q$ and $E\in T(q)$.

\emph{Part (a).} Fix $q\in Q$. Suppose first that experiments $E,E'\in T(q)$ have linearly independent block values $y=\phi(E,q)$ and $y'=\phi(E',q)$. Both are nonzero and nonnegative, so $u=y/(\one\cdot y)$ and $u'=y'/(\one\cdot y')$ are well defined, and $u\neq u'$ because $y$ and $y'$ are not proportional. Let $\lambda=\one+\varepsilon(u-u')$ with $0<\varepsilon<1$. Every component of $u-u'$ lies in $[-1,1]$, because $u$ and $u'$ are nonnegative with components that sum to one, so every component of $\lambda$ is positive. The ranking condition at $q$ would require $\lambda\cdot y=\nu\,\one\cdot y$ and $\lambda\cdot y'=\nu\,\one\cdot y'$, that is, $\lambda\cdot u=\nu=\lambda\cdot u'$. But
\[
\lambda\cdot u-\lambda\cdot u'=\one\cdot(u-u')+\varepsilon(u-u')\cdot(u-u')=\varepsilon\|u-u'\|^2>0,
\]
where $\|\cdot\|$ is the Euclidean norm and $\one\cdot(u-u')=1-1=0$. So an objective that is aligned on the range with these scales gives $E$ and $E'$ different average scales at $q$. Conversely, suppose that the block values of the experiments in $T(q)$ are pairwise linearly dependent. If the block values $\bar y$ of some experiment are nonzero, the block values of every experiment are a multiple of $\bar y$, and the multiple is nonnegative because both vectors are nonnegative. For any positive scales, the ranking condition then holds at $q$ with $\nu(q)=(\lambda\cdot\bar y)/(\one\cdot\bar y)>0$. If all the block values are zero, it holds with any $\nu(q)>0$.

\emph{Part (b).} Fix $q\in Q$ and let $\Psi(q)$ be the span of the differences $u(E,q)-u(E',q)$ over pairs of experiments in $T(q)$ with positive value to the principal. We show that a vector $\lambda$ of positive scales satisfies the ranking condition at $q$ if and only if it is orthogonal to $\Psi(q)$. If no experiment in $T(q)$ has positive value, every block value at $q$ is zero, because block values are nonnegative and sum to the value, so the condition holds for every $\lambda$ with any $\nu(q)>0$, and $\Psi(q)=\{0\}$. Otherwise some $E^*\in T(q)$ has $I_V(E^*,q)>0$. If the condition holds with $\nu=\nu(q)$, then $\lambda\cdot\phi(E,q)=\nu\,I_V(E,q)$ for every $E\in T(q)$, and dividing by $I_V(E,q)$ gives $\lambda\cdot u(E,q)=\nu$ for every experiment with positive value, so $\lambda$ is orthogonal to every difference of value shares. Conversely, if $\lambda$ is orthogonal to $\Psi(q)$, then $\lambda\cdot u(E,q)$ takes a common value $\nu$ for all experiments with positive value, and $\nu=\lambda\cdot u(E^*,q)>0$ because $u(E^*,q)$ is nonnegative and nonzero and $\lambda$ is positive. Then $\lambda\cdot\phi(E,q)=\nu\,\one\cdot\phi(E,q)$ for every experiment with positive value, and for every experiment with zero value both sides are zero, so the condition holds with $\nu(q)=\nu$. A vector of positive scales therefore satisfies the ranking condition at every belief in $Q$ if and only if it is orthogonal to every $\Psi(q)$, that is, to their sum $\Psi$, and by Claim~\ref{claim:tasks}(a) an objective that is aligned on the range with these scales ranks evidence as the principal does exactly then.

Every difference of value shares is orthogonal to $\one$, because the components of each vector of value shares sum to one, so $\Psi\subseteq\one^\perp$. By Fact~3 and Lemma~\ref{lem:decomp}(d), a block-proportional objective is exact on the range if and only if its scales are all equal. If $\Psi=\one^\perp$, then $\Psi^\perp=\R\one$, and every positive vector orthogonal to $\Psi$ has equal components. If $\Psi\neq\one^\perp$, then $\Psi^\perp$ contains a vector $z$ that is not a multiple of $\one$, and for small $\varepsilon>0$ the vector $\one+\varepsilon z$ is positive, orthogonal to $\Psi$ and not a multiple of $\one$. Some objective that is aligned on the range has these scales, and it ranks evidence as the principal does without being exact on the range.

\emph{Part (c).} By Claim~\ref{claim:tasks}(b), an objective that is aligned on the range pays for evidence as the principal does if and only if $\lambda-\one$ is orthogonal to every $\phi(E,q)$ with $q\in Q$ and $E\in T(q)$. If these vectors span $\R^K$, then $\lambda=\one$. If they do not, some $z\neq0$ is orthogonal to all of them, and for small $\varepsilon>0$ the vector $\one+\varepsilon z$ is positive and satisfies the condition, so some aligned objective with these scales pays for evidence as the principal does without having $\lambda_X=1$ for every block.
\end{proof}

\section{When decisions settle the choice}\label{supp:choose}

We prove the claims in Section~\ref{sec:test} that the principal's decisions settle a choice among costless experiments when one experiment on the menu is at least as valuable as the others in every problem of a suitable class, which covers garblings and experiments ordered by accuracy, when the AI can run its experiments together, and when the experiments have positive value in at most one block, and that an experiment has positive value in a block only if one of its signals changes which action of the block is best. The sets $T(q)$ are those of Section~\ref{supp:corollary}.

\begin{suppclaim}\label{claim:choose}\label{claim:ordered}
Let Assumption~\ref{ass:acting} hold, let $q\in Q$, and let $M$ be a menu at $q$.
\begin{enumerate}[label=(\alph*)]
\item Let $\mathcal C$ be a class of decision problems, each a finite set of actions with a payoff function on $\Omega$, that contains, for every block $X$, the problem with the actions of $X$ and the principal's payoffs. If an experiment $E^*\in M$ is at least as valuable at $q$ as every other experiment on $M$ in every problem of $\mathcal C$, then $E^*$ is best in every block, and the principal's decisions settle the choice from $M$ at $q$. This is so, in particular, if every other experiment on $M$ is a garbling of $E^*$.
\item Suppose that the finitely many experiments in $T(q)$ have a joint distribution of signals in each state, that the AI can run any set of them and observe their signals, and that every such combination keeps beliefs on the range at $q$. On the menu of these combinations, the experiment that reports all the signals is best in every block, so the principal's decisions settle the choice.
\item An experiment $E$ on $M$ has $\phi_X(E,q)>0$ only if no action of $X$ maximizes $V(\cdot,q_s)$ over the actions of $X$ at every posterior $q_s$ of $E$. If there is at most one block in which some experiment on $M$ has positive value, an experiment that is best in that block is best in every block, and the principal's decisions settle the choice from $M$ at $q$.
\end{enumerate}
\end{suppclaim}

\begin{proof}
Each block value $\phi_X(E,q)$ is the value of $E$ at $q$ in the decision problem with the actions of $X$ and the payoff function $V$.

\emph{Part (a).} That problem belongs to $\mathcal C$, so $\phi_X(E^*,q)\geq\phi_X(E,q)$ for every $E\in M$ and every block $X$, and Theorem~\ref{thm:test}(a) applies. For garblings, let $\mathcal C$ be the class of all decision problems, in each of which a garbling of an experiment is worth no more than the experiment \citep{blackwell1953}.

\emph{Part (b).} For a set $U\subseteq T(q)$, let $E_U$ be the experiment that reports the signals of the experiments in $U$, drawn from their joint distribution, with $E_\emptyset$ revealing nothing. By hypothesis these experiments form a menu at $q$, and each is a garbling of $E_{T(q)}$, obtained by discarding the signals of the experiments outside $U$, so part (a) applies.

\emph{Part (c).} Suppose that an action $a$ of $X$ maximizes $V(\cdot,q_s)$ over the actions of $X$ at every posterior $q_s$ of $E$. The posteriors average to $q$ and $V(a,\cdot)$ is linear, so $\sum_sP(s)V^*_X(q_s)=\sum_sP(s)V(a,q_s)=V(a,q)\leq V^*_X(q)$, and $\phi_X(E,q)\leq0$. Block values are nonnegative, so $\phi_X(E,q)=0$. For the second statement, let $X$ be the only block in which some experiment on $M$ has positive value, if there is one, and let $E^*$ be best in $X$. In every other block every experiment on $M$ has value zero, so $E^*$ is best there as well, and Theorem~\ref{thm:test}(a) applies. If there is no such block, every experiment on $M$ is best in every block.
\end{proof}

\noindent Part (a) applies to any class that contains the principal's problem on $A_Q$ and every problem obtained from a member of the class by removing actions. For instance, let the states be ordered, and call a problem \emph{single crossing} if its actions can be ordered so that the difference between the payoffs of a higher and a lower action is nonpositive in every state below some state and nonnegative in every other state. Removing actions from a single crossing problem leaves a single crossing problem. For monotone decision problems and experiments whose signals are ordered by likelihood ratios, \citet{lehmann1988} shows that an experiment that is more accurate in the sense of that paper is at least as valuable, and \citet{persico2000} and \citet{quahstrulovici2009} extend the result to single crossing problems and beyond. Under the conditions of those results, if the principal's problem on $A_Q$ is single crossing and one experiment on the menu is more accurate than each of the others, the principal's decisions settle the choice, and a designer can conclude this without finding the blocks. The conclusion holds for every objective that is aligned on the range, whether or not that objective is itself single crossing. The converse of the garbling case of part (a) fails, since Theorem~\ref{thm:test} requires only $K$ inequalities for each pair of experiments, one for each block, and these can hold for experiments that the order of \citet{blackwell1953} does not rank, as for the two tests of the file example at beliefs below $1/13$.

\section{The size of the loss}\label{supp:misdirection}

This section proves the bound for a relative scale that is wrong by a limited amount and the guarantees from lotteries over experiments, all stated in Section~\ref{sec:test}, shows that the bound is attained whenever there are two blocks, bounds the loss when experiments are costly, derives the guarantee from bisection stated in Section~\ref{sec:records}, computes the share that an experiment guarantees after a record of rankings, and gives the share that a whole deployment secures, used in Section~\ref{sec:practice}. We use the notation of the proof of Theorem~\ref{thm:test} in Appendix~\ref{app:proofs}.

\begin{suppclaim}\label{claim:ratio}
Let Assumption~\ref{ass:acting} hold, fix $q\in Q$ and a menu $M$ at $q$, and let $S$ be aligned on the range. Let $\lambda_{\min}$ and $\lambda_{\max}$ be the smallest and largest of its scales among the blocks in which some experiment on $M$ has positive value. If there is such a block, every experiment $E_S$ on $M$ that is optimal for the AI and every experiment $E_V$ on $M$ that is optimal for the principal satisfy $I_V(E_S,q)\geq(\lambda_{\min}/\lambda_{\max})\,I_V(E_V,q)$.
\end{suppclaim}

\begin{proof}
For every experiment $E$ on the menu, $y(E)$ is nonnegative and vanishes outside the blocks in which some experiment on the menu has positive value, so $\lambda_{\min}\,\one\cdot y(E)\leq\lambda\cdot y(E)\leq\lambda_{\max}\,\one\cdot y(E)$. The experiments are costless, so the AI's optimality gives $\lambda\cdot y(E_S)\geq\lambda\cdot y(E_V)$, and
\[
\lambda_{\min}\,I_V(E_V,q)\leq\lambda\cdot y(E_V)\leq\lambda\cdot y(E_S)\leq\lambda_{\max}\,I_V(E_S,q).
\]
\end{proof}

\paragraph{The bound in Claim~\ref{claim:ratio} is attained whenever there are two blocks.} Let the blocks $X$ and $Y$ have scales $\lambda_X<\lambda_Y$, since with equal scales the bound is one and holds with equality. The two blocks share exactly one action $c$: the path from $X$ to $Y$ in the tree $\mathcal H$ alternates between blocks and actions, and with only two blocks it is $X,c,Y$ for some action $c$, whereas two shared actions $c$ and $c'$ would form the cycle $X,c,Y,c',X$. Take $c$ as the root of $\mathcal H$, so that $r_X=r_Y=c$, and recall that $h_X(q)=V^*_X(q)-V(r_X,q)$. By Fact~5, $c$ is the only optimal action at every belief of a nonempty set $U_c\subseteq Q$ that is open relative to $L$, and for an action $a\neq c$ of $X$, some belief $q_a\in Q$ has $a$ as its only optimal action. Fix $q\in U_c$, so that $h_X(q)=h_Y(q)=0$. At $q_a$, $h_X(q_a)=V(a,q_a)-V(c,q_a)>0$, and by Lemma~\ref{lem:structure}(d), applied with $v=a$ and the block $Y$, whose path to $a$ begins with $c$, the action $c$ is best in $Y$, so $h_Y(q_a)=0$. For small $\varepsilon>0$, the belief $q'=q+\varepsilon(q-q_a)$ lies in $U_c$, so $h_X(q')=h_Y(q')=0$. Let $\pi=\varepsilon/(1+\varepsilon)$, so that $\pi q_a+(1-\pi)q'=q$. By the splitting lemma \citep{aumannmaschler1995,kamenicagentzkow2011}, some experiment leads at $q$ to the posteriors $q_a$ and $q'$ with probabilities $\pi$ and $1-\pi$. Both posteriors lie in $Q$, and the experiment's block values are $(\pi h_X(q_a),0)$. Likewise, an action $b\neq c$ of $Y$ gives an experiment with block values $(0,y)$ for some $y>0$. Diluting one of the two experiments, we obtain experiments $E_X$ and $E_Y$ with block values $(x,0)$ and $(0,\lambda_Xx/\lambda_Y)$ for some $x>0$. On the costless menu $\{E_X,E_Y\}$ the principal chooses $E_X$, whereas the AI values both at $\lambda_Xx$, so $E_Y$ is optimal for it, and the value of $E_Y$ to the principal is $(\lambda_X/\lambda_Y)\,I_V(E_X,q)$.

\paragraph{Lotteries over the menu.} A lottery is a probability distribution $\pi$ on a menu $M$ at $q$, under which the AI runs each $E\in M$ with probability $\pi(E)$. When some experiment on $M$ has positive value, let $\mathcal B'$ be the set of blocks $X$ with $m_X>0$ and $K'\geq1$ their number, let $\hat y_X(E)=\phi_X(E,q)/m_X\in[0,1]$ for $X\in\mathcal B'$ and $E\in M$, and let
\[
G(\pi)=\min_{X\in\mathcal B'}\ \sum_{E\in M}\pi(E)\,\hat y_X(E),
\]
which is the guaranteed share $G(E)$ of Section~\ref{sec:test} when $\pi$ puts probability one on $E$.

\begin{suppclaim}\label{claim:lottery}\label{claim:bestlottery}
Let Assumption~\ref{ass:acting} hold, let $q\in Q$, and let $M$ be a menu at $q$ on which some experiment has positive value.
\begin{enumerate}[label=(\alph*)]
\item For every lottery $\pi$ on $M$ and every objective $W$ that is aligned on the range, $\sum_E\pi(E)\,I_W(E,q)\geq G(\pi)\max_{E'\in M}I_W(E',q)$, and no number larger than $G(\pi)$ has this property.
\item The largest value of $G(\pi)$ over the lotteries on $M$ is
\[
G^*=\max_{\pi}\ \min_{X\in\mathcal B'}\ \sum_{E\in M}\pi(E)\,\hat y_X(E)=\min_{w}\ \max_{E\in M}\ \sum_{X\in\mathcal B'}w_X\,\hat y_X(E),
\]
where $w$ ranges over the probability distributions on $\mathcal B'$. A lottery that puts probability $1/K'$ on an experiment that is best in $X$, for each $X\in\mathcal B'$, has $G(\pi)\geq1/K'$, so $G^*\geq1/K'$. Also $G^*\geq G(E)$ for every $E\in M$, $G^*=1$ if and only if some experiment on $M$ is best in every block, and $G^*$ is attained by a lottery that puts positive probability on at most $K'$ experiments.
\item If each experiment on $M$ has positive value in at most one block, then $G(\pi)\leq1/K'$ for every lottery $\pi$, so $G^*=1/K'$.
\item The numbers $\hat y_X(E)$, and with them $G(\pi)$ and $G^*$, are the same whether they are computed from $V$ or from an objective that is aligned on the range.
\end{enumerate}
\end{suppclaim}

\begin{proof}
\emph{Part (a).} Let $E_\pi$ be the experiment that runs each $E\in M$ with probability $\pi(E)$ and reports which it ran together with its signal. Its posteriors are those of the experiments on $M$, so it keeps beliefs on the range, and its block values are $\sum_E\pi(E)\,\phi_X(E,q)$, so that $I_W(E_\pi,q)=\sum_E\pi(E)\,I_W(E,q)$ for every objective $W$ that is aligned on the range. Adding $E_\pi$ to the menu changes neither the numbers $m_X$ nor $\max_{E'\in M}I_W(E',q)$, and on the enlarged menu $G(E_\pi)=G(\pi)$. Part (a) is therefore Theorem~\ref{thm:test}(b) applied to $E_\pi$.

\emph{Part (b).} The function $G$ is continuous on the compact set of lotteries, so it has a largest value, and the equality of the two expressions is the minimax theorem \citep{vonneumann1928} for the zero-sum game in which one player chooses $E\in M$, the other chooses $X\in\mathcal B'$, and the first receives $\hat y_X(E)$. The lottery described in part (b) has $\sum_E\pi(E)\hat y_X(E)\geq1/K'$ for every $X\in\mathcal B'$, adding the probabilities of an experiment chosen for several blocks, and $G(E)$ is the value of $G$ at the lottery that puts probability one on $E$. Since $\hat y_X(E)\leq1$, $G^*\leq1$, with equality if some experiment is best in every block. Conversely, if $G^*=1$, every experiment that a maximizing lottery uses has $\hat y_X(E)=1$ for every $X\in\mathcal B'$, and so is best in every block, since every experiment on the menu is best in the blocks outside $\mathcal B'$. Finally, $G^*$ is the value of the linear program that maximizes $g$ over $g\geq0$, $\pi(E)\geq0$ and $s_X\geq0$ subject to the $K'+1$ equations $\sum_E\pi(E)\hat y_X(E)-g-s_X=0$, $X\in\mathcal B'$, and $\sum_E\pi(E)=1$. It has a basic optimal solution, in which at most $K'+1$ variables are positive, and since $g=G^*>0$, at most $K'$ of the probabilities are positive.

\emph{Part (c).} The sets of experiments that have positive value in the different blocks of $\mathcal B'$ are disjoint, so a lottery $\pi$ puts total probability at most $1/K'$ on one of them, say the set for $X^*$. Since $\hat y_{X^*}(E)\leq1$ and is zero off that set, $G(\pi)\leq\sum_E\pi(E)\hat y_{X^*}(E)\leq1/K'$.

\emph{Part (d).} If $S$ is aligned on the range with scales $\lambda_X$, Theorem~\ref{thm:test}(c) gives $\phi^S_X(E,q)=\lambda_X\phi_X(E,q)$, so the largest value of $\phi^S_X$ on the menu is $\lambda_Xm_X$ and $\hat y_X(E)$ is unchanged.
\end{proof}

\paragraph{Costly experiments.} Fix $q\in Q$ and a finite menu of experiments that keep beliefs on the range at $q$, with costs $c(E)\geq0$ that the objective counts as the principal does, and let $S$ be aligned on the range with scales $\lambda=(\lambda_X)$. Let $E_S$ maximize $I_S(E,q)-c(E)$ and $E_V$ maximize $U(E)=I_V(E,q)-c(E)$ over the menu, and let $\chi(E)=I_S(E,q)-I_V(E,q)$ be the amount by which the objective overvalues $E$.

\begin{supplemma}\label{lem:overvalue}
Under these conditions, the principal's loss from the AI's choice is $U(E_V)-U(E_S)$, and $0\leq U(E_V)-U(E_S)\leq\chi(E_S)-\chi(E_V)$. If experiments cost nothing and there are two blocks with $\lambda_1<\lambda_2$, then $\phi_2(E_S,q)\geq\phi_2(E_V,q)$ and $\phi_1(E_S,q)\leq\phi_1(E_V,q)$.
\end{supplemma}

\begin{proof}
The AI's actions at every belief in $\cl Q$ are optimal for the principal, so when the AI runs $E$ and then acts, the principal's expected payoff is $\sum_sP(s)V^*(q_s)-c(E)=V^*(q)+U(E)$. The loss from the AI's choice is therefore $U(E_V)-U(E_S)$, which is nonnegative because $E_V$ maximizes $U$. Write $\Delta=y(E_S)-y(E_V)$. The two optimality conditions are
\begin{equation}\label{eq:optS}
\lambda\cdot\Delta\geq c(E_S)-c(E_V),
\end{equation}
because $E_S$ is optimal for the AI and $E_V$ is on the menu, and
\begin{equation}\label{eq:optV}
c(E_S)-c(E_V)\geq\one\cdot\Delta,
\end{equation}
because $E_V$ is optimal for the principal and $E_S$ is on the menu. The definition of $U$ and then \eqref{eq:optS} give
\[
U(E_V)-U(E_S)=-\one\cdot\Delta+\bigl[c(E_S)-c(E_V)\bigr]\leq-\one\cdot\Delta+\lambda\cdot\Delta=(\lambda-\one)\cdot\Delta=\chi(E_S)-\chi(E_V).
\]
Without costs, the two conditions read $\lambda\cdot\Delta\geq0\geq\one\cdot\Delta$, so $(\lambda-\nu\one)\cdot\Delta=\lambda\cdot\Delta-\nu\,\one\cdot\Delta\geq0$ for every $\nu\geq0$. With two blocks and $\lambda_1<\lambda_2$, taking $\nu=\lambda_1$ gives $(\lambda_2-\lambda_1)\Delta_2\geq0$, so $\Delta_2\geq0$, and taking $\nu=\lambda_2$ gives $(\lambda_1-\lambda_2)\Delta_1\geq0$, so $\Delta_1\leq0$ because $\lambda_1-\lambda_2<0$.
\end{proof}

Now let the AI's belief be drawn from a distribution on $Q$, with a finite menu at each belief $q$ of experiments that keep beliefs on the range at $q$, with costs $c(E)\geq0$, and let the principal's \emph{gap} $\gamma(q)$ be the amount by which, in net value, its best experiment on the menu at $q$ beats the best of those that are not optimal for it, with $\gamma(q)=\infty$ when all are optimal. We take the menus and the AI's choices to vary with the belief so that the loss and $\gamma$ are measurable. In the example of Section~\ref{sec:example}, $\gamma(p)=|2-3p|/5$ for the two tests at beliefs $p\neq2/3$ in $[1/5,4/5]$, which is small only near $p^*=2/3$, the belief at which the firm switches between them.

\begin{suppclaim}\label{claim:avgloss}
Let Assumption~\ref{ass:acting} hold, let $S$ be aligned on the range, and let the menus be as above. If the objective's overvaluations of any two experiments on any menu differ by at most $\bar\chi$, the principal's expected loss is at most $\bar\chi\Pr(\gamma(q)\leq\bar\chi)$. If also $\Pr(\gamma(q)\leq x)\leq\bar kx$ for every $x>0$, it is at most $\bar k\bar\chi^2$.
\end{suppclaim}

\begin{proof}
Fix a belief $q$, write $E_S$ and $E_V$ for the choices there, and let $\ell(q)=U(E_V)-U(E_S)$. If $E_S$ is optimal for the principal, then $\ell(q)=0$. Otherwise $\ell(q)\geq\gamma(q)$, by the definition of $\gamma$, and $\ell(q)\leq\chi(E_S)-\chi(E_V)\leq\bar\chi$, by Lemma~\ref{lem:overvalue} and the hypothesis. So whenever $\ell(q)>0$, $\gamma(q)\leq\ell(q)\leq\bar\chi$. The loss is therefore zero at every belief with $\gamma(q)>\bar\chi$ and at most $\bar\chi$ at every other belief, and taking expectations gives $\E[\ell(q)]\leq\bar\chi\Pr(\gamma(q)\leq\bar\chi)$. If also $\Pr(\gamma(q)\leq x)\leq\bar kx$ for every $x>0$, taking $x=\bar\chi$ gives $\E[\ell(q)]\leq\bar k\bar\chi^2$.
\end{proof}

\noindent Claim~\ref{claim:avgloss} bounds the expected loss by $\bar k\bar\chi^2$ when $\Pr(\gamma(q)\leq x)\leq\bar kx$ for every $x>0$. Such a constant $\bar k$ exists whenever the menu of experiments and their costs are the same at every belief, every experiment on the menu keeps beliefs on the range at every belief in $Q$, as is the case when $Q$ is the set of beliefs with full support, and the AI's belief has a density bounded by $\bar g$ with respect to Lebesgue measure on $L$. To see this, note that the value of an experiment is piecewise linear in the belief, because $P(s)V^*(q_s)=\max_a\sum_\omega q(\omega)f(s\mid\omega)V(a,\omega)$, so every difference $U(E)-U(E')$ between net values is affine on each of finitely many convex polytopes that cover the closure of $Q$. A piece of lower dimension than $L$ has probability zero, so we may take each piece to have the dimension of $L$. On a piece on which the difference is constant, the beliefs at which $0<|U(E)-U(E')|\leq x$ form the empty set unless the constant is nonzero and at most $x$ in absolute value, so they have probability at most $x$ divided by the absolute value of the constant if it is nonzero, and zero otherwise. On a piece on which it is not constant, its gradient within $L$ has some length $\varsigma>0$, and these beliefs lie in a slab of $L$ of width $2x/\varsigma$, whose intersection with the bounded set $Q$ has measure at most $2xC_Q/\varsigma$, where $C_Q$ bounds the $(\dim L-1)$-dimensional measure of the intersection of $Q$ with a hyperplane of $L$, with $C_Q=1$ when $\dim L=1$, so they have probability at most $2\bar gC_Qx/\varsigma$. The event $\gamma(q)\leq x$ lies in the union of these events over the finitely many pairs of experiments and pieces, which gives $\bar k$. In the example of Section~\ref{sec:example}, let the objective have scales $(1,\lambda_2)$ with $1\leq\lambda_2\leq8$, so that the AI prefers the confirming test at beliefs above $p_S=2/(2+\lambda_2)$, whereas the firm prefers it above $p^*=2/3$, and let the belief be uniform on $[1/5,4/5]$. The firm's expected loss is then
\[
\mathcal D=\frac53\int_{p_S}^{p^*}\frac{2-3p}{5}\,dp=\frac{(p^*-p_S)^2}{2}=\frac29\Bigl(\frac{\lambda_2-1}{\lambda_2+2}\Bigr)^2,
\]
which is of second order in $\lambda_2-1$. With $C_Q=1$, the constants are $\bar k=50/9$ and $\bar\chi=4(\lambda_2-1)/25$, which give the bound $(32/225)(\lambda_2-1)^2$.

\paragraph{The guarantee from bisection.} In the example of Section~\ref{sec:example}, let $\theta$ be the ratio of the firm's stakes on its upper margin to those on its lower one, so that the firm's shortfalls from leaving and from deleting, relative to quarantining, are proportional to $p-t_1$ and to $\theta(t_2-p)$ at beliefs between its switches. The firm is then indifferent between deleting and leaving at $p^\circ=(t_1+t_2\theta)/(1+\theta)$, and a belief of indifference $p$ corresponds to the ratio $(p-t_1)/(t_2-p)$. After $k$ questions, each asking whether the firm would rather delete or leave the file at the midpoint of the interval known to contain $p^\circ$, that interval $[p_L,p_H]$ has width $w=2^{-k}(t_2-t_1)$ and allows the ratios between $\theta_L=(p_L-t_1)/(t_2-p_L)$ and $\theta_H=(p_H-t_1)/(t_2-p_H)$. Let $\eta=\min\{p^\circ-t_1,t_2-p^\circ\}=(t_2-t_1)\min\{\theta,1\}/(1+\theta)$, and suppose that $w<\eta$, since otherwise the share claimed below is not positive. Then $p_L-t_1\geq p^\circ-t_1-w\geq\eta-w$, so $(p_H-t_1)/(p_L-t_1)=1+w/(p_L-t_1)\leq\eta/(\eta-w)$, and likewise $(t_2-p_L)/(t_2-p_H)\leq\eta/(\eta-w)$. Hence $\theta_H/\theta_L\leq\bigl(\eta/(\eta-w)\bigr)^2$. Consider the objective that has the firm's switches and the ratio of stakes $\sqrt{\theta_L\theta_H}$. Its scales relative to the firm's are in the ratio $\sqrt{\theta_L\theta_H}/\theta$, and since $\theta$ lies between $\theta_L$ and $\theta_H$, the smaller of its two scales is at least $\sqrt{\theta_L/\theta_H}$ times the larger. By Claim~\ref{claim:ratio}, on every menu of costless experiments that keep beliefs on the range, an AI with this objective secures at least the share
\[
\sqrt{\theta_L/\theta_H}\ \geq\ 1-\frac w\eta=1-2^{-k}\bigl(1+\max\{\theta,1/\theta\}\bigr)
\]
of the value of the firm's own choice. With $\theta=1$ the share is $1-2^{1-k}$, and in general the number of questions needed for a given share grows with the logarithm of $\max\{\theta,1/\theta\}$.

\paragraph{Records that bound the relative scale.} A record of rankings that includes no indifference bounds the relative scale without fixing it (Section~\ref{sec:records}), and the guaranteed share that it leaves can be computed by linear programming. Let a record consist of the principal's rankings of pairs $(a_i,b_i)$ of actions in $A_Q$ at beliefs $q_i\in Q$, for $i=1,\dots,N$, let $\delta^i=\delta(a_i,b_i;q_i)\in\R^K$ be the vectors of gains of Lemma~\ref{lem:gains}, and let $\sigma_i\in\{1,0,-1\}$ be the sign of $\one\cdot\delta^i=V(a_i,q_i)-V(b_i,q_i)$, which is what the record reports. Let $\Lambda$ be the set of vectors $\lambda\in\R^K$ with positive components such that $\lambda\cdot\delta^i$ has the sign $\sigma_i$ for every $i$. By Lemma~\ref{lem:gains}(b) and Theorem~\ref{thm:main}(a), the objectives that are aligned on the range and rank every recorded pair as the principal does are those whose scales belong to $\Lambda$, and every vector in $\Lambda$ arises. Fix $q\in Q$ and a menu $M$ at $q$ on which some experiment has positive value, and for $E\in M$ let $G_\Lambda(E)$ be the largest number $g$ such that $I_W(E,q)\geq g\max_{E'\in M}I_W(E',q)$ for every such objective $W$.

\begin{suppclaim}\label{claim:finiterecords}
Let Assumption~\ref{ass:acting} hold. For $E\in M$,
\[
G_\Lambda(E)=\min\bigl\{v(E,E'):E'\in M,\ y(E')\neq0\bigr\},
\]
where $v(E,E')$ is the value of the linear program that minimizes $\lambda\cdot y(E)$ over $\lambda\in\R^K$ subject to $\lambda\geq0$, $\lambda\cdot y(E')=1$, $\sigma_i\,\lambda\cdot\delta^i\geq0$ for every $i$ with $\sigma_i\neq0$, and $\lambda\cdot\delta^i=0$ for every $i$ with $\sigma_i=0$. Moreover, $G(E)\leq G_\Lambda(E)\leq1$, with $G_\Lambda(E)=G(E)$ when the record is empty, and $G_\Lambda(E)=1$ if and only if $E$ is optimal on $M$ at $q$ for every objective that is aligned on the range and ranks every recorded pair as the principal does. The same numbers result when $y$ and the vectors $\delta^i$ are computed from an objective that is aligned on the range in place of $V$.
\end{suppclaim}

\begin{proof}
The vectors $y(E)$ are nonnegative, and $\lambda\cdot y(E')>0$ when $\lambda$ has positive components and $y(E')\neq0$. Hence
\[
G_\Lambda(E)=\inf_{\lambda\in\Lambda}\ \frac{\lambda\cdot y(E)}{\max_{E'\in M}\lambda\cdot y(E')}=\min_{E':\,y(E')\neq0}\ \inf_{\lambda\in\Lambda}\ \frac{\lambda\cdot y(E)}{\lambda\cdot y(E')}.
\]
Fix $E'$ with $y(E')\neq0$. The ratio is unchanged when $\lambda$ is multiplied by a positive number, so the inner infimum is the infimum of $\lambda\cdot y(E)$ over the $\lambda\in\Lambda$ with $\lambda\cdot y(E')=1$. Let $\bar\Lambda$ be the set of $\lambda\geq0$ that satisfy the constraints on the vectors $\delta^i$ in the linear program, so that $\Lambda\subseteq\bar\Lambda$. The vector $\one$ belongs to $\Lambda$, because $\one\cdot\delta^i$ has the sign $\sigma_i$. If $\lambda\in\bar\Lambda$ has $\lambda\cdot y(E')=1$ and $t>0$, the vector $\lambda^t=(\lambda+t\one)/(1+t\,\one\cdot y(E'))$ has positive components and $\lambda^t\cdot y(E')=1$, and $\lambda^t\cdot\delta^i$ is a positive multiple of $\lambda\cdot\delta^i+t\,\one\cdot\delta^i$, which has the sign $\sigma_i$, so $\lambda^t\in\Lambda$. Since $\lambda^t\cdot y(E)$ tends to $\lambda\cdot y(E)$ as $t\to0$, the infimum over $\Lambda$ equals the infimum over $\bar\Lambda$, which is the value $v(E,E')$ of the linear program. The program is feasible, because $\one/(\one\cdot y(E'))$ satisfies its constraints, and its objective is bounded below by zero, so the value is attained.

Every vector of positive scales belongs to $\Lambda$ when the record is empty, and then $G_\Lambda(E)=G(E)$ by Theorem~\ref{thm:test}(b). A record makes $\Lambda$ smaller, so $G_\Lambda(E)\geq G(E)$, and $G_\Lambda(E)\leq1$ because the ratio in the first display is at most one. The ratio equals one at $\lambda$ exactly when $E$ maximizes $\lambda\cdot y(\cdot)$ on $M$, so $G_\Lambda(E)=1$ if and only if $E$ does so for every $\lambda\in\Lambda$. For the last statement, let $S$ be aligned on the range with scales $\lambda^S$. Then $\phi^S_X(E,q)=\lambda^S_X\,y_X(E)$ by Theorem~\ref{thm:test}(c), and the gain of $S$ from $a_i$ over $b_i$ within block $X$ at $q_i$ is $\lambda^S_X\,\delta^i_X$ by \eqref{eq:blocks} and Fact~3. The change of variables $\mu_X=\lambda_X/\lambda^S_X$ carries the feasible set of each program onto that of the program written with these numbers and preserves its objective, so the values are the same.
\end{proof}

\noindent The signs $\sigma_i$ are the principal's answers, and everything else in the programs comes from the objective that the designer has.

\paragraph{The share that a deployment secures.} The guaranteed share of Section~\ref{sec:test} concerns one belief. Over a deployment, let the AI's belief be drawn from a distribution $\mu$ on $Q$, and let the AI run the experiment $E(q)$ from a menu $M(q)$ at each belief $q$, where each $M(q)$ is a menu at $q$ in the sense of Section~\ref{sec:test}. We take the menus and the AI's choices to vary with the belief so that the expectations below exist. For each block $X$, let $\bar y_X=\E_\mu[\phi_X(E(q),q)]$ and $\bar m_X=\E_\mu[\max_{E\in M(q)}\phi_X(E,q)]$, and let
\[
\bar G=\min\bigl\{\bar y_X/\bar m_X:\ \bar m_X>0\bigr\},
\]
with $\bar G=1$ if every $\bar m_X$ is zero.

\begin{suppclaim}\label{claim:deployment}
Let Assumption~\ref{ass:acting} hold. For every objective $W$ that is aligned on the range,
\[
\E_\mu\bigl[I_W(E(q),q)\bigr]\geq\bar G\,\E_\mu\Bigl[\max_{E'\in M(q)}I_W(E',q)\Bigr],
\]
and if some $\bar m_X$ is positive, no number larger than $\bar G$ has this property. The number $\bar G$ is the same whether it is computed from $V$ or from an objective that is aligned on the range.
\end{suppclaim}

\begin{proof}
Let $W$ be aligned on the range with scales $w_X$, which do not depend on the belief. By Theorem~\ref{thm:main}(b), $\E_\mu[I_W(E(q),q)]=\sum_Xw_X\bar y_X$, and by the definition of $\bar G$, $\bar y_X\geq\bar G\,\bar m_X$ for every block $X$, trivially when $\bar m_X=0$. At every belief, $\max_{E'\in M(q)}I_W(E',q)\leq\sum_Xw_X\max_{E\in M(q)}\phi_X(E,q)$, and taking expectations gives $\E_\mu[\max_{E'\in M(q)}I_W(E',q)]\leq\sum_Xw_X\bar m_X$, which proves the inequality. For the second statement, let $X^*$ be a block with $\bar m_{X^*}>0$ and $\bar y_{X^*}=\bar G\,\bar m_{X^*}$, and let $W^{X^*,\varepsilon}$ be the objectives of the proof of Theorem~\ref{thm:test}. As $\varepsilon\to0$, the left side of the inequality tends to $\bar y_{X^*}=\bar G\,\bar m_{X^*}$, whereas the expectation on the right is at least $\bar m_{X^*}$ for every $\varepsilon$. So for any $g>\bar G$, the inequality with $g$ in place of $\bar G$ fails once $\varepsilon$ is small enough. Finally, by Theorem~\ref{thm:test}(c), an objective that is aligned on the range with scales $\lambda_X$ multiplies every $\phi_X$, and so $\bar y_X$ and $\bar m_X$, by $\lambda_X$, which leaves $\bar G$ unchanged.
\end{proof}

\noindent The average over the beliefs of the guaranteed shares carries no such guarantee. If the AI's choice is best in every block at the frequent beliefs at which little is at stake and has no value in one block at the rare beliefs at which much is, the average is close to one while $\bar G$ is close to zero.

\section{Evidence gathered in sequence}\label{supp:sequential}

We state precisely and prove the claim in Section~\ref{sec:test} that the decomposition of Theorem~\ref{thm:main} extends to evidence gathered in sequence. A \emph{plan} $\sigma$ prescribes, after each history of experiments and signals, either stopping or a further experiment, and $q_\sigma$ is the belief at which it stops.

\begin{suppclaim}\label{claim:sequential}
Let Assumption~\ref{ass:acting} hold and let $S$ be aligned on the range with scales $\lambda_X$. Let a plan $\sigma$ stop in finite time with probability one, at beliefs in the closure of $Q$ only. Then the AI's value of $\sigma$ at $q\in Q$, $\E_\sigma[S^*(q_\sigma)]-S^*(q)$, is $\sum_X\lambda_X\phi_X(\sigma,q)$, where $\phi_X(\sigma,q)=\E_\sigma[h_X(q_\sigma)]-h_X(q)\geq0$, and the principal's value is $\sum_X\phi_X(\sigma,q)$.
\end{suppclaim}

\begin{proof}
Fix $q\in Q$ and a plan $\sigma$, and let $q_0=q,q_1,q_2,\dots$ be the AI's beliefs after successive experiments, held fixed once the plan stops, so that $q_\sigma$ is the limit of $q_t$. We first show that $\E_\sigma[q_\sigma]=q$. This is standard, since beliefs form a martingale with values in the bounded set $\Delta(\Omega)$ and the plan stops in finite time with probability one.

By hypothesis, $q_\sigma$ lies in the closure of $Q$. Lemma~\ref{lem:decomp}(c) gives $S^*(q')=S(r,q')+\sum_X\lambda_Xh_X(q')$ at every $q'$ in the closure of $Q$, and in particular at $q_\sigma$ and at $q$. Each $h_X$ is continuous, hence bounded on the compact closure of $Q$, so the expectations below exist. Because $S(r,\cdot)$ is linear, $\E_\sigma[S(r,q_\sigma)]=S(r,\E_\sigma[q_\sigma])=S(r,q)$. Subtracting $S^*(q)$,
\[
\E_\sigma[S^*(q_\sigma)]-S^*(q)=\sum_X\lambda_X\bigl(\E_\sigma[h_X(q_\sigma)]-h_X(q)\bigr)=\sum_X\lambda_X\phi_X(\sigma,q).
\]
Each $h_X$ is convex, being the maximum of finitely many linear functions minus a linear function, so Jensen's inequality gives $\E_\sigma[h_X(q_\sigma)]\geq h_X(\E_\sigma[q_\sigma])=h_X(q)$, that is, $\phi_X(\sigma,q)\geq0$. The same argument with $V^*(q')=V(r,q')+\sum_Xh_X(q')$, which Lemma~\ref{lem:decomp}(c) also gives, yields the principal's value.
\end{proof}

If each experiment costs $c$, which the objective counts as the principal does, and $N_\sigma$ is the number of experiments that $\sigma$ runs, with $\E_\sigma[N_\sigma]<\infty$, the AI's net value of $\sigma$ is $\sum_X\lambda_X\phi_X(\sigma,q)-c\,\E_\sigma[N_\sigma]$ and the principal's is $\sum_X\phi_X(\sigma,q)-c\,\E_\sigma[N_\sigma]$. Lemma~\ref{lem:overvalue} and Claim~\ref{claim:avgloss} therefore hold for menus of plans, with $c\,\E_\sigma[N_\sigma]$ in place of $c(E)$, and Theorem~\ref{thm:test}(a) and~(b) and Claims~\ref{claim:ratio} and~\ref{claim:lottery} hold for menus of costless plans, because their proofs use only that each option on the menu has block values and a cost. The dilution $\sigma^\alpha$, which runs $\sigma$ with probability $\alpha$ and otherwise stops at $q$, has $\phi_X(\sigma^\alpha,q)=\alpha\,\phi_X(\sigma,q)$, so the proofs of Claim~\ref{claim:tasks} and of Claim~\ref{claim:psi}(a) and~(b) apply with costless plans in place of experiments.

\section{Evidence on one question}\label{supp:fileexample}

We prove the claims in Section~\ref{sec:onequestion} about a range that is a segment, and derive the beliefs at which the firm's decisions settle the choice between the clearing and the confirming test, as stated in Sections~\ref{sec:example} and~\ref{sec:onequestion}.

\paragraph{The segment.} Let Assumption~\ref{ass:acting} hold on a range $Q$ whose affine hull $L$ has dimension one, let $q^0$ and $q^1$ be the ends of the segment $\cl Q$, and let $p\in[0,1]$ be the position of the belief $q=(1-p)\,q^0+p\,q^1$. Expected payoffs are linear in the belief, so $V(a,q)$ is an affine function of $p$ for each action $a$, and the posteriors of an experiment that keeps beliefs on the range at $q$ have positions $p_s$ that average to $p$. For each $a\in A_Q$, the positions at which $a$ is optimal form a closed interval, because $\mathcal R_a$ is closed and convex, and by (R1) the interval has positive length. Two distinct actions of $A_Q$ are not both optimal on an interval of positive length, because their expected payoffs would then be equal at every belief in $L$, contrary to (R2), and by Assumption~\ref{ass:acting}(iii) some action of $A_Q$ is optimal at every position. The intervals of the $n$ actions in $A_Q$ therefore cover $[0,1]$ and have disjoint interiors, and we list the actions as $a_1,\dots,a_n$ in the order of their intervals from $q^0$ to $q^1$. The actions $a_i$ and $a_{i+1}$ are the only actions of $A_Q$ that are optimal where their intervals meet, since the interval of any other action would overlap the interior of one of theirs, so they form the $i$th margin, at a position $t_i\in(0,1)$, and these are all the margins. By Lemma~\ref{lem:triple}(c), no three actions are optimal at any belief in $Q$, so every margin is a block. Every action other than $a_1$ and $a_n$ is optimal on an interval with beliefs of $Q$ on both sides, so the beliefs in $Q$ at which it is not optimal form a disconnected set, and it separates the range by Lemma~\ref{lem:blocks}(c). The difference $V(a_{i+1},q)-V(a_i,q)$ vanishes at $t_i$ and is negative on the interior of the interval of $a_i$, so it equals $d_i\,(p-t_i)$ with $d_i>0$, the increase in the slope of $V^*$ at $t_i$, and for $i<j$
\[
V(a_j,q)-V(a_i,q)=\sum_{m=i}^{j-1}d_m\,(p-t_m),
\]
an affine function of $p$ with a positive slope. The principal's expected payoffs thus have strictly increasing differences in the action, in the order of the list, and the position of the belief. This is the cardinal form of the single crossing property of \citet{milgromshannon1994}, and the principal's choice moves up the list as the position rises, as in the monotone decision problems of \citet{karlinrubin1956}. No assumption on the principal's payoffs in each state is needed, because the beliefs on a segment are ordered by their position. The margin graph is the path that joins each $a_i$ to $a_{i+1}$, so by Claim~\ref{claim:peaked}(b) the expected payoffs are also concave in a suitably spaced action. Finally, the value of acting in the block of the $i$th margin is $\max\{V(a_i,q),V(a_{i+1},q)\}=V(a_i,q)+d_i\,(p-t_i)^+$, and the affine term drops out of the value of an experiment, so
\[
\phi_i(E,p)=d_i\,x_E(t_i),\qquad\text{where}\qquad x_E(t)=\sum_sP(s)\,(p_s-t)^+-(p-t)^+,
\]
as in \eqref{eq:kinks}.

\paragraph{The file example.} The firm switches at $t_1=1/10$ and $t_2=9/10$, with $d_1=d_2=10$, so the value of a test in the block of the $i$th margin is $\phi_i=10\,x_E(t_i)$.

At a belief $p$, the clearing test reports ``clear'' with probability $2(1-p)/5$, which leads to the posterior $0$, and otherwise leads to the posterior $5p/(3+2p)$, which is at least $p$. This posterior exceeds $t_1$ when $p>1/16$ and exceeds $t_2$ when $p>27/32$. Hence
\[
\phi_1=\begin{cases}0&\text{if }p\leq\frac1{16},\\[2pt]\frac{48p-3}{5}&\text{if }\frac1{16}\leq p\leq\frac1{10},\\[2pt]\frac{2(1-p)}{5}&\text{if }p\geq\frac1{10},\end{cases}
\qquad
\phi_2=\begin{cases}0&\text{if }p\leq\frac{27}{32},\\[2pt]\frac{32p-27}{5}&\text{if }\frac{27}{32}\leq p\leq\frac9{10},\\[2pt]\frac{18(1-p)}{5}&\text{if }p\geq\frac9{10}.\end{cases}
\]
The confirming test reports ``confirm'' with probability $p/5$, which leads to the posterior $1$, and otherwise leads to the posterior $4p/(5-p)$, which is at most $p$. This posterior is below $t_1$ when $p<5/41$ and below $t_2$ when $p<45/49$. Hence
\[
\phi_1=\begin{cases}\frac{9p}{5}&\text{if }p\leq\frac1{10},\\[2pt]\frac{5-41p}{5}&\text{if }\frac1{10}\leq p\leq\frac5{41},\\[2pt]0&\text{if }p\geq\frac5{41},\end{cases}
\qquad
\phi_2=\begin{cases}\frac p5&\text{if }p\leq\frac9{10},\\[2pt]\frac{45-49p}{5}&\text{if }\frac9{10}\leq p\leq\frac{45}{49},\\[2pt]0&\text{if }p\geq\frac{45}{49}.\end{cases}
\]
In the block of the lower margin, the confirming test is at least as good as the clearing test exactly when $p\leq1/13$: below $1/16$ the clearing test is worth nothing, between $1/16$ and $1/10$ the comparison is $9p\geq48p-3$, and above $1/10$ the clearing test is strictly better, because $2(1-p)>5-41p$ when $p>1/13$. In the block of the upper margin, the clearing test is at least as good as the confirming test exactly when $p\geq27/31$: below $27/32$ the clearing test is worth nothing, between $27/32$ and $9/10$ the comparison is $32p-27\geq p$, and above $9/10$ it is $18(1-p)\geq45-49p$, which holds because $p\geq27/31$ there. The confirming test is therefore best in both blocks when $p\leq1/13$, the clearing test is best in both when $p\geq27/31$, and at beliefs strictly between $1/13$ and $27/31$ the clearing test is strictly better in the block of the lower margin and the confirming test strictly better in the block of the upper one. By Theorem~\ref{thm:test}(a), the firm's decisions settle the choice exactly when $p\leq1/13$ or $p\geq27/31$.

\paragraph{A continuum of ordered actions.} The text takes the actions to be finite in number. Suppose instead that the action is a number $a$ in an interval $A=[\underline a,\bar a]$ of positive length, and that for each state $\omega$ the payoff $V(a,\omega)$ is continuously differentiable and strictly concave in $a$, with derivative $V_a(a,\omega)$. Write $V_a(a,q)=\sum_\omega q(\omega)V_a(a,\omega)$. For each belief $q$, the expected payoff $V(\cdot,q)$ is strictly concave, so it has a single maximizer $a^*(q)$, and $V_a(z,q)$ is positive for $z<a^*(q)$ and negative for $z>a^*(q)$. The maximizer is continuous in $q$ by the maximum theorem.

\begin{suppclaim}\label{claim:continuum}
Under these conditions:
\begin{enumerate}[label=(\alph*)]
\item If $a'<a<a''$, every connected set of beliefs that contains a belief at which $a'$ is optimal for the principal and a belief at which $a''$ is optimal contains a belief at which $a$ is optimal.
\item For every continuous function $\mu:A\to(0,\infty)$, the objective $S(a,\omega)=\int_{\underline a}^a\mu(z)\,V_a(z,\omega)\,dz$ has $\BR_S(q)=\BR_V(q)$ at every belief $q$, and every experiment $E$ has, at every belief $q$,
\[
I_S(E,q)=\int_A\mu(z)\,\phi_z(E,q)\,dz\qquad\text{and}\qquad I_V(E,q)=\int_A\phi_z(E,q)\,dz,
\]
where $\phi_z(E,q)=\sum_sP(s)\,V_a(z,q_s)^+-V_a(z,q)^+\geq0$.
\item If an experiment $E^*$ in a finite set $M$ of experiments is optimal on $M$ at $q$ for every objective of part (b), then $\phi_z(E^*,q)\geq\phi_z(E,q)$ for every $E\in M$ and every $z\in A$.
\end{enumerate}
\end{suppclaim}

\begin{proof}
\emph{Part (a).} The function $a^*$ is continuous, so the optimal actions at the beliefs of a connected set form a connected subset of $A$, that is, an interval. It contains $a'$ and $a''$, and hence $a$.

\emph{Part (b).} Fix a belief $q$. Expected payoffs are linear in the belief, so $S(a,q)=\int_{\underline a}^a\mu(z)\,V_a(z,q)\,dz$, and $S(\cdot,q)$ has the derivative $\mu(a)V_a(a,q)$, which is positive for $a<a^*(q)$ and negative for $a>a^*(q)$ because $\mu$ is positive. So $S(\cdot,q)$ is strictly increasing below $a^*(q)$ and strictly decreasing above it, and $a^*(q)$ is its only maximizer. For the same reason,
\[
S^*(q)=\int_{\underline a}^{a^*(q)}\mu(z)\,V_a(z,q)\,dz=\int_A\mu(z)\,V_a(z,q)^+\,dz ,
\]
and substituting this into \eqref{eq:value} gives the expression for $I_S$. Each $\phi_z(E,q)$ is nonnegative because $x\mapsto x^+$ is convex, $V_a(z,\cdot)$ is linear in the belief and the posteriors average to $q$. With $\mu=1$ the objective is $V(a,\omega)-V(\underline a,\omega)$, which values every experiment as $V$ does, and this gives the expression for $I_V$.

\emph{Part (c).} Suppose, toward a contradiction, that $\phi_{z_0}(E^*,q)<\phi_{z_0}(E,q)$ for some $E\in M$ and $z_0\in A$. The function $D(z)=\phi_z(E,q)-\phi_z(E^*,q)$ is continuous, because $V_a(z,\omega)$ is continuous in $z$. So there are $\delta>0$ and $r>0$ such that $D(z)\geq\delta$ at every $z\in A$ within $r$ of $z_0$, and $|D|$ is bounded on $A$ by some number $B$. Let $h(z)=\max\{0,1-|z-z_0|/r\}$, a continuous function with values in $[0,1]$ that vanishes at distances of $r$ or more from $z_0$, and for $\varepsilon>0$ let $\mu_\varepsilon=\varepsilon+h$, which is continuous and positive. By part (b), the objective with the weight $\mu_\varepsilon$ values $E$ above $E^*$ by
\[
\int_A\mu_\varepsilon(z)\,D(z)\,dz\geq\delta\int_Ah(z)\,dz-\varepsilon B\,(\bar a-\underline a),
\]
which is positive for small $\varepsilon$, since $\int_Ah>0$. This contradicts the optimality of $E^*$.
\end{proof}

\noindent Part (a) says that every action separates those below it from those above. Part (b) says that weighting the principal's marginal payoff $V_a(a,\cdot)$ at each action by any positive amount $\mu(a)$ changes no decision, which is the single-agent case of Proposition~8 of \citet{morrisui2004}. The weight plays the role of the scale of a block, and decisions leave it free at every action. With two states, $V_a(z,p)=d(z)\,[p-t(z)]$ whenever $d(z)=V_a(z,\omega_1)-V_a(z,\omega_2)$ is not zero, where $t(z)$ is the value of $p$ at which the marginal payoff at $z$ is zero, the counterpart of a switching belief. As in the calculation of block values above, $\phi_z(E,p)=|d(z)|\,x_E(t(z))$, which is zero unless $t(z)\in(0,1)$.\footnote{For $d(z)<0$, use $V_a(z,p)^+=|d(z)|\,[(p-t(z))^+-(p-t(z))]$, whose last term is linear in $p$ and drops out.} Suppose that the principal's optimal action is interior at every belief in $(0,1)$ and is not the same at all of them. Then every belief $t\in(0,1)$ is $t(z)$ for the action $z=a^*(t)$, because $V_a(z,t)=0$ and $d(z)\neq0$, since $d(z)=0$ would make $V_a(z,\cdot)$ vanish at every belief and $z$ optimal at all of them. By part (c), an experiment that is optimal on a menu at a belief $p\in(0,1)$ for every objective of part (b) then has the largest $x_E(t)$ on the menu at every $t\in[0,1]$, and so is at least as informative as every other experiment on the menu in the sense of \citet{blackwell1953} (Section~\ref{sec:onequestion}). Conversely, such an experiment is optimal in every decision problem. The comparison at $K$ switching beliefs in Section~\ref{sec:onequestion} thus becomes a comparison at every belief. For a reported probability rewarded by a proper scoring rule, part (b) is the representation of \citet{schervish1989}.

\section{Right answers}\label{supp:rules}

We state precisely and prove the claim in Section~\ref{sec:middle} that an AI rewarded only for right answers never takes a cautious action. A \emph{right-answer rule} is an objective of the form $S(a,\omega)=\ell(\omega)+w(\omega)\,\one\{a\in R(\omega)\}$, where the weight $w(\omega)$ is positive, each $R(\omega)$ is nonempty and $R(\omega)\subseteq\BR_V(\delta_\omega)$ for every $\omega$. A state in which the principal is indifferent among all actions is covered by $R(\omega)=A$.

\begin{suppclaim}\label{claim:rules}
Let $S$ be a right-answer rule. At every point mass, $\BR_S(\delta_\omega)=R(\omega)\subseteq\BR_V(\delta_\omega)$, so the AI chooses an action that is optimal for the principal at $\delta_\omega$, a right answer, whenever its evidence reveals the state. No cautious action is optimal for the AI at any belief, so the AI is misaligned at every belief at which a cautious action is the principal's unique optimal action.
\end{suppclaim}

\begin{proof}
At a point mass $\delta_\omega$, $S(a,\delta_\omega)=S(a,\omega)$, which equals $\ell(\omega)+w(\omega)$ if $a\in R(\omega)$ and $\ell(\omega)<\ell(\omega)+w(\omega)$ otherwise. Since $R(\omega)$ is nonempty, $\BR_S(\delta_\omega)=R(\omega)$, which is contained in $\BR_V(\delta_\omega)$ because $S$ is a right-answer rule. Now let $c$ be a cautious action. Because $c$ is optimal for the principal at no point mass, $c\notin\BR_V(\delta_\omega)\supseteq R(\omega)$ for every $\omega$, so $S(c,q)=\sum_\omega q(\omega)\ell(\omega)$ at every belief $q$. Fix $q$, and choose $\omega'$ with $q(\omega')>0$ and an action $b\in R(\omega')$. Then
\begin{align*}
S(b,q)&=\sum_\omega q(\omega)\ell(\omega)+\sum_\omega q(\omega)w(\omega)\,\one\{b\in R(\omega)\}\\
&\geq\sum_\omega q(\omega)\ell(\omega)+q(\omega')w(\omega')>S(c,q),
\end{align*}
where the first inequality drops nonnegative terms and the second uses $q(\omega')w(\omega')>0$. So $c\notin\BR_S(q)$. If $\BR_V(q)=\{c\}$, the nonempty set $\BR_S(q)$ excludes $c$ and is not contained in $\BR_V(q)$, so the AI is misaligned at $q$.
\end{proof}

\section{When caution separates, and how common several blocks are}\label{supp:prevalence}

This section states the condition of Section~\ref{sec:middle} under which a middle course separates two alternatives when they are the only actions, which follows from Claim~\ref{claim:between}, applies it to the three-state example, proves the claim that a cautious action that costs little enough separates the range, and reports how often randomly drawn problems have several blocks.

\begin{suppclaim}\label{claim:middle}
Let Assumption~\ref{ass:acting} hold with $A_Q=\{a,b,c\}$.
\begin{enumerate}[label=(\alph*)]
\item The action $c$ separates $a$ and $b$ if and only if $V(c,q)>V(a,q)$ at every belief $q\in Q$ at which $V(a,q)=V(b,q)$.
\item The action $c$ separates $a$ and $b$ if and only if there is $\alpha\in(0,1)$ such that $V(c,q)\geq\alpha V(a,q)+(1-\alpha)V(b,q)$ at every belief $q\in Q$. When $Q$ is the set of beliefs with full support, the condition is that $V(c,\omega)\geq\alpha V(a,\omega)+(1-\alpha)V(b,\omega)$ in every state $\omega$.
\end{enumerate}
\end{suppclaim}

\begin{proof}
\emph{Part (a).} Let $c$ separate $a$ and $b$, and let $V(a,q)=V(b,q)$ at $q\in Q$. By Claim~\ref{claim:between}(c), $c$ beats $a$ or $b$ at $q$, and so both. Conversely, suppose that $V(c,q)>V(a,q)$ at every belief in $Q$ at which $a$ and $b$ tie. Then $a$ and $b$ are not both optimal at any belief in $Q$, so $\{a,b\}$ is not a margin. The graph $\mathcal G_Q$ is connected by Lemma~\ref{lem:structure}(a) and its vertices are $a$, $b$ and $c$, so its margins are $\{a,c\}$ and $\{c,b\}$. Every path from $a$ to $b$ in $\mathcal G_Q$ therefore visits $c$, and $c$ separates $a$ and $b$ by Lemma~\ref{lem:blocks}(b).

\emph{Part (b).} The only disjoint sets $A_1$ and $A_2$ with union $\{a,b\}$ that contain $a$ and $b$ respectively are $\{a\}$ and $\{b\}$, so by Claim~\ref{claim:between}(b), $c$ separates $a$ and $b$ if and only if some $\alpha\in(0,1)$ has $V(c,q)\geq\alpha V(a,q)+(1-\alpha)V(b,q)$ at every belief in $\cl Q$. By continuity, it holds at every belief in $\cl Q$ if and only if it holds at every belief in $Q$. When $Q$ is the set of beliefs with full support, $\cl Q=\Delta(\Omega)$, and a linear function is nonnegative on $\Delta(\Omega)$ if and only if it is nonnegative at every point mass, which gives the condition in every state.
\end{proof}

\noindent Part (a) says that caution separates when it costs less than either mistake wherever the two mistakes cost the same. In the three-state example of Section~\ref{sec:example}, the firm's payoffs in the states in which the file is malicious, critical and routine are $(-10,0,0)$ for leaving, $(0,-30,-5)$ for deleting and $(-\kappa,-\kappa,-\kappa)$ for quarantining. A lottery that deletes with probability $\beta$ and otherwise leaves costs $10(1-\beta)$, $30\beta$ and $5\beta$ in the three states, so quarantining is at least as good as it in every state if and only if $\kappa\leq\min\{10(1-\beta),5\beta\}$, which is largest, at $10/3$, when $\beta=2/3$. By part (b), quarantining separates deleting and leaving on the beliefs with full support when $\kappa\leq10/3$, and it does not when $\kappa>10/3$ and all three actions are in $A_Q$.

The next claim does not restrict the number of actions. We call an action \emph{right} in a state if it is optimal for the principal when that state is known for certain.

\begin{suppclaim}\label{claim:cheapcaution}
Let $Q$ be the set of beliefs with full support, let $c$ be a cautious action, and suppose that each state $\omega$ has a single right action $a(\omega)$. Let $s=\max_\omega[V(a(\omega),\omega)-V(c,\omega)]$ be the most by which $c$ falls short of the right action. There is $\bar s>0$, which depends only on the payoffs of the actions other than $c$, such that $c$ separates the range whenever $s<\bar s$.
\end{suppclaim}

\begin{proof}
Because $c$ is cautious, it is right in no state, so $a(\omega)\neq c$ for every $\omega$. For $q\in\Delta(\Omega)$, let
\[
J(q)=\sum_\omega q(\omega)V(a(\omega),\omega)-\max_{b\neq c}V(b,q),
\]
the value at $q$ of an experiment that reveals the state when $c$ is unavailable. It is continuous in $q$, and it is nonnegative because $V(b,q)=\sum_\omega q(\omega)V(b,\omega)\leq\sum_\omega q(\omega)V(a(\omega),\omega)$ for every action $b$. It is zero exactly when some action $b\neq c$ has $V(b,\omega)=V(a(\omega),\omega)$ for every $\omega$ with $q(\omega)>0$, that is, since $a(\omega)$ is the only right action in $\omega$, exactly when $q$ belongs to one of the sets $F_b=\{q\in\Delta(\Omega):a(\omega)=b\text{ whenever }q(\omega)>0\}$. Let $B$ be the set of actions that are right in some state. For $b\in B$ the set $F_b$ is a nonempty face of the simplex, and these faces are closed and pairwise disjoint. The set $B$ has at least two actions. Otherwise a single action $b\neq c$ would be the only right action in every state, so that $V(b,\omega)>V(c,\omega)$ for every $\omega$, and $c$ would be optimal at no belief, whereas a cautious action is optimal at some belief in $Q$.

Let $3\eta>0$ be the smallest distance between two of the finitely many disjoint compact sets $F_b$, $b\in B$, and let $N_b$ be the set of beliefs within distance $\eta$ of $F_b$, so that the sets $N_b$ are pairwise disjoint and open relative to $\Delta(\Omega)$. The function $J$ is positive on the compact set $\Delta(\Omega)\setminus\bigcup_{b\in B}N_b$, so it is at least some $\bar s>0$ there, where $\bar s$ is any positive number if that set is empty. Neither $J$ nor the sets $N_b$ involve the payoffs of $c$.

Suppose that $s<\bar s$. At a belief $q\in Q$ at which $c$ is not optimal, some action $b\neq c$ has
\[
V(b,q)>V(c,q)\geq\sum_\omega q(\omega)V(a(\omega),\omega)-s,
\]
so $J(q)<s<\bar s$ and $q\in\bigcup_{b\in B}N_b$. The beliefs in $Q$ at which $c$ is not optimal therefore lie in the union of the pairwise disjoint open sets $N_b$. They meet at least two of these sets. For $b\in B$, choose $\omega$ with $a(\omega)=b$. Since $V(b,\omega)>V(c,\omega)$, continuity gives $V(b,q)>V(c,q)$ at every belief $q$ close enough to the point mass $\delta_\omega\in F_b$, and some such belief has full support and lies within $\eta$ of $\delta_\omega$, hence in $N_b\cap Q$. A set that is covered by pairwise disjoint open sets and meets two of them is disconnected. By the last step of the proof of Lemma~\ref{lem:blocks}(c), which does not use Assumption~\ref{ass:acting}, $c$ then separates two actions in $A_Q$, so it separates the range.
\end{proof}

\noindent Without a single right action in each state, cheap caution need not separate. With $V(a,\cdot)=(0,0,-1)$, $V(b,\cdot)=(0,-1,0)$ and $V(c,\cdot)=(-\kappa,-\kappa,-\kappa)$ for $0<\kappa<1/2$, the three actions tie at $(1-2\kappa,\kappa,\kappa)$ and form a single block, because $a$ and $b$ are both right in the first state.

\paragraph{Random payoffs.} Table~\ref{tab:prevalence} reports the share of randomly drawn problems that have two or more blocks. For each number of states $n$, dimension $d$ of the range and number of actions $m$, we draw the payoffs $V(a,\omega)$ independently from the standard normal distribution, and a range equal to the relative interior of the intersection of the simplex with an affine subspace of dimension $d$. The subspace passes through a belief drawn uniformly from the simplex, and its direction space is drawn uniformly from the $d$-dimensional subspaces of vectors whose coordinates sum to zero. For $d=n-1$ the range is the set of beliefs with full support. Each cell uses $3{,}000$ draws, of which we keep those with at least three actions in $A_Q$. We find $A_Q$ and the margins by linear programming and count the blocks as the biconnected components of the margin graph (Claim~\ref{claim:biconnected}). The code is in the replication files.

\begin{table}[ht]
\centering
\small
\begin{tabular}{@{}ccccc@{}}
\toprule
States & Dimension of & \multicolumn{3}{c}{Share with two or more blocks (problems kept)}\\
\cmidrule(l){3-5}
$n$ & the range $d$ & $m=6$ actions & & $m=10$ actions\\
\midrule
3 & 1 & 1.00 (479) & & 1.00 (754)\\
3 & 2 & 0.34 (2{,}295) & & 0.35 (2{,}670)\\
4 & 1 & 1.00 (314) & & 1.00 (490)\\
4 & 2 & 0.37 (1{,}679) & & 0.37 (2{,}145)\\
4 & 3 & 0.17 (2{,}794) & & 0.18 (2{,}945)\\
6 & 1 & 1.00 (138) & & 1.00 (230)\\
6 & 2 & 0.40 (1{,}009) & & 0.37 (1{,}328)\\
6 & 3 & 0.21 (2{,}094) & & 0.23 (2{,}435)\\
6 & 4 & 0.12 (2{,}823) & & 0.12 (2{,}930)\\
6 & 5 & 0.05 (2{,}990) & & 0.04 (2{,}998)\\
\bottomrule
\end{tabular}
\caption{The share of random problems with at least three actions in $A_Q$ that have two or more blocks, by the number of states, the dimension of the range and the number of actions. The number of problems kept out of $3{,}000$ is in parentheses.}
\label{tab:prevalence}
\end{table}

On a segment there are always several blocks when $A_Q$ has at least three actions, by Lemma~\ref{lem:triple}. The share falls with the dimension of the range and, for a given dimension, depends little on the number of states or of actions. When the range is the set of beliefs with full support, several blocks are common with three states and rare with six. Payoffs drawn independently have no structure, and the two kinds of structure that give several scales whatever the dimension of the range are a cautious action that costs little (Claim~\ref{claim:cheapcaution}) and additivity across decisions (Section~\ref{supp:separable}).

\section{Declining to answer}\label{supp:answers}

We state precisely and prove the claims in Section~\ref{sec:answers}. There are $n\geq3$ states $\omega_1,\dots,\omega_n$ and $n+1$ actions. The action $a_i$ has payoff $r_i>0$ in state $\omega_i$ and $-w_i<0$ in every other state, the action $c$ has payoff zero in every state, and $\tau_i=w_i/(r_i+w_i)$.

\begin{suppclaim}\label{claim:answers}
Let $Q$ be the set of beliefs with full support and let $\sum_i\tau_i>1$. Then Assumption~\ref{ass:acting} holds, every block contains $c$, and the actions $a_i$ and $a_j$ belong to a common block if and only if $i$ and $j$ are joined by a chain of answers in which consecutive answers $k$ and $l$ have $\tau_k+\tau_l<1$. In particular, if $\tau_i+\tau_j\geq1$ whenever $i\neq j$, there are $n$ blocks, each made up of one action $a_i$ and $c$.
\end{suppclaim}

\begin{proof}
Write $q_i=q(\omega_i)$ and $\sigma_i=r_i+w_i>0$, so that $V(a_i,q)=\sigma_i(q_i-\tau_i)$ and $V(c,q)=0$, with $0<\tau_i<1$. The range $Q$ is the set of beliefs with full support, $L=\aff Q$ is the set of vectors in $\R^\Omega$ whose components sum to one, and $\cl Q=\Delta(\Omega)$. The action $c$ is optimal at $q$ exactly when $q_i\leq\tau_i$ for every $i$, and $a_i$ is optimal exactly when $q_i\geq\tau_i$ and $\sigma_i(q_i-\tau_i)\geq\sigma_j(q_j-\tau_j)$ for every $j$. For a set $J$ of answers, write $\tau_J=\sum_{i\in J}\tau_i$, with $\tau_\emptyset=0$.

\emph{Step 1: beliefs with given optimal actions.} Let $J$ be a set of answers with $\tau_J<1$. We show that some belief in $Q$ has exactly the actions $a_i$, $i\in J$, and $c$ as optimal actions, and that if $J$ is nonempty, some belief in $Q$ has exactly the actions $a_i$, $i\in J$, as optimal actions. Since $\sum_i\tau_i>1$, some answer lies outside $J$. For $\eta\geq0$, let $q_i=\tau_i+\eta/\sigma_i$ for $i\in J$ and $q_k=\vartheta\,(\tau_k+\eta/\sigma_k)$ for $k\notin J$, where
\[
\vartheta=\frac{1-\sum_{i\in J}(\tau_i+\eta/\sigma_i)}{\sum_{k\notin J}(\tau_k+\eta/\sigma_k)},
\]
so that the components of $q$ sum to one. For $\eta$ small enough the numerator is positive, because $\tau_J<1$, and $\vartheta<1$, because the numerator is less than the denominator exactly when $1<\sum_i\tau_i+\eta\sum_i1/\sigma_i$. Then $q$ has full support, $\sigma_i(q_i-\tau_i)=\eta$ for $i\in J$, and $\sigma_k(q_k-\tau_k)<\eta$ for $k\notin J$, because $q_k<\tau_k+\eta/\sigma_k$. With $\eta=0$, the actions $a_i$, $i\in J$, and $c$ all have expected payoff zero and every other action has a negative one. With $\eta>0$ and $J$ nonempty, the actions $a_i$, $i\in J$, have expected payoff $\eta>0$ and every other action has less.

Conversely, if the actions $a_i$, $i\in J$, are all optimal at a belief $q\in Q$, for a nonempty set $J$, then $\tau_J<1$. Indeed $q_i\geq\tau_i$ for $i\in J$, so $\tau_J\leq\sum_{i\in J}q_i\leq1$, and the last inequality is strict unless $J$ contains every answer, in which case $\tau_J=\sum_i\tau_i>1$ would contradict it.

\emph{Step 2: Assumption~\ref{ass:acting} holds.} By Step~1 with $J=\emptyset$ and $\eta=0$, the action $c$ is the only optimal action at some belief in $Q$, and by Step~1 with $J=\{i\}$ and $\eta>0$, so is each $a_i$, since $\tau_i<1$. Expected payoffs are continuous, so each action is the only optimal one on a subset of $Q$ that is open relative to $L$. Hence $A_Q=A$, (R1) holds and the third condition holds. For (R2), fix $i$ and an answer $k\neq i$, and for $j\neq i$ choose $k\notin\{i,j\}$, which is possible because $n\geq3$. Moving a point of $L$ in the direction that raises its $i$th component and lowers its $k$th by the same amount stays in $L$ and changes $V(a_i,\cdot)-V(c,\cdot)=\sigma_i(q_i-\tau_i)$ and $V(a_i,\cdot)-V(a_j,\cdot)=\sigma_i(q_i-\tau_i)-\sigma_j(q_j-\tau_j)$ at the rate $\sigma_i\neq0$, so neither difference vanishes throughout $L$. For the second condition, let two or three actions be optimal at a belief in $Q$, and let $J$ be the set of answers $i$ for which $a_i$ is among them. If $J$ is nonempty, then $\tau_J<1$ by Step~1, and if it is empty, $\tau_J=0$. In either case Step~1 gives a belief in $Q$ at which exactly those actions are optimal, with $\eta=0$ if $c$ is among them and $\eta>0$ if it is not.

\emph{Step 3: the margins.} By Step~1 with $J=\{i\}$ and $\eta=0$, $\{a_i,c\}$ is a margin for every $i$. For $i\neq j$, Step~1 shows that $\{a_i,a_j\}$ is a margin if $\tau_i+\tau_j<1$, and that $a_i$ and $a_j$ are both optimal at no belief in $Q$ otherwise.

\emph{Step 4: the blocks.} By Lemma~\ref{lem:blocks}, two distinct actions belong to a common block if and only if no action separates them, and an action separates two others if and only if every path between them in $\mathcal G_Q$ visits it. The actions $a_i$ and $c$ form a margin, so no action separates them and they belong to a common block. Let $i\neq j$. The path $a_i,c,a_j$ visits no action $a_k$, so none of these separates $a_i$ and $a_j$. A path from $a_i$ to $a_j$ in $\mathcal G_Q$ that does not visit $c$ passes only through margins $\{a_k,a_l\}$, which by Step~3 are the pairs with $\tau_k+\tau_l<1$. So $c$ does not separate $a_i$ and $a_j$, and the two belong to a common block, if and only if $i$ and $j$ are joined by a chain of answers in which consecutive answers $k$ and $l$ have $\tau_k+\tau_l<1$. Every block contains $c$: otherwise its actions would be actions $a_i$ no two of which are separated by another action, and adding $c$, which no action separates from any $a_i$, would give a larger set with this property. If $\tau_i+\tau_j\geq1$ whenever $i\neq j$, no two of the actions $a_i$ belong to a common block, so each block is made up of $c$ and a single $a_i$, and every $a_i$ belongs to such a block, which gives $n$ blocks.
\end{proof}

\noindent For the value of verifying an answer, let $q\in Q$ have $q_k\leq\tau_k$ for every $k$, so that the principal declines, and let $q_k/(1-q_i)\leq\tau_k$ for all $k\neq i$, so that it would still decline if answer $i$ were ruled out. The experiment that reveals whether answer $i$ is correct leads to the posterior $\delta_{\omega_i}$ with probability $q_i$ and otherwise to the posterior that puts probability $q_k/(1-q_i)$ on each $\omega_k$ with $k\neq i$. In a block that contains $a_i$, the value of acting is zero at $q$ and at the second posterior, where no answer of the block reaches its threshold, and it is $\sigma_i(1-\tau_i)=r_i$ at the first, so the experiment's value in this block is $q_i\,r_i$. In a block that does not contain $a_i$, the value of acting is zero at $q$ and at both posteriors, so the experiment's value there is zero.

\section{Comparisons and noise}\label{supp:records}

This section proves the claims in Section~\ref{sec:records} about which comparisons add to the principal's decisions, and the claims in Section~\ref{sec:robust} about noisy records. It ends with the rankings that decisions fix and with the principal's decisions when an action is unavailable. We say that an objective \emph{ranks $a$ and $b$ at $q$ as the principal does} if $S(a,q)-S(b,q)$ and $V(a,q)-V(b,q)$ are both positive, both zero or both negative.

\begin{suppclaim}\label{claim:informative}
Let Assumption~\ref{ass:acting} hold, let $a,b\in A_Q$ and $q\in Q$, and let $\delta=\delta(a,b;q)\in\R^K$ be the vector of the principal's gains from $a$ over $b$ within blocks (Lemma~\ref{lem:gains}). Every objective that is aligned on the range ranks $a$ and $b$ at $q$ as the principal does if and only if the components of $\delta$ are all nonnegative or all nonpositive. This is the case, in particular, if $a$ or $b$ is optimal for the principal at $q$, and if $a$ and $b$ belong to the same block. Consequently, two distinct actions in $A_Q$ belong to a common block if and only if every objective that is aligned on the range ranks them as the principal does at every belief in $Q$.
\end{suppclaim}

\begin{proof}
By Lemma~\ref{lem:gains}(a) and~(b), an objective that is aligned on the range with scales $\lambda$ has $S(a,q)-S(b,q)=\lambda\cdot\delta$, whereas $V(a,q)-V(b,q)=\one\cdot\delta$. If $\delta\geq0$, then $\one\cdot\delta\geq0$ and $\lambda\cdot\delta\geq0$, and since every $\lambda_X$ is positive, each is zero exactly when $\delta=0$. So every aligned objective ranks $a$ and $b$ as the principal does. The case $\delta\leq0$ is the same with the signs reversed. Conversely, suppose that $\delta_X>0>\delta_Y$ for some blocks $X$ and $Y$, and let $u_X$ be the vector with a one in the component of $X$ and zeros elsewhere. For small $\varepsilon>0$, the scales $u_X+\varepsilon\one$ give $\lambda\cdot\delta=\delta_X+\varepsilon\,\one\cdot\delta>0$ and the scales $u_Y+\varepsilon\one$ give $\lambda\cdot\delta=\delta_Y+\varepsilon\,\one\cdot\delta<0$. Whichever way the principal ranks $a$ and $b$, an aligned objective with one of these vectors of scales ranks them differently. Finally, if $a$ is optimal for the principal at $q$, Lemma~\ref{lem:gains}(c) gives $\delta\geq0$. If $b$ is, it gives $\delta(b,a;q)\geq0$, and $\delta(a,b;q)=-\delta(b,a;q)$ by Lemma~\ref{lem:gains}(a). If $a$ and $b$ belong to the same block $X$, the path from $a$ to $b$ in $\mathcal H$ is $a,X,b$, so $\delta$ has at most one nonzero component. For the last statement, it remains to consider distinct actions $a$ and $b$ that belong to no common block. The path from $a$ to $b$ in $\mathcal H$ is then $a=z_0,X_1,z_1,X_2,z_2,\dots$, with at least two blocks. By Fact~5, $z_1$ is the principal's only optimal action at some belief $q\in Q$, where $\delta_{X_1}(a,b;q)=V(z_0,q)-V(z_1,q)<0$ and $\delta_{X_2}(a,b;q)=V(z_1,q)-V(z_2,q)>0$. By the first part of the claim, some objective that is aligned on the range does not rank $a$ and $b$ at $q$ as the principal does.
\end{proof}

\noindent A comparison therefore restricts the relative scale only if the principal's choice between the two options trades a gain within one block against a loss within another, which requires that neither option be the principal's own choice at that belief. As an AI's decisions improve, the comparisons of its own responses that are used to fit its reward are more often comparisons in which one response is the principal's choice, and these add nothing to decisions.

\subsection*{Noisy records}

Records of human choices are noisy, and how often people choose one action over another depends on how much is at stake between them. Let the \emph{temperature} $\tau>0$ measure the noise in logit choices.

\begin{suppclaim}\label{claim:noise}
Let Assumption~\ref{ass:acting} hold.
\begin{enumerate}[label=(\alph*)]
\item If at each belief $q\in Q$ the principal chooses each action $a\in A$ with probability proportional to $\exp(V(a,q)/\tau)$, an objective under which $\exp(S(a,q))/\sum_{b\in A}\exp(S(b,q))$ equals these probabilities at every belief in $Q$ satisfies $S(a,q)=V(a,q)/\tau+\beta\cdot q$ for every $a\in A$ and $q\in Q$, for some $\beta:\Omega\to\R$.
\item Suppose that, for actions $a$ and $b$ in the same block $X$, the principal chooses $a$ over $b$ at $q$ with probability $1/(1+\exp(-g_{ab}\cdot q/\tau_X))$, and that no comparison involves actions of different blocks. An objective that matches these probabilities with the logit form $1/(1+\exp(-e_{ab}\cdot q))$ at every belief in $Q$ satisfies $e_{ab}\cdot q=g_{ab}\cdot q/\tau_X$ for all actions $a$ and $b$ of each block $X$ and every $q\in Q$, and such objectives exist. If no action outside $A_Q$ is optimal for it on the closure of $Q$, it is block-proportional with $\lambda_X=1/\tau_X$.
\end{enumerate}
\end{suppclaim}

\begin{proof}
\emph{Part (a).} If the objective's logit probabilities equal the principal's at $q\in Q$, then for any two actions $a$ and $b$ the ratio of their probabilities gives $\exp(S(a,q)-S(b,q))=\exp([V(a,q)-V(b,q)]/\tau)$, so $S(a,q)-S(b,q)=[V(a,q)-V(b,q)]/\tau$. Fix $r\in A$ and let $\beta=S(r,\cdot)-V(r,\cdot)/\tau$. For every $a\in A$ and $q\in Q$, $S(a,q)=S(r,q)+[V(a,q)-V(r,q)]/\tau=V(a,q)/\tau+\beta\cdot q$.

\emph{Part (b).} The function $x\mapsto1/(1+e^{-x})$ is strictly increasing, so matching the probabilities at $q$ means $e_{ab}\cdot q=g_{ab}\cdot q/\tau_X$. By Theorem~\ref{thm:main}(a), some objective is block-proportional with the scales $1/\tau_X$, and by Fact~3 it has $e_{ab}\cdot q=g_{ab}\cdot q/\tau_X$ for all actions $a$ and $b$ of each block $X$ and every $q\in Q$, so it matches all the probabilities. An objective that matches them and makes no action outside $A_Q$ optimal on the closure of $Q$ satisfies \eqref{eq:blocks} with $\lambda_X=1/\tau_X$ for every block, so it is block-proportional.
\end{proof}

\noindent In part (a) the fitted objective is exact, so with noise that is the same for every comparison, the frequencies of choices reveal the relative scale. In part (b) the fitted scale of each block is the reciprocal of its temperature, so the fitted relative scale of two blocks is the principal's, which is one, multiplied by the ratio of the precisions $1/\tau_X$ with which the principal compares their actions.

\subsection*{Rankings that decisions fix, and decisions with an action unavailable}

The next claim describes the comparisons of two actions that decisions fix at a belief at which the principal has a single optimal action, and proves the statement in Section~\ref{sec:records} on a path of actions.

\begin{suppclaim}\label{claim:sides}
Let Assumption~\ref{ass:acting} hold, let $q\in Q$, let $v$ be the principal's only optimal action at $q$, and let $a$ and $b$ be distinct actions in $A_Q$.
\begin{enumerate}[label=(\alph*)]
\item If $a=v$, or if $a$ lies on the path from $v$ to $b$ in the tree $\mathcal H$, then $V(a,q)>V(b,q)$, and every objective that is aligned on the range strictly prefers $a$ to $b$ at $q$.
\item If $v$ differs from $a$ and $b$ and lies on the path from $a$ to $b$ in $\mathcal H$, some objective that is aligned on the range strictly prefers $a$ to $b$ at $q$, and another strictly prefers $b$ to $a$.
\item Let the margin graph be the path that joins each $a_i$ to $a_{i+1}$ in a list $a_1,\dots,a_n$ of the actions in $A_Q$, and let $v=a_j$. For $i<k$, part (a) applies to $a_k$ and $a_i$ if $k\leq j$ and to $a_i$ and $a_k$ if $j\leq i$, and part (b) applies to $a_i$ and $a_k$ if $i<j<k$.
\end{enumerate}
\end{suppclaim}

\begin{proof}
Let $a=z_0,X_1,z_1,\dots,X_m,z_m=b$ be the path from $a$ to $b$ in $\mathcal H$. By Lemma~\ref{lem:gains}, $\delta_{X_l}(a,b;q)=V(z_{l-1},q)-V(z_l,q)$ for $l=1,\dots,m$, the other gains are zero, $V(a,q)-V(b,q)$ is the sum of the gains, and an objective that is aligned on the range with scales $\lambda_X$ has $S(a,q)-S(b,q)=\sum_l\lambda_{X_l}\,\delta_{X_l}(a,b;q)$. We use the following fact. \emph{If an action $x\neq v$ lies on the path from $v$ to an action $y\neq x$ in $\mathcal H$, then $V(x,q)>V(y,q)$.} Indeed, $x$ separates $v$ and $y$ by Lemma~\ref{lem:blocks}(b), so it beats $v$ or $y$ at $q$ by Claim~\ref{claim:between}(c), and it does not beat $v$.

\emph{Part (a).} Under either hypothesis, the path from $v$ to $b$ in $\mathcal H$ ends with the path from $a$ to $b$. For each $l$, the action $z_{l-1}$ therefore equals $v$ or lies on the path from $v$ to $z_l$, so $V(z_{l-1},q)>V(z_l,q)$, in the first case because $v$ is the only optimal action and in the second by the fact. Every gain $\delta_{X_l}(a,b;q)$ is thus positive, and both differences are positive.

\emph{Part (b).} Let $v=z_j$ with $0<j<m$. For $l>j$, the action $z_{l-1}$ equals $v$ or lies on the path from $v$ to $z_l$, so $\delta_{X_l}(a,b;q)>0$ as in part (a). For $l\leq j$, the action $z_l$ equals $v$ or lies on the path from $v$ to $z_{l-1}$, so $V(z_l,q)>V(z_{l-1},q)$ and $\delta_{X_l}(a,b;q)<0$. The gains within blocks therefore take both signs, and the proof of Claim~\ref{claim:informative} gives one objective that is aligned on the range and strictly prefers $a$ to $b$ at $q$, and another that strictly prefers $b$ to $a$.

\emph{Part (c).} By Claim~\ref{claim:peaked}(a), every margin is a block, so the blocks are the pairs $\{a_i,a_{i+1}\}$ and the tree $\mathcal H$ is the path $a_1,\{a_1,a_2\},a_2,\dots,a_n$. If $i<k\leq j$, then $a_k$ equals $v$ or lies on the path from $v$ to $a_i$. If $j\leq i<k$, then $a_i$ equals $v$ or lies on the path from $v$ to $a_k$. If $i<j<k$, then $v$ lies on the path from $a_i$ to $a_k$.
\end{proof}

\noindent Decisions thus fix how the principal ranks the actions along every path in $\mathcal H$ that leads away from its choice. When the margin graph is a path, they fix how it ranks any two actions on the same side of its choice and leave open every comparison between an action on one side and an action on the other. These comparisons are the choices between mistakes of Proposition~\ref{prop:records}.

The last claim of this section concerns the principal's decisions when one of its actions is unavailable, the record in the footnote on conveying a belief in Section~\ref{sec:practice}. For $c\in A_Q$, we say that an objective $S$ \emph{makes the principal's decisions without $c$} if, at every belief $q\in Q$, every action that maximizes $S(\cdot,q)$ over $A_Q\setminus\{c\}$ also maximizes $V(\cdot,q)$ over $A_Q\setminus\{c\}$.

\begin{suppclaim}\label{claim:unavailable}
Let Assumption~\ref{ass:acting} hold, let $A_Q$ contain at least three actions, let $S$ be aligned on the range with scales $\lambda_X$, and let $c\in A_Q$.
\begin{enumerate}[label=(\alph*)]
\item At every belief $q\in Q$ at which $c$ is optimal for the principal, every action that maximizes $V(\cdot,q)$ over $A_Q\setminus\{c\}$ belongs to a block that contains $c$.
\item The objective $S$ makes the principal's decisions without $c$ if and only if $\lambda_X=\lambda_Y$ for all blocks $X$ and $Y$ that contain $c$. In particular, every objective that is aligned on the range does so when $c$ does not separate the range.
\item The objective $S$ makes the principal's decisions without $c$ for every action $c$ that separates the range if and only if $S$ is exact on the range. For each action $c_0$ that separates the range, some objective that is aligned on the range makes the principal's decisions without every action other than $c_0$ and is not exact on the range.
\end{enumerate}
\end{suppclaim}

\begin{proof}
\emph{Part (a).} Let $q\in Q$ have $c\in\BR_V(q)$, and let $b\in A_Q\setminus\{c\}$ belong to no block that contains $c$. The path from $c$ to $b$ in $\mathcal H$ is $c=z_0,X_1,z_1,\dots,X_m,z_m=b$ with $m\geq2$. The action $z_1$ differs from $c$ and $b$ and lies on this path, so it separates $c$ and $b$ (Lemma~\ref{lem:blocks}(b)) and beats $c$ or $b$ at $q$ (Claim~\ref{claim:between}(c)). It does not beat $c$, so $V(z_1,q)>V(b,q)$, and $b$ does not maximize $V(\cdot,q)$ over $A_Q\setminus\{c\}$.

\emph{Part (b), sufficiency.} Let every block that contains $c$ have the scale $\lambda$, and fix $q\in Q$. If $c\notin\BR_V(q)$, then $\BR_S(q)=\BR_V(q)\subseteq A_Q\setminus\{c\}$ by Theorem~\ref{thm:main}(a), so the maximizers of $S(\cdot,q)$ and of $V(\cdot,q)$ over $A_Q\setminus\{c\}$ are the same. Let $c\in\BR_V(q)$, and let $b^*$ maximize $V(\cdot,q)$ over $A_Q\setminus\{c\}$. By part (a), $b^*$ and $c$ belong to a common block, so $S(c,q)-S(b^*,q)=\lambda\,[V(c,q)-V(b^*,q)]$ by \eqref{eq:rep}. For any $b\in A_Q\setminus\{c\}$, let $c=z_0,X_1,z_1,\dots,X_m,z_m=b$ be the path from $c$ to $b$ in $\mathcal H$. The gains $\delta_{X_l}(c,b;q)=V(z_{l-1},q)-V(z_l,q)$ are nonnegative by Lemma~\ref{lem:gains}(c), and $X_1$ contains $c$, so Lemma~\ref{lem:gains}(b) gives
\begin{align*}
S(c,q)-S(b,q)&=\lambda\,\bigl[V(c,q)-V(z_1,q)\bigr]+\sum_{l\geq2}\lambda_{X_l}\,\delta_{X_l}(c,b;q)\\
&\geq\lambda\,\bigl[V(c,q)-V(b^*,q)\bigr]=S(c,q)-S(b^*,q),
\end{align*}
where the inequality uses $V(z_1,q)\leq V(b^*,q)$. So $b^*$ maximizes $S(\cdot,q)$ over $A_Q\setminus\{c\}$. If $b$ does so too, the inequality holds with equality, so $V(z_1,q)=V(b^*,q)$ and every gain with $l\geq2$ is zero, which gives $V(b,q)=V(z_1,q)=V(b^*,q)$. So $b$ maximizes $V(\cdot,q)$ over $A_Q\setminus\{c\}$.

\emph{Part (b), necessity.} Let $S$ make the principal's decisions without $c$, and let $X$ and $Y$ be distinct blocks that contain $c$. The set $U_c$ of beliefs in $Q$ at which $c$ is the principal's only optimal action is nonempty by Fact~5, and it is convex and open relative to $L$, being the intersection of $Q$ with the open half-spaces $\{\xi:g_{cd}\cdot\xi>0\}$, $d\in A_Q\setminus\{c\}$. For $q\in U_c$, let $B(q)$ be the set of maximizers of $V(\cdot,q)$ over $A_Q\setminus\{c\}$. By part (a), every action in $B(q)$ belongs to a block that contains $c$, and to only one, because two blocks that contain $c$ share no other action, as $\mathcal H$ has no cycle.

We first find, for each block $Z$ that contains $c$, a belief in $U_c$ at which $B$ is a single action of $Z$. Every action of a block belongs to one of its margins, so $Z$ has a margin $\{a_Z,c\}$, and some $q_Z\in Q$ has $\BR_V(q_Z)=\{a_Z,c\}$. As in the proof of Lemma~\ref{lem:strict}, some $z\in L_0$ has $g_{ca_Z}\cdot z>0$. For small $\varepsilon>0$, the belief $q_Z+\varepsilon z$ lies in $Q$, the action $c$ beats $a_Z$ there, and $a_Z$ beats every other action of $A_Q\setminus\{c\}$ there, as it does at $q_Z$. So $q_Z+\varepsilon z\in U_c$ and $B(q_Z+\varepsilon z)=\{a_Z\}$.

For distinct $x,y\in A_Q\setminus\{c\}$, the set $H_{xy}=\{\xi\in L:g_{xy}\cdot\xi=0\}$ is an affine subspace of $L$ of smaller dimension, because $g_{xy}$ is not zero on $L_0$. For distinct $x,y,w\in A_Q\setminus\{c\}$, the set $M_{xyw}=H_{xy}\cap H_{yw}$, at whose points the three actions have equal expected payoffs, is empty or, by Lemma~\ref{lem:triple}(b), an affine subspace of dimension $\dim L-2$. The beliefs of $U_c$ at which $B$ equals $\{a_X\}$ form a nonempty set that is open relative to $L$, and so do those at which $B$ equals $\{a_Y\}$. By Fact~4, the first set contains a belief $u$ that lies in no $H_{xy}$. The beliefs $u'$ for which the segment from $u$ to $u'$ meets a given $M_{xyw}$ lie in the affine hull of $u$ and $M_{xyw}$, which has dimension at most $\dim L-1$, so by Fact~4 the second set contains a belief $u'$ such that the segment from $u$ to $u'$ meets no $M_{xyw}$.

Let $\sigma(t)=(1-t)\,u+t\,u'$ for $t\in[0,1]$, which lies in the convex set $U_c$. For distinct $x,y\in A_Q\setminus\{c\}$, the function $t\mapsto g_{xy}\cdot\sigma(t)$ is affine and is not zero at $t=0$, so it vanishes at most once. The set $B(\sigma(t))$ therefore has more than one action at only finitely many $t$, say $t_1<\dots<t_r$, and it is a single action, the same one, throughout each of the intervals between them. At $t_l$ the set $B(\sigma(t_l))$ contains, by continuity, the actions $x_{l-1}$ and $x_l$ that maximize just before and just after $t_l$. It contains no third action, since $\sigma(t_l)$ lies in no $M_{xyw}$, and $x_{l-1}\neq x_l$, since the payoffs of the two actions of $B(\sigma(t_l))$ are distinct affine functions of $t$ that agree at $t_l$, so that each is the larger on one side of $t_l$. Thus $x_0=a_X$, $x_r=a_Y$, and $B(\sigma(t_l))=\{x_{l-1},x_l\}$ for $l=1,\dots,r$.

Fix $l$, write $x=x_{l-1}$, $x'=x_l$ and $q_l=\sigma(t_l)$, and let $Z$ and $Z'$ be the blocks that contain $c$ together with $x$ and with $x'$. In the decision problem whose actions are those of $A_Q\setminus\{c\}$, with the payoffs $V$, the only optimal actions at $q_l$ are $x$ and $x'$, the stake $g_{xx'}$ is not zero on $L_0$, and the objective $S$ is aligned at every belief in $Q$, because it makes the principal's decisions without $c$. Lemma~\ref{lem:local}, applied to that problem, gives $\lambda>0$ with $e_{xx'}\cdot\xi=\lambda\,g_{xx'}\cdot\xi$ for every $\xi\in\spn Q$. Since $x$ and $c$ belong to $Z$ and $c$ and $x'$ belong to $Z'$, \eqref{eq:blocks} and Fact~3 give $e_{xc}\cdot\xi=\lambda_Z\,g_{xc}\cdot\xi$ and $e_{cx'}\cdot\xi=\lambda_{Z'}\,g_{cx'}\cdot\xi$ for every $\xi\in\spn Q$. Because $e_{xx'}=e_{xc}+e_{cx'}$ and $g_{xx'}=g_{xc}+g_{cx'}$,
\[
(\lambda_Z-\lambda)\,g_{xc}\cdot\xi+(\lambda_{Z'}-\lambda)\,g_{cx'}\cdot\xi=0\qquad\text{for every }\xi\in\spn Q.
\]
The two maps are linearly independent on $\spn Q$ by Lemma~\ref{lem:triple}(a), so $\lambda_Z=\lambda=\lambda_{Z'}$. Hence the blocks that contain $c$ together with $x_0,x_1,\dots,x_r$ all have the same scale, and the first of them is $X$ and the last is $Y$. If $c$ does not separate the range, it belongs to a single block (Lemma~\ref{lem:blocks}(c)), and the condition holds for every aligned objective.

\emph{Part (c).} If $S$ is exact on the range, its scales are equal (Lemma~\ref{lem:decomp}(d)), and it makes the principal's decisions without every action by part (b). Conversely, let $S$ make the principal's decisions without every action that separates the range. Two blocks that share an action share one that belongs to at least two blocks and so separates the range (Lemma~\ref{lem:blocks}(c)), and by part (b) they have the same scale. Any two blocks are joined by a path in the tree $\mathcal H$, on which consecutive blocks share the action between them, so all the scales are equal and $S$ is exact on the range by Lemma~\ref{lem:decomp}(d). For the last statement, let $c_0$ separate the range. Deleting $c_0$ from $\mathcal H$ leaves at least two components, one for each block that contains $c_0$. Give the scale one to the blocks of one component and the scale two to the others, and let $S'$ be an objective that is aligned on the range with these scales, which exists by Theorem~\ref{thm:main}(a). For every action $c\neq c_0$, the blocks that contain $c$ are joined to one another through $c$ and so lie in one component, and they have the same scale. By part (b), $S'$ makes the principal's decisions without $c$. Its scales are not all equal, so it is not exact on the range (Lemma~\ref{lem:decomp}(d)).
\end{proof}

\noindent Part (b) is the counterpart for decisions of Proposition~\ref{prop:records}(c), in which the cost of an action changes. The principal's decisions with one separating action unavailable, among the other actions of $A_Q$, fix the relative scale of every block around it and of no others, and by part (c) none of the separating actions can be left out. The restriction to $A_Q$ matters, because an action that is never optimal can become the fallback when $c$ is unavailable. If the file example had a fourth action that costs $2$ whatever the file, so that quarantining dominates it, the tripled objective extended to value that action at $-14/3$ when the file is malicious and at $-4/3$ when it is legitimate would remain aligned, and with quarantining unavailable it would choose exactly as the firm does, leaving the file below $p=1/5$, taking the fourth action between $1/5$ and $4/5$ and deleting the file above $4/5$. Read the other way, part (b) says that an aligned objective whose scales differ around a separating action can take a wrong action when that action is unavailable. In the file example, the firm deletes a file when $p\geq1/2$ if it cannot quarantine. The tripled objective deletes only when $p\geq7/10$, and between these beliefs it leaves a file that the firm would delete, at a loss to the firm of $20p-10$, which approaches $4$.

\section{Fitting decisions by maximizing the margin}\label{supp:maxmargin}

We state precisely and prove the claim in Section~\ref{sec:robust} about a procedure that fits decisions by maximizing the classification margin. Let $n=2$ and $A=A_Q=\{a_1,\dots,a_k\}$ with $k\geq3$, ordered as $p$ rises, with switching beliefs $t_i$ and slope increases $d_i$ as in \eqref{eq:kinks}. Let a finite record of the principal's decisions contain at least one belief in the interior of each action's optimal region and no switching belief. For each $i<k$, let $p_i^-$ be the largest recorded belief at which the principal chooses $a_i$, let $p_i^+$ be the smallest at which it chooses $a_{i+1}$, and let $\bar p_i=(p_i^-+p_i^+)/2$. The \emph{max-margin objective} minimizes $\sum_{a,\omega}S(a,\omega)^2$ among the objectives under which, at each recorded belief, the recorded action beats every other action by at least one. It is the direction to which gradient descent on the cross-entropy loss of a linear model converges on separable data, under a nondegeneracy condition that \citet{soudry2018} conjecture holds for almost all data sets (their Theorem~7, as corrected in the 2024 arXiv version). Beliefs $(p,1-p)$ lie on a line, so that condition may fail, and we treat the max-margin objective as a benchmark.

\begin{suppclaim}\label{claim:maxmargin}
The max-margin objective has
\[
S(a_{i+1},p)-S(a_i,p)=\frac{2\,(p-\bar p_i)}{p_i^+-p_i^-}\quad\text{for each }i<k,
\]
so it switches from $a_i$ to $a_{i+1}$ at $\bar p_i$ with kink sharpness $2/(p_i^+-p_i^-)$. Let $\lambda_i=2/[(p_i^+-p_i^-)d_i]$ be the ratio of this sharpness to the principal's. If the record consists of $N$ independent draws from a distribution with a density $f$ that is continuous and positive at each switching belief, then as $N\to\infty$ each $\bar p_i$ converges in probability to $t_i$, and for $i\neq j$
\[
\frac{\lambda_i}{\lambda_j}\ \longrightarrow\ \frac{f(t_i)\,d_j}{f(t_j)\,d_i}\,F\quad\text{in distribution},
\]
where $F$ has the $F$-distribution with $(4,4)$ degrees of freedom.
\end{suppclaim}

\begin{proof}
Write $S(a)=S(a,\cdot)\in\R^2$, with first component the payoff in state $\omega_1$, and $\bar S=(1/k)\sum_aS(a)$. The proof has five steps.

\emph{Step 1: reduction.} Each constraint compares two actions at the same belief, so adding the same vector $\beta\in\R^2$ to every $S(a)$ leaves the feasible set unchanged. Since $\sum_a\|S(a)\|^2=\sum_a\|S(a)-\bar S\|^2+k\|\bar S\|^2$, subtracting $\bar S$ from every action lowers the objective unless $\bar S=0$, so the max-margin objective has $\bar S=0$ and minimizes $\Xi(S)=\sum_a\|S(a)-\bar S\|^2$ over the feasible set, and any minimizer of $\Xi$ with $\bar S=0$ is the max-margin objective. Let $e_i=S(a_{i+1})-S(a_i)$ for $i<k$. The identity $\sum_a\|x_a-\bar x\|^2=(1/k)\sum_{j<l}\|x_l-x_j\|^2$ for vectors $x_1,\dots,x_k$ with mean $\bar x$ gives
\[
\Xi(S)=\frac1k\sum_{1\leq j<l\leq k}\Bigl\|\sum_{i=j}^{l-1}e_i\Bigr\|^2.
\]

\emph{Step 2: a lower bound.} For $i<k$, let $\ell_i(p)=e_i\cdot(p,1-p)=S(a_{i+1},p)-S(a_i,p)$, an affine function with $\ell_i(1)=e_i(\omega_1)$ and $\ell_i(0)=e_i(\omega_2)$, and let $w_i=p_i^+-p_i^-$. At $p_i^-$ the principal chooses $a_i$, so feasibility requires $\ell_i(p_i^-)\leq-1$, and at $p_i^+$ it chooses $a_{i+1}$, so $\ell_i(p_i^+)\geq1$. The slope $s_i=e_i(\omega_1)-e_i(\omega_2)$ of $\ell_i$ therefore satisfies $s_iw_i=\ell_i(p_i^+)-\ell_i(p_i^-)\geq2$. Since $1-p_i^+\geq0$ and $p_i^-\geq0$,
\[
\begin{aligned}
e_i(\omega_1)&=\ell_i(p_i^+)+s_i(1-p_i^+)\geq1+\frac{2(1-p_i^+)}{w_i}=\frac{2(1-\bar p_i)}{w_i}>0,\\
e_i(\omega_2)&=\ell_i(p_i^-)-s_ip_i^-\leq-1-\frac{2p_i^-}{w_i}=-\frac{2\bar p_i}{w_i}<0,
\end{aligned}
\]
where the strict inequalities use $0<\bar p_i<1$. Let $e_i^*=(2/w_i)(1-\bar p_i,-\bar p_i)$. For $j<l$, the first component of $\sum_{i=j}^{l-1}e_i$ is a sum of positive terms, each at least the first component of the corresponding $e_i^*$, and the second is a sum of negative terms, each at most the second component of $e_i^*$. So $\|\sum_{i=j}^{l-1}e_i\|^2\geq\|\sum_{i=j}^{l-1}e_i^*\|^2$ for every $j<l$, and $\Xi(S)$ is at least its value at any objective with differences $e^*$. For $l=j+1$ the inequality is strict unless $e_j=e_j^*$, so equality requires $e=e^*$.

\emph{Step 3: the differences $e^*$ are feasible.} Let $S^*$ have differences $e^*$, so that $\ell_i^*(p)=S^*(a_{i+1},p)-S^*(a_i,p)=2(p-\bar p_i)/w_i$, which is increasing, with $\ell_i^*(p_i^-)=-1$ and $\ell_i^*(p_i^+)=1$. Let $p$ be a recorded belief at which the principal chooses $a_j$. For $i\geq j$ we have $p\leq p_i^-$: for $i=j$ by the definition of $p_j^-$, and for $i>j$ because the principal chooses $a_j$ only at beliefs up to $t_j$, whereas $p_i^-$ is a recorded belief at which it chooses $a_i$, which exceeds $t_{i-1}\geq t_j$, since the record contains no switching belief. Hence $\ell_i^*(p)\leq-1$ for every $i\geq j$. Likewise $p\geq p_i^+$ and $\ell_i^*(p)\geq1$ for every $i<j$. For $l>j$, $S^*(a_l,p)-S^*(a_j,p)=\sum_{i=j}^{l-1}\ell_i^*(p)\leq-(l-j)\leq-1$, and for $l<j$, $S^*(a_j,p)-S^*(a_l,p)=\sum_{i=l}^{j-1}\ell_i^*(p)\geq j-l\geq1$. So the recorded action beats every other action by at least one at every recorded belief.

\emph{Step 4: the max-margin objective.} By Steps~2 and~3, the minimizers of $\Xi$ over the feasible set are exactly the feasible objectives with differences $e^*$, and by Step~1 the max-margin objective is the one among them with $\bar S=0$. It has $S(a_{i+1},p)-S(a_i,p)=\ell_i^*(p)$. Because $\bar p_i<p_i^+\leq p_{i+1}^-<\bar p_{i+1}$, the zeros of the $\ell_i^*$ increase with $i$, so the objective's best action is $a_j$ between $\bar p_{j-1}$ and $\bar p_j$. It therefore switches from $a_i$ to $a_{i+1}$ at $\bar p_i$, with kink sharpness $2/w_i$, the slope of $\ell_i^*$.

\emph{Step 5: the limit.} Let the record be $N$ independent draws from a distribution with density $f$, continuous and positive at each $t_i$. With probability one no draw equals a switching belief. Because $f$ is positive near each $t_i$, each action's region has positive probability, so the probability that the record contains a belief in the interior of each region tends to one. For large $N$, with probability tending to one, $p_i^-$ is the largest draw below $t_i$ and $p_i^+$ the smallest draw above it. Fix $x,y\geq0$. The event that $N(t_i-p_i^-)>x$ and $N(p_i^+-t_i)>y$ is the event that no draw falls in $(t_i-x/N,t_i+y/N)$, whose probability is
\[
\Bigl(1-\int_{t_i-x/N}^{t_i+y/N}f\Bigr)^N\longrightarrow e^{-f(t_i)(x+y)},
\]
because $N\int_{t_i-x/N}^{t_i+y/N}f\to f(t_i)(x+y)$ by the continuity of $f$ at $t_i$. So $N(t_i-p_i^-)$ and $N(p_i^+-t_i)$ converge jointly in distribution to independent exponential random variables with rate $f(t_i)$, and $Nw_i$ converges in distribution to $G_i/f(t_i)$, where $G_i$, the sum of two independent unit exponentials, has the Gamma$(2,1)$ distribution. In particular $|\bar p_i-t_i|\leq w_i$ converges to zero in probability. For distinct $t_i$ and $t_j$, the same computation for the union of two disjoint intervals, one around each, gives joint convergence with $G_i$ and $G_j$ independent. Since $\lambda_i/\lambda_j=(w_jd_j)/(w_id_i)=(Nw_j)d_j/[(Nw_i)d_i]$, the continuous mapping theorem gives convergence in distribution to $[f(t_i)d_j/(f(t_j)d_i)]\,G_j/G_i$. Finally, $2G_i$ and $2G_j$ are independent chi-squared variables with four degrees of freedom, so $G_j/G_i=(2G_j/4)/(2G_i/4)$ has the $F$-distribution with $(4,4)$ degrees of freedom.
\end{proof}

In the example of Section~\ref{sec:example}, $d_1=d_2=10$, and a density symmetric about $1/2$ has $f(1/10)=f(9/10)$, so $\lambda_2/\lambda_1$ converges in distribution to $F=G_1/G_2$. Let $\Upsilon=G_2/(G_1+G_2)$, which has the Beta$(2,2)$ distribution, with density $6b(1-b)$, so that $F=(1-\Upsilon)/\Upsilon$. Then $F>2$ exactly when $\Upsilon<1/3$, which has probability $\int_0^{1/3}6b(1-b)\,db=7/27$, and by symmetry $F<1/2$ has the same probability, so the limiting relative scale is more than twice the firm's, or less than half of it, with probability $14/27$. The share that Claim~\ref{claim:ratio} guarantees to an objective with the firm's switches and this relative scale, $\min\{F,1/F\}=\min\{\Upsilon,1-\Upsilon\}/\max\{\Upsilon,1-\Upsilon\}$, has mean $2\int_0^{1/2}[b/(1-b)]\,6b(1-b)\,db=12\int_0^{1/2}b^2\,db=1/2$, and since $x\mapsto\min\{x,1/x\}$ is bounded and continuous, the mean for a record of $N$ beliefs converges to $1/2$.

\section{Beliefs within a capability bound}\label{supp:range}

We state precisely and prove the claim in Section~\ref{sec:capability} on the beliefs that an AI can hold under a capability bound.

\begin{suppclaim}\label{claim:range}
Let $q^0$ be a prior with full support, let $B$ be an experiment whose signals all have positive probability at $q^0$, and let $P_B$ be the set of its posteriors at $q^0$. The beliefs that arise with positive probability under some garbling of $B$ are exactly the beliefs in the convex hull $\conv P_B$. If $B$ is fully informative, in that each of its signals has positive probability in at most one state, then $\conv P_B=\Delta(\Omega)$.
\end{suppclaim}

\begin{proof}
We show first that every posterior of a garbling lies in $\conv P_B$, then that every belief in $\conv P_B$ is such a posterior, and finally treat the fully informative case. Let $B$ have signals $s\in\Sigma$, with probabilities $P(s)>0$ and posteriors $q_s$. A garbling of $B$ is described by a Markov kernel $\mathcal K(s'\mid s)$ from $\Sigma$ to a finite set $\Sigma'$ of new signals, applied independently of the state given $s$.

\emph{Step 1: every posterior of a garbling lies in $\conv P_B$.} Fix a new signal $s'$ with probability $P'(s')=\sum_sP(s)\mathcal K(s'\mid s)>0$. Because the new signal is independent of the state given $s$, the joint probability of $\omega$, $s$ and $s'$ is $q^0(\omega)f(s\mid\omega)\mathcal K(s'\mid s)$, and $q^0(\omega)f(s\mid\omega)=P(s)q_s(\omega)$ by the definition of the posterior $q_s$. Summing over $s$ and dividing by $P'(s')$, Bayes' rule gives the posterior after $s'$ as
\[
q_{s'}(\omega)=\sum_{s\in\Sigma}\frac{P(s)\mathcal K(s'\mid s)}{P'(s')}\,q_s(\omega).
\]
The weights $P(s)\mathcal K(s'\mid s)/P'(s')$ are nonnegative and sum to one, so $q_{s'}\in\conv P_B$.

\emph{Step 2: every belief in $\conv P_B$ is a posterior of a garbling.} Let $q=\sum_s\alpha_sq_s$ with $\alpha_s\geq0$ and $\sum_s\alpha_s=1$. Let $\varepsilon=\min\{1,\min_sP(s)\}$, which is positive because every $P(s)$ is positive, so that $\varepsilon\alpha_s\leq P(s)$ for every $s$. Define a garbling with two new signals by $\mathcal K(1\mid s)=\varepsilon\alpha_s/P(s)$, which lies in $[0,1]$, and $\mathcal K(2\mid s)=1-\mathcal K(1\mid s)$. Then $P'(1)=\sum_sP(s)\varepsilon\alpha_s/P(s)=\varepsilon>0$, and the weights in the display of Step~1 are $P(s)\mathcal K(1\mid s)/P'(1)=\alpha_s$, so the posterior after signal~$1$ is $q$.

\emph{Step 3: the fully informative case.} If every signal of $B$ reveals the state, each signal with positive probability leads to a point mass. Every state $\omega$ is revealed by some signal $s$ with $f(s\mid\omega)>0$, because $\sum_sf(s\mid\omega)=1$, and that signal has $P(s)\geq q^0(\omega)f(s\mid\omega)>0$ because the prior has full support. Hence $P_B=\{\delta_\omega:\omega\in\Omega\}$, whose convex hull is $\Delta(\Omega)$. Steps~1 and~2 do not use the full support of the prior, so they hold with any belief in its place.
\end{proof}

\section{States the AI cannot tell apart}\label{supp:classes}

We state precisely and prove the claim behind the scanner example in Section~\ref{sec:capability}: evidence that has the same distribution in every state of a class keeps the AI's beliefs at the prior's relative weights within each class, so the principal's stakes need to be known only as averages over each class.

\begin{suppclaim}\label{claim:classes}
Let the states be partitioned into classes $\Omega_1,\dots,\Omega_J$, let the prior $q^0$ have full support, and suppose that the AI's capability bound $B$ has the same distribution of signals $f_j$ in every state of class $\Omega_j$. For a payoff function $W$, let $\bar W(a,j)=\sum_{\omega\in\Omega_j}q^0(\omega\mid\Omega_j)W(a,\omega)$ be its average over class $j$ with the prior's weights. Then every belief $q$ that the AI can hold satisfies $q(\omega)=q(\Omega_j)\,q^0(\omega\mid\Omega_j)$ for every $j$ and every $\omega\in\Omega_j$, and at every such belief $W(a,q)=\sum_jq(\Omega_j)\bar W(a,j)$ for every payoff function $W$ and action $a$. Hence the AI is aligned at $q$ if and only if it is aligned at the belief $(q(\Omega_1),\dots,q(\Omega_J))$ in the problem whose states are the classes, with payoff function $\bar V$ and objective $\bar S$.
\end{suppclaim}

\begin{proof}
We first show that the AI's beliefs keep the prior's relative weights within each class and then compute expected payoffs at such beliefs. After a signal $s$ of positive probability $P(s)$, Bayes' rule gives
\[
q_s(\omega)=\frac{q^0(\omega)f_j(s)}{P(s)}\qquad\text{for }\omega\in\Omega_j.
\]
Summing over $\omega\in\Omega_j$ gives $q_s(\Omega_j)=q^0(\Omega_j)f_j(s)/P(s)$, and substituting $q^0(\omega)=q^0(\Omega_j)\,q^0(\omega\mid\Omega_j)$ into the display gives $q_s(\omega)=q_s(\Omega_j)\,q^0(\omega\mid\Omega_j)$. These equations are linear in $q$, so they hold at every belief in $\conv P_B$, which by Claim~\ref{claim:range} is the set of beliefs the AI can hold. Now fix such a belief $q$, a payoff function $W$ and an action $a$. Grouping the states by class,
\[
W(a,q)=\sum_j\sum_{\omega\in\Omega_j}q(\Omega_j)\,q^0(\omega\mid\Omega_j)W(a,\omega)=\sum_jq(\Omega_j)\bar W(a,j).
\]
Applying this to $W=V$ and to $W=S$ shows that $\BR_V(q)$ and $\BR_S(q)$ are the sets of best actions at the belief $(q(\Omega_1),\dots,q(\Omega_J))$ under $\bar V$ and under $\bar S$, so the AI is aligned at $q$ exactly when it is aligned at that belief in the problem whose states are the classes.
\end{proof}

\section{The loss from greater capability}\label{supp:capbound}

We state precisely and prove the bound in Section~\ref{sec:capability} on what an objective that is aligned on a range $Q$ can lose at beliefs outside it. Let Assumption~\ref{ass:acting} hold on $Q$, with at least two actions in $A_Q$, and let $S$ be aligned on $Q$ with scales $\lambda_X$, the largest $\lambda_{\max}$ and the smallest $\lambda_{\min}$. For a belief $q\in\Delta(\Omega)$, let
\[
\Gamma(q)=\sum_XV^*_X(q)-\sum_{c\in A_Q}(n_c-1)V(c,q)-V^*(q),
\]
computed with the blocks of $Q$, where $n_c$ is the number of blocks that contain $c$, and let $d(q)$ be the Euclidean distance from $q$ to $\spn Q$. We write $\delta_X(a,b;\cdot)$ for the vector in $\R^\Omega$ whose inner product with each $\xi$ is the gain $\delta_X(a,b;\xi)$ of Lemma~\ref{lem:gains}, and $\|\cdot\|$ for the Euclidean norm.

\begin{suppclaim}\label{claim:capbound}
At every belief $q\in\Delta(\Omega)$ at which the AI's optimal actions, and some optimal action of the principal, lie in $A_Q$, the principal's loss $V^*(q)-V(x,q)$ from each action $x$ that is optimal for the AI is at most
\[
\Bigl(\frac{\lambda_{\max}}{\lambda_{\min}}-1\Bigr)\Gamma(q)+\frac{d(q)\,\bar\varepsilon}{\lambda_{\min}},
\qquad\text{where }\bar\varepsilon=\max_{a,b\in A_Q}\Bigl\|S(a,\cdot)-S(b,\cdot)-\sum_X\lambda_X\delta_X(a,b;\cdot)\Bigr\|,
\]
and $\Gamma(q)\geq0$.
\end{suppclaim}

\begin{proof}
Fix such a belief $q$, an action $x\in\BR_S(q)$, which belongs to $A_Q$ by hypothesis, and an action $a^*\in A_Q$ that is optimal for the principal at $q$, so that $V(a^*,q)=V^*(q)$. Root $\mathcal H$ at an action $r\in A_Q$, and for $v\in A_Q$ and each block $X$ let $D_X(v)=V(x^v_X,q)-V(r_X,q)$ if $X\in\mathcal P(v)$ and $D_X(v)=0$ otherwise. We first show that $\delta_X(a,b;q)=D_X(a)-D_X(b)$ for all $a,b\in A_Q$ and every block $X$. The paths from $r$ to $a$ and from $r$ to $b$ share an initial segment, which ends at a vertex $u$, and the path from $a$ to $b$ consists of the path from $a$ to $u$ followed by the path from $u$ to $b$. A block before $u$ on the shared segment is followed on both paths by the same action, so $D_X(a)=D_X(b)$, and the block is not on the path from $a$ to $b$, so $\delta_X(a,b;q)=0$. A block $X$ after $u$ on the path from $r$ to $a$ belongs to $\mathcal P(a)$ but not $\mathcal P(b)$, and the path from $a$ to $b$ crosses it from $x^a_X$ to $r_X$, so $\delta_X(a,b;q)=V(x^a_X,q)-V(r_X,q)=D_X(a)-D_X(b)$. The case of a block after $u$ on the path from $r$ to $b$ is the same with the roles of $a$ and $b$ exchanged and the signs reversed. If $u$ is a block, the path from $a$ to $b$ crosses it from $x^a_u$ to $x^b_u$, so $\delta_u(a,b;q)=V(x^a_u,q)-V(x^b_u,q)=D_u(a)-D_u(b)$. A block on neither path has $D_X(a)=D_X(b)=0=\delta_X(a,b;q)$.

Let $\Delta_X=\delta_X(x,a^*;q)=D_X(x)-D_X(a^*)$. By Lemma~\ref{lem:gains}(a), the loss from $x$ is $V(a^*,q)-V(x,q)=-\sum_X\Delta_X$. Let $\varepsilon(x,a^*)=S(x,\cdot)-S(a^*,\cdot)-\sum_X\lambda_X\delta_X(x,a^*;\cdot)$. By Theorem~\ref{thm:main}(a), $S$ is block-proportional, so Lemma~\ref{lem:gains}(b) gives $\varepsilon(x,a^*)\cdot\xi=0$ for every $\xi\in\spn Q$. Let $\bar q$ be the orthogonal projection of $q$ onto $\spn Q$, so that $\|q-\bar q\|=d(q)$ and $\varepsilon(x,a^*)\cdot q=\varepsilon(x,a^*)\cdot(q-\bar q)$, and the Cauchy--Schwarz inequality gives $|\varepsilon(x,a^*)\cdot q|\leq\bar\varepsilon\,d(q)$. Because $x$ is optimal for the AI, $0\leq S(x,q)-S(a^*,q)=\sum_X\lambda_X\Delta_X+\varepsilon(x,a^*)\cdot q$, so $\sum_X\lambda_X\Delta_X\geq-\bar\varepsilon\,d(q)$. With $\Delta_X^+=\max\{\Delta_X,0\}$ and $\Delta_X^-=\max\{-\Delta_X,0\}$, this gives $\lambda_{\min}\sum_X\Delta_X^-\leq\lambda_{\max}\sum_X\Delta_X^++\bar\varepsilon\,d(q)$, so the loss, $\sum_X\Delta_X^--\sum_X\Delta_X^+$, is at most $(\lambda_{\max}/\lambda_{\min}-1)\sum_X\Delta_X^++\bar\varepsilon\,d(q)/\lambda_{\min}$.

It remains to show that $\sum_X\Delta_X^+\leq\Gamma(q)$. Each $D_X(x)$ is at most $h_X(q)=V^*_X(q)-V(r_X,q)$, because $x^x_X$ belongs to $X$, and each $h_X(q)$ is nonnegative, because $r_X$ belongs to $X$. If $X\in\mathcal P(a^*)$, then $\Delta_X\leq V^*_X(q)-V(x^{a^*}_X,q)$, and otherwise $\Delta_X=D_X(x)\leq h_X(q)$. Both bounds are nonnegative, so they bound $\Delta_X^+$. Writing the path from $r$ to $a^*$ as $r=y_0,X_1,y_1,\dots,X_m,y_m=a^*$, with $y_{j-1}=r_{X_j}$ and $y_j=x^{a^*}_{X_j}$, the sum $\sum_{X\in\mathcal P(a^*)}[V(x^{a^*}_X,q)-V(r_X,q)]$ telescopes to $V(a^*,q)-V(r,q)$ at every belief, so the sum of the bounds over the blocks is
\[
\sum_Xh_X(q)-\sum_{X\in\mathcal P(a^*)}\bigl[V(x^{a^*}_X,q)-V(r_X,q)\bigr]=\sum_Xh_X(q)-V(a^*,q)+V(r,q).
\]
By Lemma~\ref{lem:decomp}(e), $V(r,q)+\sum_Xh_X(q)=\sum_XV^*_X(q)-\sum_c(n_c-1)V(c,q)$, so the sum is $\Gamma(q)$, which is therefore nonnegative.
\end{proof}

\noindent The claim says that, among the actions that are optimal somewhere on the old range, a more capable AI takes wrong actions only at beliefs where the principal's own choice trades one old block against another, or where it depends on directions in which the old evidence never moved the AI's beliefs. By the last step of the proof, $\Gamma(q)$ is the sum over the blocks of $Q$ of the amounts by which the action through which each block is joined to the principal's best action falls short of the best action in that block. By Lemma~\ref{lem:decomp}(a) these amounts are all zero on $\cl Q$. At a belief outside it, one of them is positive when the principal's choice gives up a gain within one old block for a larger gain within another. With two blocks that share an action $c$, $\Gamma(q)$ is the smaller of the gains over $c$ on its two sides, which is positive only where actions on both sides of $c$ beat it. The second term covers objectives whose errors lie in directions orthogonal to the old range, such as the objective in Section~\ref{sec:capability} that puts the cost of deleting any legitimate file at the average over critical and routine files, and such errors matter in proportion to the distance from the old range. Actions that are optimal only at the new beliefs are constrained by decisions on the old range only in never being chosen there, so the claim does not cover them.

In the three-state example with $\kappa=5$ and the scanner of Section~\ref{sec:capability}, the objective that triples the stakes between quarantining and deleting keeps the firm's payoffs $(-10,0,0)$ for leaving and $(-5,-5,-5)$ for quarantining and values deleting at $(10,-80,-5)$, where the states are ordered as malicious, critical and routine. Its scales are $1$ and $3$, and $\bar\varepsilon=0$. At the belief $(2/5,1/20,11/20)$ the firm's expected costs are $4$ for leaving, $17/4$ for deleting and $5$ for quarantining, so it leaves the file, whereas the objective values deleting at $-11/4$ and leaving at $-4$, so the AI deletes it, at a loss of $1/4$. There $\Gamma=\min\{5-4,5-17/4\}=3/4$, and the bound is $3/2$. At $(1/5,0,4/5)$ the firm's costs are $2$, $4$ and $5$, the objective is indifferent between deleting and leaving, both of which it values at $-2$, and $\Gamma=\min\{3,1\}=1$, so the loss of $2$ from deleting equals the bound.

\paragraph{An upgrade that lowers the firm's payoff.} In the same example, let the scanner's reading be $p$, so that the belief is $(p,(1-p)/2,(1-p)/2)$, and let the AI also learn whether the file is in use by a critical process, which is the case exactly when the file is critical. With probability $(1-p)/2$ it is, the belief becomes $(0,1,0)$, and the firm and the tripled objective both leave the file. Otherwise the belief becomes $(\pi,0,1-\pi)$ with $\pi=2p/(1+p)$. There the firm's expected costs are $10\pi$ for leaving, $5(1-\pi)$ for deleting and $5$ for quarantining, so the firm leaves the file when $\pi\leq1/3$, that is, when $p\leq1/5$, and deletes it otherwise. The tripled objective values leaving at $-10\pi$, deleting at $15\pi-5$ and quarantining at $-5$, so it deletes the file when $\pi>1/5$, that is, when $p>1/9$, and leaves it when $p<1/9$. At readings below $1/9$ and at readings of at least $1/5$, the more capable AI therefore takes an action that is optimal for the firm after either signal. At readings strictly between $1/9$ and $1/5$, it deletes a file that the firm would leave. The firm's expected cost with the inspection is then $\frac{1+p}{2}\cdot5(1-\pi)=\frac52(1-p)$, whereas without it the AI leaves the file at an expected cost of $10p$, and $\frac52(1-p)>10p$ because $p<1/5$. At the belief $(1/4,0,3/4)$ that the inspection then produces, $\Gamma=\min\{5-5/2,\,5-15/4\}=5/4$, and the loss of $5/4$ from deleting is within the bound $(\lambda_{\max}/\lambda_{\min}-1)\Gamma=5/2$ of Claim~\ref{claim:capbound}. The tripled objective takes the right action in each state, deleting a malicious file and leaving a critical or a routine one, so an AI that learned the state would act as the firm would.

\section{Acting under a penalty on divergence from a reference policy}\label{supp:penalty}

The text takes the AI to maximize its objective. This section supports the remark in Section~\ref{sec:model} that the scales that decisions leave free matter for acting when training penalizes divergence from a reference policy. Let Assumption~\ref{ass:acting} hold with two states and $Q=(0,1)$, and let $A=A_Q=\{a_1,\dots,a_n\}$ with $n\geq2$, ordered so that the principal switches from $a_i$ to $a_{i+1}$ at the belief $t_i$, where $0<t_1<\dots<t_{n-1}<1$. As in Section~\ref{supp:fileexample}, $V(a_{i+1},p)-V(a_i,p)=d_i\,(p-t_i)$ with $d_i>0$, every margin is a block, and by Theorem~\ref{thm:main}(a) an objective $S$ that is aligned on the range has $S(a_{i+1},p)-S(a_i,p)=\lambda_i\,d_i\,(p-t_i)$ for scales $\lambda_i>0$. Fix a probability distribution $r$ on $A$ with $r(a)>0$ for every $a$, the reference policy, and a weight $\tau>0$. At a belief $p$, let the AI choose the distribution $\mu$ over actions that maximizes
\[
\sum_a\mu(a)\,S(a,p)-\tau\sum_a\mu(a)\log\frac{\mu(a)}{r(a)},
\]
its expected payoff under the objective less $\tau$ times the Kullback--Leibler divergence of its policy from the reference policy. The maximand is strictly concave in $\mu$, and the first-order conditions give the policy $\mu_\tau(a\mid p)$ that is proportional to $r(a)\exp(S(a,p)/\tau)$. Let $c_i=\log\bigl(r(a_{i+1})/r(a_i)\bigr)$, and let
\[
\ell_\tau(p)=\sum_a\mu_\tau(a\mid p)\,\bigl[V^*(p)-V(a,p)\bigr]
\]
be the principal's expected loss from the AI's action at the belief $p$.

\begin{suppclaim}\label{claim:penalty}
In this setting, let $S$ be aligned on the range with scales $\lambda_i$.
\begin{enumerate}[label=(\alph*)]
\item For every $i$ and every $p\in[0,1]$, $\mu_\tau(a_{i+1}\mid p)/\mu_\tau(a_i\mid p)=\exp\bigl(c_i+\lambda_i\,d_i\,(p-t_i)/\tau\bigr)$. The AI is therefore as likely to take $a_{i+1}$ as $a_i$ at the belief $t_i-\tau\,c_i/(\lambda_i\,d_i)$, when that belief lies in $[0,1]$, and more likely at higher beliefs.
\item If the AI's belief has a continuous density $f$ on $[0,1]$, then
\[
\lim_{\tau\to0}\ \frac1{\tau^2}\int_0^1f(p)\,\ell_\tau(p)\,dp=\sum_{i=1}^{n-1}\frac{f(t_i)}{\lambda_i^2\,d_i}\Bigl(\frac{\pi^2}{6}+\frac{c_i^2}{2}\Bigr).
\]
\item Let $f(t_i)>0$ for every $i$, and let $w_i>0$ and $B>0$ be given. Among the vectors of positive scales with $\sum_iw_i\lambda_i=B$, the limit in part (b) is smallest when $\lambda_i$ is proportional to $\bigl[f(t_i)\,(\pi^2/3+c_i^2)/(w_i\,d_i)\bigr]^{1/3}$. These scales are equal across margins only if the numbers in brackets are.
\end{enumerate}
\end{suppclaim}

\begin{proof}
\emph{Part (a).} The ratio of the two probabilities is $[r(a_{i+1})/r(a_i)]\exp\bigl([S(a_{i+1},p)-S(a_i,p)]/\tau\bigr)$, which is the expression stated. It equals one where $c_i+\lambda_i\,d_i\,(p-t_i)/\tau=0$ and is increasing in $p$.

\emph{Part (b).} Let $\rho$ be the largest ratio $r(a)/r(a')$ over pairs of actions, let $\bar v$ be the largest value of $V^*(p)-V(a,p)$ over actions and beliefs, and fix $\eta>0$ smaller than half of the distance between any two of the points $0,t_1,\dots,t_{n-1},1$. Let $N_i=[t_i-\eta,t_i+\eta]$, and let $C$ be the set of beliefs in $[0,1]$ that lie in the interior of no $N_i$.

On $C$, a compact set that contains no switching belief, the principal has a single optimal action $a^*(p)$ at each belief, and it is the only action that maximizes $S(\cdot,p)$ (Theorem~\ref{thm:main}(a)). By continuity there is $\kappa>0$ with $S(a^*(p),p)-S(a,p)\geq\kappa$ for every $p\in C$ and every $a\neq a^*(p)$. Then $\mu_\tau(a\mid p)\leq\rho\,e^{-\kappa/\tau}$ for such $a$, because $\mu_\tau(a\mid p)\leq\mu_\tau(a\mid p)/\mu_\tau(a^*(p)\mid p)$, and so $\ell_\tau(p)\leq n\,\rho\,\bar v\,e^{-\kappa/\tau}$ on $C$. The integral of $f\,\ell_\tau$ over $C$ is therefore at most $n\,\rho\,\bar v\,e^{-\kappa/\tau}$, which is of smaller order than $\tau^2$.

Fix $i$. On $N_i$, the principal's optimal action is $a_i$ at beliefs below $t_i$ and $a_{i+1}$ at beliefs above, and no action other than these two maximizes $S(\cdot,p)$. By continuity there is $\kappa_i>0$ with $\max\{S(a_i,p),S(a_{i+1},p)\}-S(a,p)\geq\kappa_i$ for every $p\in N_i$ and every $a\notin\{a_i,a_{i+1}\}$, so the total probability $\epsilon_\tau(p)$ of such actions is at most $n\,\rho\,e^{-\kappa_i/\tau}$. Write $s=\lambda_i\,d_i\,(p-t_i)/\tau$. The loss from the wrong one of the two actions is $d_i\,|p-t_i|=\tau\,|s|/\lambda_i$. By part (a), that action has the share $h_i(s)$ of the probability $1-\epsilon_\tau(p)$ of the two, where $h_i(s)=1/(1+e^{c_i+s})$ for $s>0$, when it is $a_i$, and $h_i(s)=1/(1+e^{-c_i-s})$ for $s<0$, when it is $a_{i+1}$. Hence, for $p\in N_i$,
\[
\ell_\tau(p)=\frac{\tau\,|s|}{\lambda_i}\,\bigl(1-\epsilon_\tau(p)\bigr)\,h_i(s)+R_\tau(p),\qquad0\leq R_\tau(p)\leq n\,\rho\,\bar v\,e^{-\kappa_i/\tau}.
\]
The integral of $f\,R_\tau$ over $N_i$ is of smaller order than $\tau^2$. Changing the variable of integration from $p$ to $s$, with $dp=\tau\,ds/(\lambda_i\,d_i)$,
\[
\int_{N_i}f(p)\,\frac{\tau\,|s|}{\lambda_i}\,\bigl(1-\epsilon_\tau(p)\bigr)\,h_i(s)\,dp=\frac{\tau^2}{\lambda_i^2\,d_i}\int_{-\eta\lambda_id_i/\tau}^{\eta\lambda_id_i/\tau}f\Bigl(t_i+\frac{\tau s}{\lambda_i d_i}\Bigr)\bigl(1-\epsilon_\tau\bigr)\,|s|\,h_i(s)\,ds.
\]
Since $h_i(s)\leq e^{|c_i|-|s|}$, the integrand on the right is at most $\max f$ times $|s|\,e^{|c_i|-|s|}$, which is integrable on $\R$, and as $\tau\to0$ it converges at each $s$ to $f(t_i)\,|s|\,h_i(s)$, because $f$ is continuous and $\epsilon_\tau\to0$. By dominated convergence the integral on the right converges to $f(t_i)\,J(c_i)$, where
\[
J(c)=\int_{-\infty}^\infty|s|\,h(s)\,ds=\int_0^\infty\frac{s}{1+e^{s+c}}\,ds+\int_0^\infty\frac{s}{1+e^{s-c}}\,ds,
\]
with $h$ defined as $h_i$ is but with $c$ in place of $c_i$. Summing over $i$ and adding the integral over $C$ gives the limit $\sum_if(t_i)J(c_i)/(\lambda_i^2d_i)$.

It remains to show that $J(c)=\pi^2/6+c^2/2$. Let $F(u)=1/(1+e^u)$, so that $J(c)=\int_0^\infty s\,[F(s+c)+F(s-c)]\,ds$ and $F(u)+F(-u)=1$. Differentiating under the integral sign, which is justified because the derivatives of the integrand are bounded by an integrable function of $s$ on every bounded set of values of $c$, and then integrating by parts,
\[
J'(c)=\int_0^\infty s\,\bigl[F'(s+c)-F'(s-c)\bigr]\,ds=-\int_0^\infty F(s+c)\,ds+\int_0^\infty F(s-c)\,ds=\int_{-c}^{c}F(u)\,du=c,
\]
where the boundary terms vanish because $s\,F(s\pm c)\to0$ as $s\to\infty$, and the last step uses $F(u)+F(-u)=1$. Moreover $J(0)=2\int_0^\infty s/(1+e^s)\,ds=2\sum_{k\geq1}(-1)^{k+1}/k^2=\pi^2/6$, by expanding $1/(1+e^s)=\sum_{k\geq1}(-1)^{k+1}e^{-ks}$ and integrating term by term. Hence $J(c)=\pi^2/6+c^2/2$.

\emph{Part (c).} Write $\Gamma_i=f(t_i)\,(\pi^2/6+c_i^2/2)/d_i>0$, so that the limit is $\sum_i\Gamma_i/\lambda_i^2$, a strictly convex function of the scales that grows without bound as any scale tends to zero. On the set of positive scales with $\sum_iw_i\lambda_i=B$ it therefore has a single minimizer, at which $-2\Gamma_i/\lambda_i^3+\nu\,w_i=0$ for a multiplier $\nu>0$, that is, $\lambda_i=(2\Gamma_i/(\nu\,w_i))^{1/3}$.
\end{proof}

\noindent Part (a) says that a reference policy that takes a cautious action less often than the actions on either side of it narrows the set of beliefs at which the AI is most likely to be cautious, from both sides, and by more where the scale is small. Part (b) says that the principal's loss from the AI's actions is of second order in the weight on the penalty and inversely proportional to the square of each scale, so the scales that decisions leave free matter for acting. Part (c) says that the scales that are best for acting under such a penalty, for a given size of the objective, are in general not the principal's, which are the scales that choosing evidence requires when the AI maximizes its objective.

\begingroup
\setstretch{1}\setlength{\bibsep}{0.2em plus 0.1em}\def\sloppy{\tolerance 9999 \emergencystretch 3em \hfuzz 1pt \vfuzz\hfuzz}\sloppy
\renewcommand{\refname}{References for the Supplemental Appendix}
\putbib[paper2A-v18-references]
\endgroup
\label{pg:suppend}
\clearpage
\end{bibunit}

\end{document}